\documentclass[12pt,preprint]{aastex}

\begin{document}

\title{Identification, Classifications, and Absolute Properties of\\ 773 Eclipsing Binaries Found in the TrES Survey}

\author{Jonathan Devor\altaffilmark{1,2}, David Charbonneau\altaffilmark{1,3},\\ Francis T. O'Donovan\altaffilmark{4}, Georgi Mandushev\altaffilmark{5}, and Guillermo Torres\altaffilmark{1}}

\altaffiltext{1}{Harvard-Smithsonian Center for Astrophysics, 60 Garden Street, Cambridge, MA 02138}
\altaffiltext{2}{Email: jdevor@cfa.harvard.edu}
\altaffiltext{3}{Alfred P. Sloan Research Fellow}
\altaffiltext{4}{California Institute of Technology, 1200 East California Boulevard, Pasadena, CA 91125}
\altaffiltext{5}{Lowell Observatory, 1400 West Mars Hill Road, Flagstaff, AZ 86001}

\begin{abstract}

In recent years we have witnessed an explosion of photometric
time-series data, collected for the purpose of finding a small
number of rare sources, such as transiting extrasolar planets and
gravitational microlenses. Once combed, these data are often set
aside, and are not further searched for the many other variable
sources that they undoubtedly contain. To this end, we describe a
pipeline that is designed to systematically analyze such data,
while requiring minimal user interaction. We ran our pipeline on a
subset of the Trans-Atlantic Exoplanet Survey dataset, and used it
to identify and model 773 eclipsing binary systems. For each
system we conducted a joint analysis of its light curve, colors,
and theoretical isochrones. This analysis provided us with
estimates of the binary's absolute physical properties, including
the masses and ages of their stellar components, as well as their
physical separations and distances. We identified three types of
eclipsing binaries that are of particular interest and merit
further observations. The first category includes 11 low-mass
candidates, which may assist current efforts to explain the
discrepancies between the observation and the models of stars at
the bottom of the main-sequence. The other two categories include
34 binaries with eccentric orbits, and 20 binaries with abnormal
light curves. Finally, this uniform catalog enabled us to identify
a number of relations that provide further constraints on binary
population models and tidal circularization theory.

\end{abstract}


\section{Introduction}

Since the mid 1990s there has been an explosion of large scale
photometric variability surveys. The search for gravitational
microlensing events, which were predicted by \citet{Paczynski86},
motivated the first wave of surveys [e.g. OGLE: \citet{Udalski94};
EROS: \citet{Beaulieu95}; DUO: \citet{Alard97}; MACHO:
\citet{Alcock98}]. Encouraged by their success, additional
surveys, searching for Gamma-Ray Bursts [e.g. ROTSE:
\citet{Akerlof00}] and general photometric variabilities [e.g.
ASAS: \citet{Pojmanski97}] soon followed.

Shortly thereafter, with the discovery of the first transiting
extrasolar planet \citep{Charbonneau00, Henry00, Mazeh00}, a
second wave of photometric surveys ensued [e.g. OGLE-III:
\citet{Udalski03}; TrES: \citet{Alonso04}; HAT: \citet{Bakos04};
SuperWASP: \citet{Christian06}; XO: \citet{McCullough06}; for
 a review, see \citet{Charbonneau07}]. Each
of these projects involved intensive efforts to locate a few
proverbial needles hidden in a very large data haystack. With few
exceptions, once the ``needles'' were found, thus fulfilling the
survey's original purpose, the many gigabytes of photometric light
curves (LCs) collected were not made use of in any other way. In
this paper we demonstrate how one can extract a great deal more
information from these survey datasets, with comparably little
additional effort, using automated pipelines. To this end, we have
made all the software tools described in this paper freely
available (see web links to the source code and working examples),
and they are designed to be used with any LC dataset.

In the upcoming decade, a third wave of ultra-large ground-based
synoptic surveys [e.g. Pan-STARRS: \citet{Kaiser02}; LSST:
\citet{Tyson02}], and ultra-sensitive space-based surveys [e.g.
KEPLER: \citet{Borucki97}; COROT: \citet{Baglin98}; GAIA:
\citet{Gilmore98}] are expected to come online. These surveys are
designed to produce photometric datasets that will dwarf all
preceding efforts. To make any efficient use of such large
quantities of data, it will become imperative to have in place a
large infrastructure of automated pipelines for performing even
the most casual data mining query.

In this paper, we focus exclusively on the identification and
analysis of eclipsing binary (EB) systems. EBs provide favorable
targets, as they are abundant and can be well modeled using
existing modeling programs [e.g. WD: \citet{Wilson71}; EBOP:
\citet{Popper81}]. Once modeled, EBs can provide a wealth of
useful astrophysical information, including constraints on binary
component mass distributions, mass-radius-luminosity relations,
and theories describing tidal circularization and synchronization.
These findings, in turn, will likely have a direct impact on our
understanding of star formation, stellar structure, and stellar
dynamics. These physical distributions of close binaries may even
help solve open questions relating to the progenitors of Type Ia
supernovae \citep{Iben84}. In additional to these, EBs can be used
as tools; both as distance indicators \citep{Stebbing10,
Paczynski97} and as sensitive detectors for tertiary companions
via eclipse timing \citep{Deeg00, Holman05, Agol05}.

In order to transform such large quantities of data into useful
information, one must construct a robust and computationally
efficient automated pipeline. Each step along the pipeline will
either measure some property of the LC, or filter out LCs that do
not belong, so as to reduce the congestion in the following, more
computationally intensive steps. One can achieve substantial gains
in speed by dividing the data into subsets, and processing them in
parallel on multiple CPUs. The bottlenecks of the analysis are the
steps that require user interaction. In our pipeline, we reduce
user interaction to essentially yes/no decisions regarding the
success of the EB models, and eliminate any need for interaction
in all but two stages. We feel that this level of interaction
provides good quality control, while minimizing its detrimental
subjective effects.

The data that we analyzed originate from 10 fields of the
Trans-atlantic Exoplanet Survey [TrES ; \citet{Alonso04}]. TrES
employs a network of three automated telescopes to survey $6^\circ
\times 6^\circ$ fields-of-view. To avoid potential systematic
noise we use the data from only one telescope, Sleuth, located at
the Palomar Observatory in Southern California \citep{ODonovan04}.
This telescope has a 10 cm physical aperture and a photometric
aperture of radius of 30". The number of LCs in each field ranges
from 10405 to 26495 (see Table \ref{tableFieldsObs}), for a total
of 185445 LCs. The LCs consist of $\sim$2000 $r$-band photometric
measurements at a 9-minute cadence. These measurements were
created by binning the image-subtraction results of 5 consecutive
90-second observations, thus improving their non-systematic
photometric noise. As a result $\sim$16\% of the LCs have an RMS
$<$1\%, and $\sim$38\% of the LCs have an RMS $<$2\% (see Table
\ref{tableFieldsYield}). The calibration of TrES images,
identification of stars therein, extraction, and decorrelation of
the LCs is described elsewhere \citep{Dunham04, Mandushev05,
ODonovan06, ODonovan07}. TrES is currently an active survey that
is continuously observing new fields, though for this paper we
have limited ourselves to these 10 fields.

\section{Method}
\label{secMethod}

The pipeline we have developed is an extended version of the
pipeline described by \citet{Devor05}. At the heart of this
analysis lie two computational routines that we have described in
earlier papers: the Detached Eclipsing Binary Light curve
fitter\footnote{The DEBiL source code, utilities, and running
example files are available online at:\newline
http://www.cfa.harvard.edu/$\sim$jdevor/DEBiL.html} [DEBiL ;
\citet{Devor05}], and the Method for Eclipsing Component
Identification\footnote{The MECI source code and running examples
are available online at:\newline
http://www.cfa.harvard.edu/$\sim$jdevor/MECI.html} [MECI ;
\citet{Devor06a, Devor06b}]. DEBiL fits each LC to a
$\it{geometric}$ model of a detached EB (steps 3 and 5 below).
This model consists of two luminous, limb-darkened spheres that
orbit in a Newtonian 2-body orbit. MECI restricts the DEBiL fit
along theoretical isochrones, and is thus able to create a
$\it{physical}$ model of each EB (step 9). This second model
describes the masses and absolute magnitudes of the EB's stellar
components, which are then used to determine the EB's distance and
absolute separation.

The pipeline consists of 10 steps. We elaborate on each of these
steps below:

\begin{enumerate}
\item Determine the period.
\item If a distinct secondary eclipse is not observed, an entry with twice the period is added.
\item Fit the orbital parameters with DEBiL.
\item Fine-tune the period using eclipse timing.
\item Refine the orbital parameters with DEBiL using the revised period.
\item Remove contaminated LCs.
\item Visually assess the quality of the EB models.
\item Match the LC sources with external databases.
\item Estimate the absolute physical properties of the binary components using MECI.
\item Classify the resulting systems using both automatic and manual criteria.
\end{enumerate}

We use the same filtering criteria as described in
\citet{Devor05}, both for removing LCs that are not periodic (step
1) and then for removing non-EB LCs (step 3). Together, these
automated filters remove approximately 97\% of the input LCs. In
addition to these filters, we perform stringent manual inspections
(steps 7 and 10) whereby we removed all the LCs we were not
confident were EBs. These inspections ultimately removed
approximately 86\% of the remaining LCs. Thus only 1 out of every
240 input LCs, were included in the final catalog.

In step (1) we use both the Box-fitting Least Squares (BLS) period
finder \citep{Kovacs02}, and a version of the analysis of
variances (AoV) period finder \citep{SchwarzenbergCzerny89,
SchwarzenbergCzerny96} to identify the periodic LCs within the
dataset and to measure their periods. In our AoV implementation,
we scan periods from 0.1 days up to the duration of each LC. We
then select the period that minimizes the variance of a linear fit
within 8 phase bins. We removed all systems with weak
periodicities [see \citet{Devor05} for details], and with one
exception (T-Lyr1-14413), all the systems whose optimal period was
found to be longer than half their LC duration. In this way we
were able to filter out many of the non-periodic variables.

\begin{figure}
\plotone{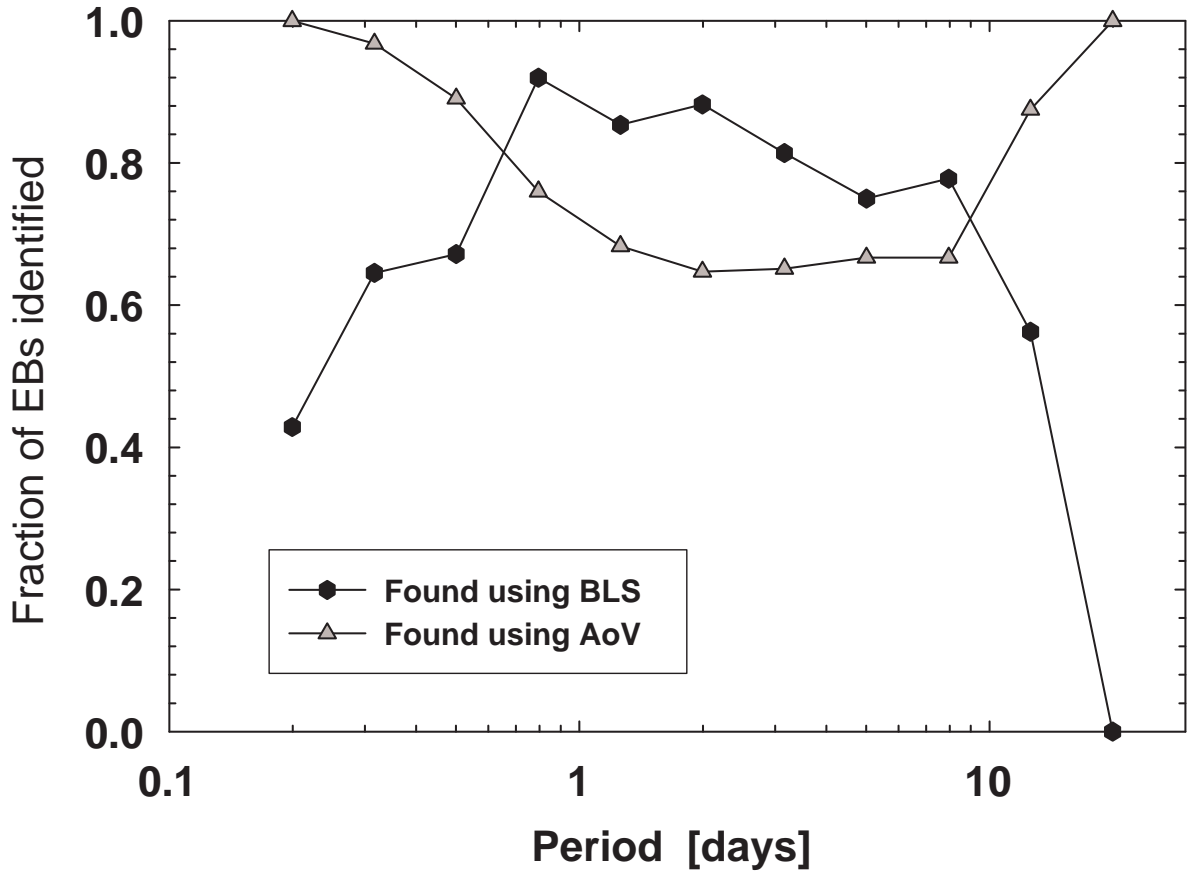}
\caption{The fraction of the EBs in the final catalog found using
the BLS algorithm and the AoV algorithm. The number
of EBs in each bin is shown in Figure \ref{figPeriodDistrib}. The
BLS method excelled at identifying EBs with short-duration eclipses
(compared to the orbital period), which predominately occur at periods
 $> 0.75$ days. The AoV method fared better with EBs that have
long-duration eclipses, which predominately occur in sub-day periods. The AoV
method also does well with EBs with period longer than $10$ days,
which may be dominated by giant-giant binaries \citep{Derekas07}, and
so also have broad eclipses. This plot demonstrates the importance
of using multiple independent methods of identifying EB, otherwise
the results will have a significant selection effect that may
bias any statistical results.}
\label{periodDiffFrac}
\end{figure}

The AoV algorithm is most effective in identifying the periods of
LCs with long duration features, such as semi-detached EBs and
pulsating stars. The BLS algorithm, in contrast, is effective at
identifying periodic systems whose features span only a brief
portion of the period, such as detached EBs and transiting planets
(see Figure \ref{periodDiffFrac}). However, the BLS algorithm is
easily fooled by outlier data points, identifying them as short
duration features. For this reason the BLS algorithm has a
significantly higher rate of false positives than AoV, especially
for long periods, which have only a few cycles over the duration
of the observations. Therefore we limit the search range of the
BLS algorithm to periods shorter than 12 days, although as Figure
\ref{periodDiffFrac} illustrates, its efficiency at locating EBs
rapidly declines at periods greater than 10 days.

\placefigure{periodDiffFrac}

In step (2), we address the ambiguity between EBs with identical
components in a circular orbit, and EBs with extremely disparate
components. The phased LC of EBs with identical components
contains two identical eclipses, whereas the phased LC of EBs with
disparate components will have a secondary eclipse below the
photometric noise level. These two cases are degenerate, since
doubling the period of a disparate system will result in a LC that
looks like an equal-component system. In the pipeline, we handle
this problem by doubling such entries; one with the period found
in step (1), and another with twice that period. Both of these
entries proceed through the pipeline independently. In many cases,
after additional processing by the following steps, one of these
entries will emerge as being far less likely than the other (see
appendix \ref{appendixSingleEclipse}), at which point it is
removed. But in cases where photometry alone cannot determine
which is correct, one needs to perform spectroscopic follow up to
break the ambiguity. In particular, a double-lined spectrum would
support the equal-component hypothesis.

Step (3) is performed using DEBiL, which fits the fractional radii
($r_{1,2}$) and observed magnitudes ($mag_{1,2}$) of the EB's
stellar components, their orbital inclination ($i$) and
eccentricity ($e$), and their epoch ($t_0$) and argument of
periastron ($\omega$). DEBiL first produces an initial guess for
these parameters, and then iteratively improves the fit using the
downhill simplex method \citep{Nelder65} with simulated annealing
\citep{Kirkpatrick83, Press92}.

\begin{figure}
\plotone{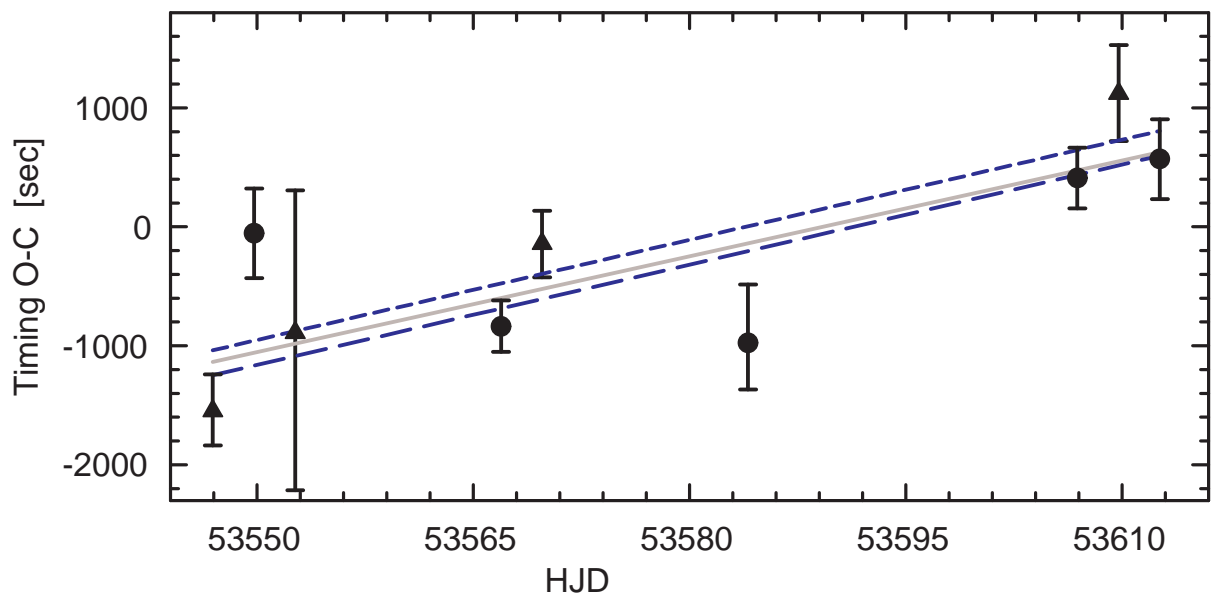}
\caption{An eclipse timing plot produced
in step (4), showing the $O-C$ residuals of the primary eclipses
(circles) and the secondary eclipses (triangles). Here, T-Lyr1-14962
is shown with an assumed period of $5.710660\: days$, as measured
with an AoV periodogram. The slope of the residuals indicates that
the assumed period is inaccurate. The grey solid line is predicted
by the best circular-orbit model, whereas the dashed lines are
predicted by the best eccentric-orbit model (compare to Figure
\ref{figTimingVariationsEcc}). After correction, we get a
fine-tuned period of $5.712516\: days$. This 0.03\% correction is
small but significant in that without having had this correction,
the eclipses would have smeared out and widened.}
\label{figTimingVariations}
\end{figure}

In step (4), we fine-tune the period ($P$) using a method based on
eclipse timing\footnote{The source code and running examples are
available online at\newline
http://www.cfa.harvard.edu/$\sim$jdevor/Timing.html}, which we
describe below. In order to produce an accurate EB model in step
(9) it is necessary to know the system's period with greater
accuracy than that produced in step (1). If we neglect to
fine-tune the period, the eclipses may be out of phase with
respect to one another, and so the phased eclipses will appear
broadened. Our timing method employs the DEBiL model produced in
step (3), and uses it to find the difference between the observed
and calculated ($O-C$) eclipse epochs. This is done by minimizing
the chi-squared fit of the model to the data points in each
eclipse, while varying only the model's epoch of periastron. When
the period estimate is off by a small quantity ($\Delta P$), the
$O-C$ difference increases by $\Delta P$ each period. This change
in the $O-C$ over time can be measured from the slope of the
linear regression, which is expected to equal $\Delta P / P$. Thus
measuring such an $O-C$ slope will yield the desired period
correction (see Figure \ref{figTimingVariations}).

\begin{figure}
\plotone{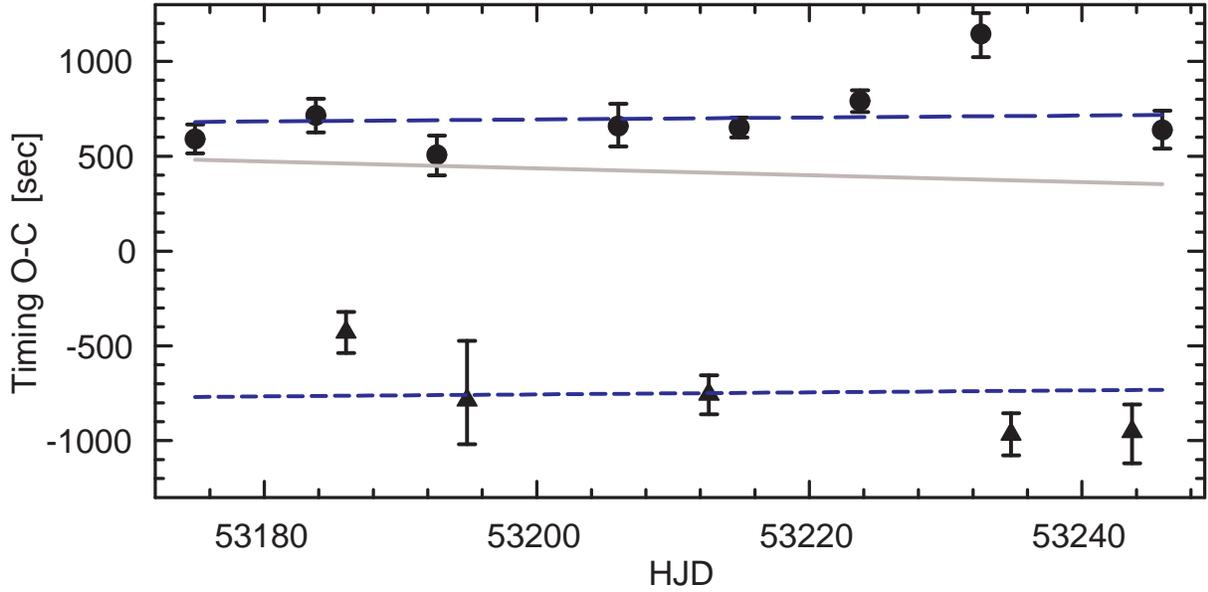}
\caption{An eclipse timing plot for
T-Cyg1-01373, with an assumed period of $4.436013\: days$. In contrast
to Figure \ref{figTimingVariations}, the slope here is consistent
with zero, thus indicating that the period does not need to be
fine-tuned. However, the $O-C$ offset between the primary
(circles) and secondary (triangle) eclipses is significant
($1449\: seconds$), indicating that this EB has an eccentric
orbit. The reduced chi-squared of the best circular-orbit model
(grey solid line) is $\chi^2_\nu = 12.9$, while the reduced
chi-squared of the best eccentric-orbit model (dashed lines) is
$\chi^2_\nu = 0.95$. Applying the $O-C$ timing offset to equation
\ref{eqOmC} provides a lower limit to the binary's orbital
eccentricity: $e \geq | e \cos \omega | \simeq 0.00594$.}
\label{figTimingVariationsEcc}
\end{figure}

If the EB has an eccentric orbit, the primary and secondary
eclipse will separate on the $O-C$ plot, and form two parallel
lines with a vertical offset of $\Delta t$ (see Figure
\ref{figTimingVariationsEcc}). We measure this offset and use it
as a sensitive method to detect orbital eccentricities. In
particular, the value of $\Delta t$ constrains $e \cos \omega$,
which in turn provides a lower limit for the system's eccentricity
\citep{Tsesevich73}:

\begin{equation}
\label{eqOmC}
e \cos \omega \simeq \frac{\pi}{2} \frac{\Delta t}{P}
\end{equation}

This formula assumes an orbital inclination of $i=90^\circ$,
making it a good approximation for eclipsing binaries. We use this
method, in combination with DEBiL, to identify the eccentric EBs
in the catalog (see Table \ref{tableEccentric}). However, in cases
where the eclipse timing measures $|e \cos \omega| < 0.005$, or
when the eccentricity is consistent with zero, we assume that the
EB is non-eccentric, and model it using a circular orbit. We
further discuss the physics of these systems in
\S\ref{subsecEccentricEBs}.

\placefigure{figTimingVariations}
\placefigure{figTimingVariationsEcc}

Step (5) is identical to step (3), except that it uses the revised
period from step (4). This step provides an improved fit to the
LCs, as evidenced by an improved chi-squared value in over 70\% of
the cases.

In step (6) we locate and remove non-EB sources that seem to be
periodic due to photometric contamination by true EBs. Such
contaminations result from overlapping point spread functions
(PSF) that cause each source to partially blend into the other.
These cases can be easily identified with a program that scans
through pairs of targets\footnote{We ran a brute force scan, which
required O($N^2$) iterations. But by employing a data structure
that can restrict the scan to nearby pairs, it is possible to
perform this scan in only O($N$) iterations, assuming that such
pairs are rare.}, and selects the ones that both have similar
periods (see description below) and are separated by an angle that
is smaller than twice the PSF. We found 14 such pairs, all of
which were separated by less than $41"$, which is well within
twice the TrES PSF ($60"$), while the remaining pairs with similar
periods were separated by over $450"$. Upon inspection, all 14 of
the pairs we found had similar eclipse shapes, indicating that we
had no false positives. Among each pair, we identify the LC with
shallower eclipses (in magnitudes) as being contaminated and
remove it from the catalog.

We define periods as being similar if the difference between them
is smaller than their combined uncertainty. We estimate the period
uncertainty using the relation: $\varepsilon_P \propto P^2 / T$,
where $T$ is the time interval between the initial and the final
observations. One arrives at this relation by noticing that when
phasing the LC, the effect of any perturbation from the true
period will grow linearly with the number of periods in the LC
(see step 4). This amplified effect will become evident once it
reaches some fraction of the period itself, in other words, when
$\varepsilon_P (T / P) \propto P$. A typical TrES LC with a
revised period will have a proportionality constant of
approximately $1/1000$. In order to avoid missing contaminated
pairs (false negatives) we adopt in this step, the extremely
liberal proportionality constant of unity.

In step (7) we conduct a visual inspection of all the LC fits.
Most EBs were successfully modeled and were included into the
catalog as is. About $1\%$ of the LCs analyzed had misidentified
periods, as a result of failures of the period finding method of
step (1). In most of these cases the period finder indicated
either a harmonic of the true period or a rational multiple of a
solar or sidereal day. In such cases we use an interactive
periodogram\footnote{LC, created by Grzegorz Pojmanski.} to find
the correct period and then reprocess the LCs through the
pipeline. Some entries were misidentified at step (2) as being
ambiguous, even though they have a detectable secondary eclipse or
have slightly unequal eclipses. In these cases the erroneous
doubled entry was removed. Lastly, some of the EBs were not fit
sufficiently well with DEBiL in step (5). These cases were
typically due to clustered outlier data points, systematic noise,
or severe activity of a stellar component (e.g. flares or spots),
which caused DEBiL to produce erroneous initial model parameters.
These cases were typically handled by having DEBiL produce the
initial model parameters from a more smoothed version of the LC.

In step (8) we match each system, through its coordinates, with
the corresponding source in the Two Micron All Sky Survey catalog
[2MASS ; \citep{Skrutskie06}]. This was done to obtain both
accurate target positions and observational magnitudes. These
magnitude measurements are then used to derive the colors of each
EB, which are incorporated into the MECI analysis, as well as to
estimate the EB's distance modulus (step 9). To this end, 2MASS
provides a unique combination of high astrometric accuracy
($\sim$0.1") together with high photometric accuracy ($\sim$0.015
mag) at multiple near-infrared bands, all while maintaining a
decent photometric resolving power ($\sim$3"). By employing these
near-infrared bands we both inherently reduce the detrimental
effects of stellar reddening, and are able to correct for much of
the remaining extinction by fitting for the Galactic interstellar
absorbtion.

In order to use the measurements from the 2MASS custom $J$, $H$,
and $K_s$ filters, we converted them to the equivalent ESO-filter
values so that they could be compared to the isochrone table
values used in the MECI analysis. This conversion was done using
approximate linear transformations \citep{Carpenter01}. However,
the colors of three EBs (T-And0-10336, T-Cyg1-02304, and
T-Per1-05205) were so anomalous that they did not permit a
reasonable model solution, thus we chose not to include any color
information in their MECI analyses.

\begin{figure}
\plotone{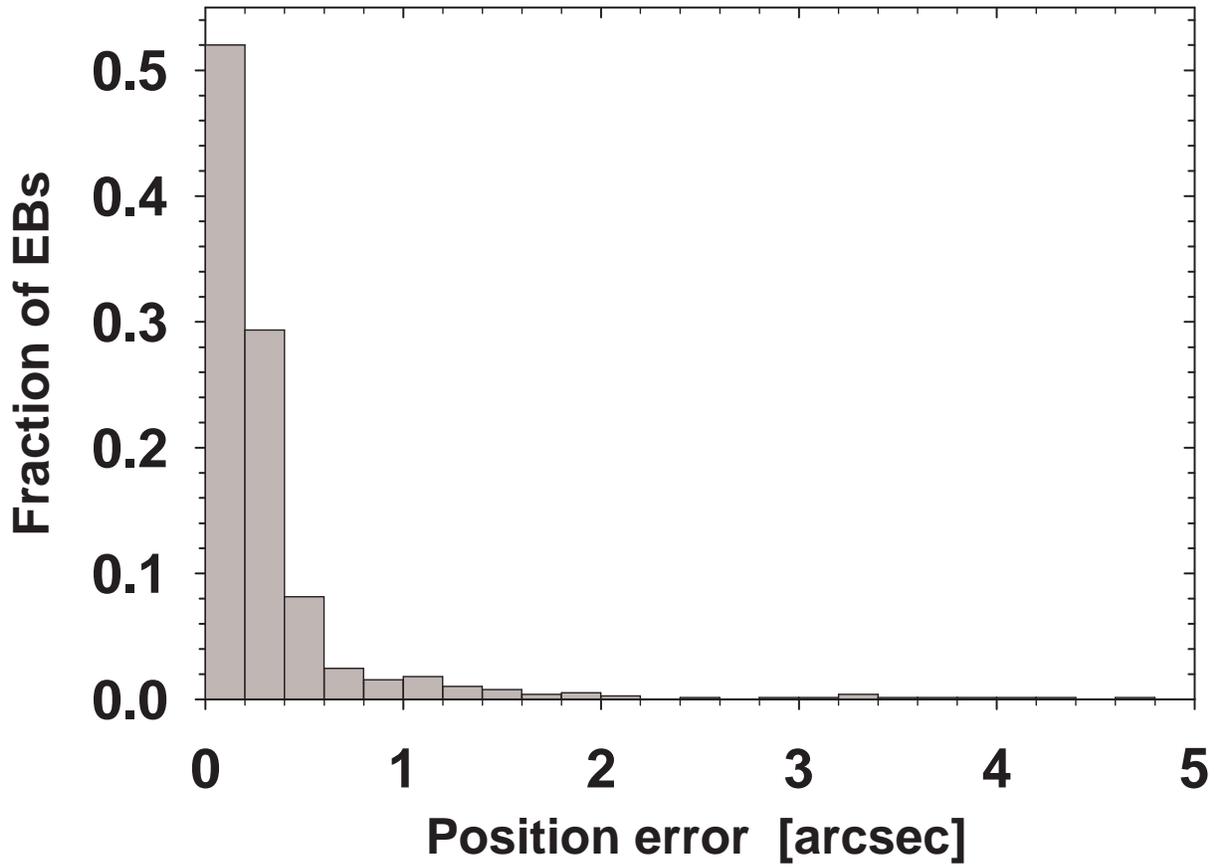}
\caption{The distribution of the catalog position errors when
matching targets to the proper motion databases. In some cases,
the position errors are dominated by the
motion of the EB during the intervening years.}
\label{figPosErr}
\end{figure}

In addition to its brightness, we also look up each EB's proper
motion. Although proper motion is not required for any of the
pipeline analyses, it provides a useful verification for low-mass
candidates (see \S\ref{subsecLowMassEBs}). These systems are
expected to have large proper motions, since they must be nearby
to be observable in this magnitude-limited survey. The most
extreme such case in the catalog is CM Draconis (T-Dra0-01363),
which has a proper motion of over 1300 mas/yr \citep{Salim03}, and
is probably the lowest mass system in our catalog. To this end, we
match each system to the Second U.S. Naval Observatory CCD
Astrograph Catalog [UCAC release 2.4 ; \citet{Zacharias04}]. When
there was no match with UCAC, we use the more comprehensive but
less accurate U.S. Naval Observatory photographic sky survey
[USNO-B release 1.0 ; \citet{Monet03}]. These matches were made
using the more accurate aforementioned adopted 2MASS coordinates.
However, because of their increased observational depth, and the
fact that some high-proper motion targets are expected to have
moved multiple arcseconds in the intervening decades, we chose to
match each target to the brightest ($R$-band) source within 7.5".
It should be noted that the position of CM Draconis shifted by
more than 22" and had to be matched manually, though 90\% of the
matches were separated by less than 0.6", and 98\% were separated
by less than 2" (see Figure \ref{figPosErr}).

The proper motions garnered from these databases can be combined
with distance estimates ($D$), to calculate the absolute
transverse velocity ($v_{tr}$) of a given EB:

\begin{equation}
v_{tr} \simeq 4.741\: km/sec \left(\frac{PM}{1\: mas/yr}\right)\left(\frac{D}{1\: kpc}\right),
\end{equation}

where $PM$ is the system's angular proper motion. In the catalog
we list the right ascension and declination components
($PM_{\alpha}$ and $PM_{\delta}$, respectively), so as to allow
one to compute the system's direction of motion in the sky. The
value of $PM$ can be computed from its components, using: $PM^2 =
PM_{\delta}^2 + PM_{\alpha}^2 \cos^2 \delta$, where $\delta$ is
the system's declination. When applying this formula, one should
be aware that USNO-B folds the $\cos \delta$ coefficient into its
listed $PM_{\alpha}$, while UCAC does not.

\begin{figure}
\plotone{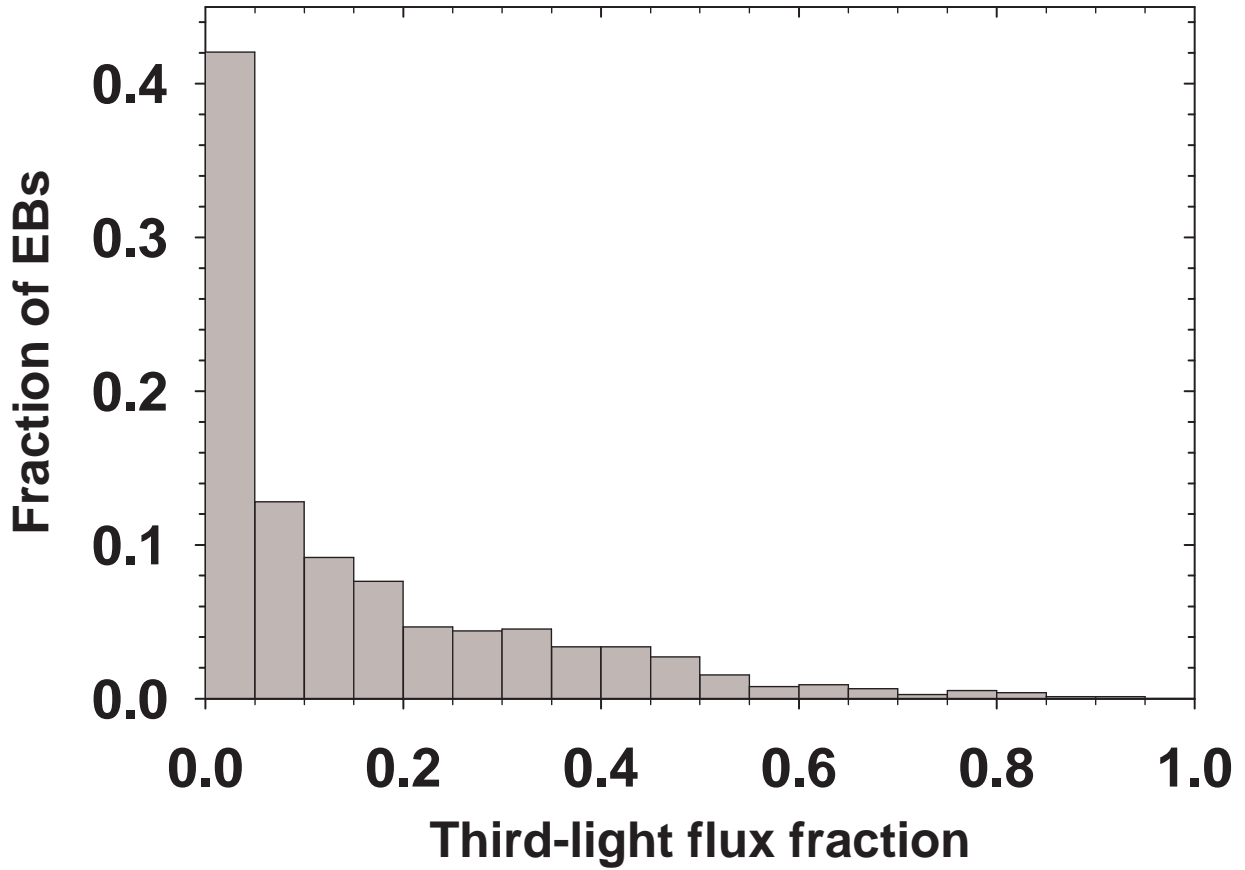}
\caption{The distribution of the $R$-band third-light flux fraction in
the catalog LCs. This fraction was calculated by summing
the fluxes of all the USNO-B sources within 30" of the target, excluding the
target, and dividing this value by the total flux within 30", including the
target. The resulting fraction ranges from 0 to 1.}
\label{figBlending}
\end{figure}

Finally, we incorporate the USNO-B photometric $B$- and
$R$-magnitude measurements into our catalog to provide a rough
estimate of the optical brightness of each target. USNO-B lists
two independent measurements in each of these filter, however in
some cases one or both of these measurements failed. When both
measurements are available, we average them for improved accuracy.
However, each measurement has a large photometric uncertainty of
$\sim$0.3 mag, thus even these averaged values will have errors
that are over an order of magnitude larger that the photometric
measurements of 2MASS. For this reason, and because of the
increased effect of stellar reddening, we chose not to incorporate
these data into the MECI analysis. However, USNO-B's high
photometric resolution ($\sim$1") enabled us to detect many
sources that blended with our targets in the TrES exposures. By
summing the $R$-band fluxes of all the USNO-B sources within 30"
of each target, we estimated the fraction of third-light included
in each LC (see Figure \ref{figBlending}). Note that this measure
provides only a lower bound to the true third-light fraction, as
some EBs are expected to have additional close hierarchical
components that would not be resolved by USNO-B. For most of the
catalog targets, the third-light flux fraction was found to be
small ($<$10\%). We therefore conclude that stellar blending will
usually have only a minor effect on the MECI analysis results,
however users should be aware of the potential biases in the
calculated properties of highly blended targets. Though it was not
applied to this catalog, in principal, given a third-light flux
fraction at a well-determined LC phase, one could correct for the
effects of blending.

\placefigure{figPosErr}
\placefigure{figBlending}

\begin{figure}
\plottwo{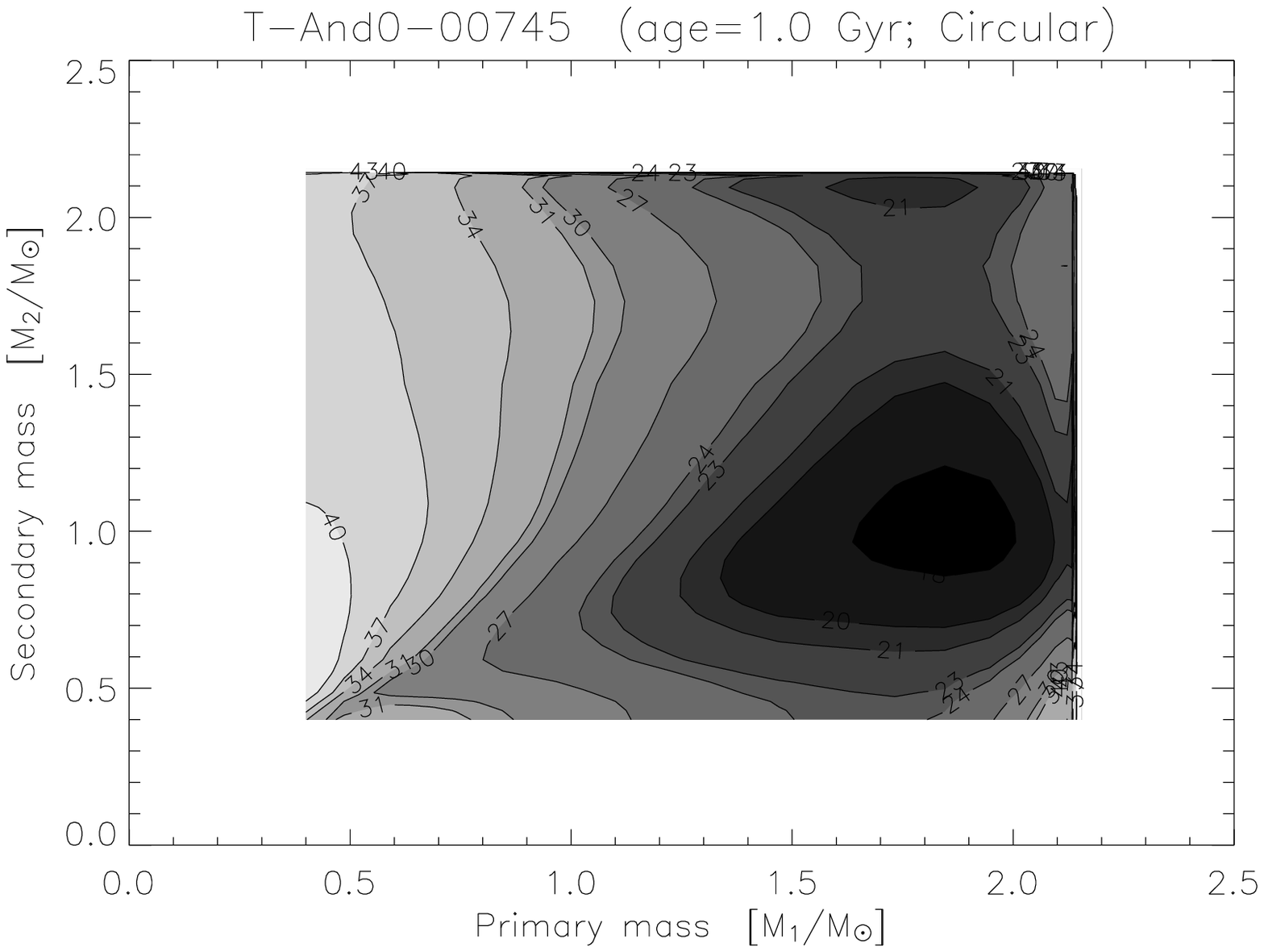}{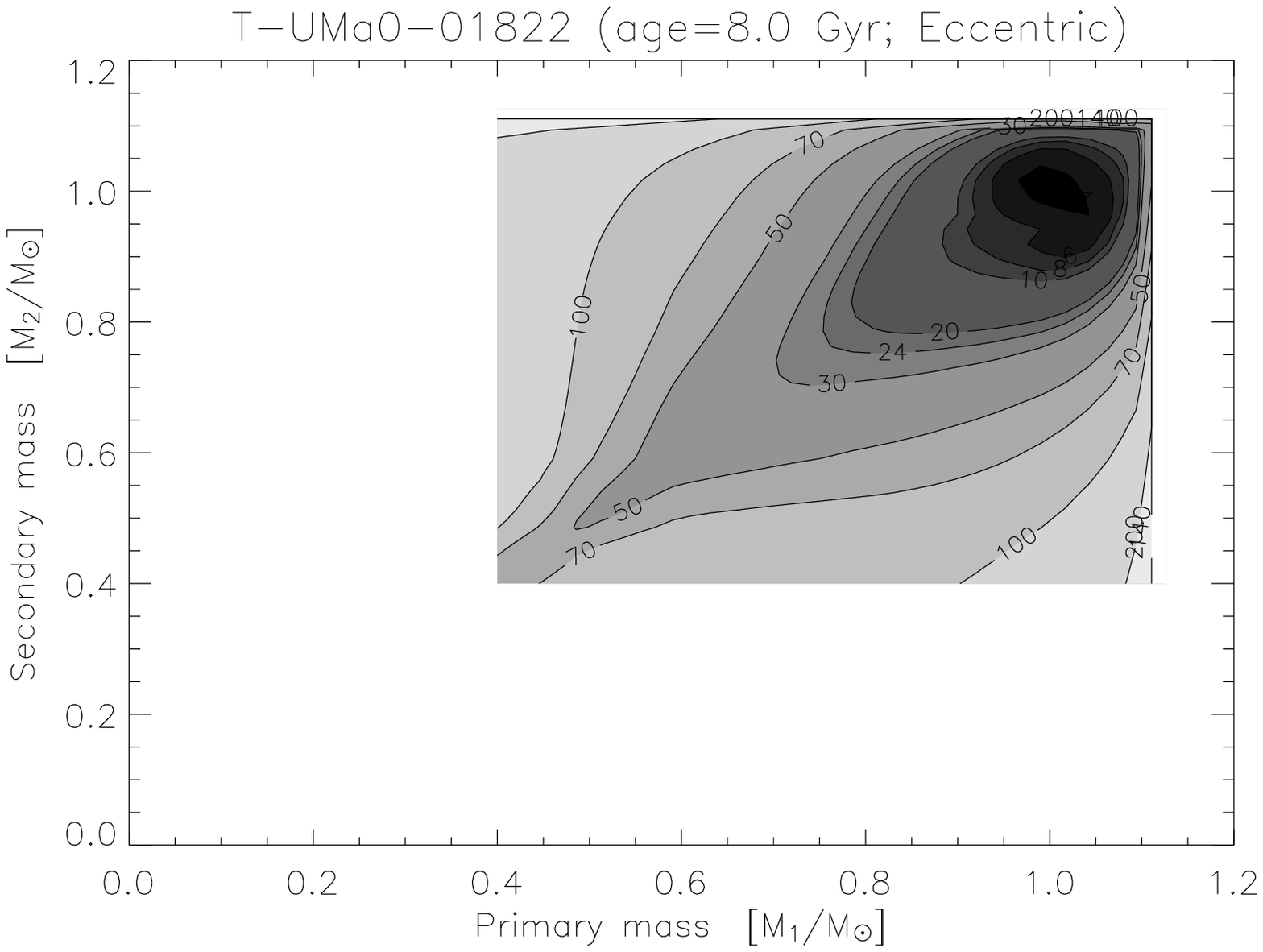}
\caption{MECI likelihood contour plots of a typical circular-orbit
EB (T-And0-00745 ; left) and eccentric-orbit EB (T-UMa0-01822 ;
right). There is no significant difference in the way MECI handles
these cases, and both usually have a single contour
minimum. The plots shown here have the ages set to the values
that produced the lowest MECI minima.}
\label{figMECI1}
\end{figure}

\begin{figure}
\plottwo{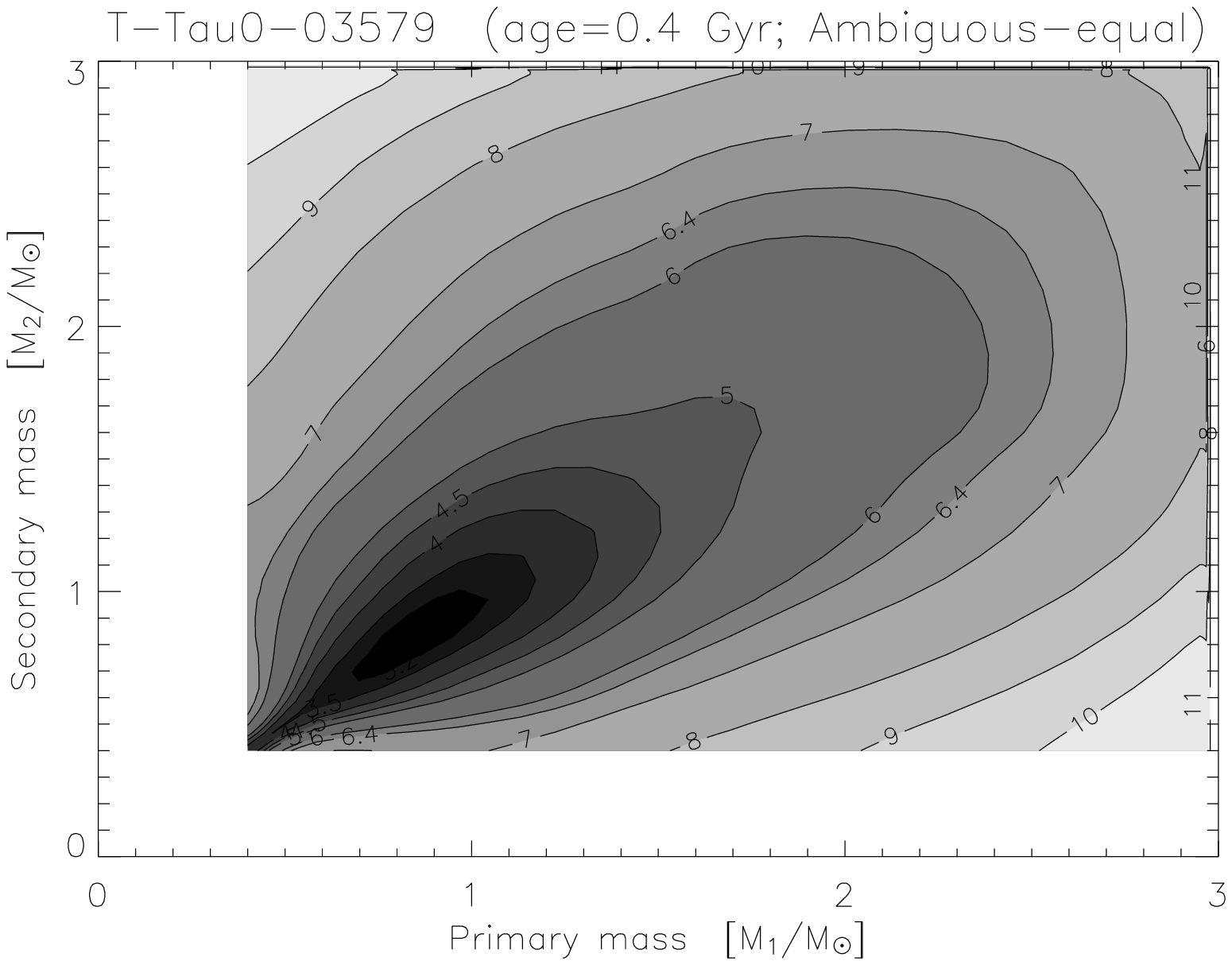}{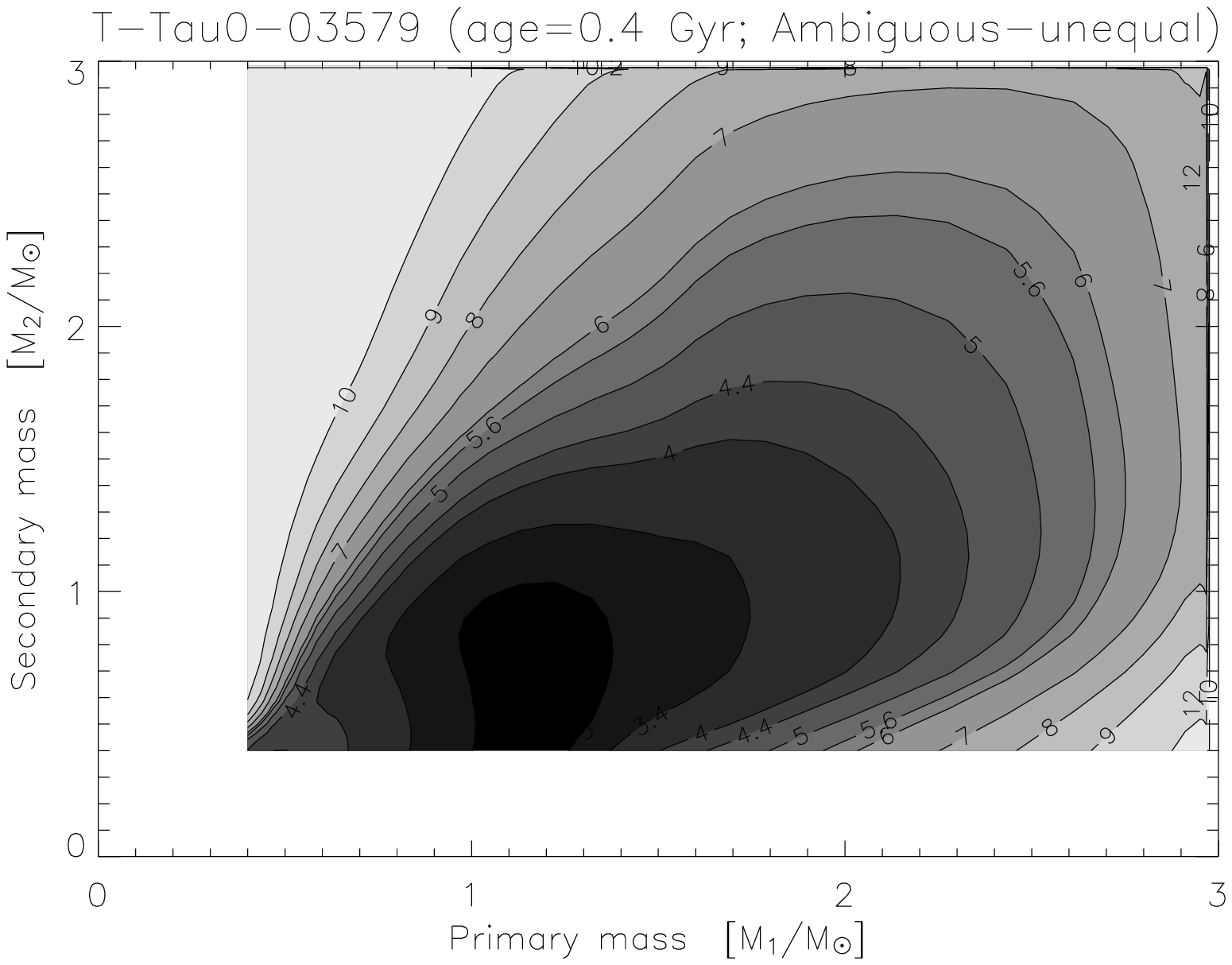}
\caption{MECI likelihood contour plots of a typical ambiguous EB
(T-Tau0-03579). These plots show the effect of assuming that the
binary components are equal (left) or unequal (right). Note that
the equal-component solution will have a nearly symmetric contour
around the diagonal, while the unequal-component solution can
provide only an upper limit to the secondary component's mass,
in this case $M_2 \lesssim 1\:M_\sun$. The plots shown here have
the ages set to the values that produced the lowest MECI minima.}
\label{figMECI2}
\end{figure}

In step (9) we analyze the LCs with MECI. We refer the reader to
the full description of this method in \citet{Devor06a, Devor06b},
and provide here only a brief outline. Given an observed EB LC and
out-of-eclipse colors, MECI will iterate through a range of values
for the EB age and the masses of its two components. By looking up
their radii and luminosities in theoretical isochrone tables, MECI
simulates the expected LC and combined colors, and selects the
model that best matches the observations, as measured by the
chi-squared statistic. Or more concisely, MECI searches the ($M_1,
M_2, age$)-parameter space for the chi-squared global minimum of
each EB. Figures \ref{figMECI1} and \ref{figMECI2} show
constant-age slices through such a parameter space. Once found,
the curvature of the global minimum along the parameter space axes
is used to determine the uncertainties of the corresponding
parameters.

The MECI analysis makes two important assumptions. The first is
that EB stellar components are coeval, which has been shown to
generally hold for close binaries \citep{Claret02}. When this
assumption is violated, MECI will often not be able to find an EB
model that successfully reproduces the LC eclipses. Such systems,
which may be of interest in their own right, make up $\sim$3\% of
the catalog and are further discussed later in this section. The
second assumption is that there is no significant reddening, or
third-light blended into the observations (i.e. from a photometric
binary or hierarchical triple). Such blending in the LC will make
the eclipses shallower, which produces an effect very similar to
that of the EB having a grazing orbit. Thus, it will cause the
measured orbital inclination to be erroneous, although it should
rarely otherwise affect the results of the MECI analysis
significantly. However, the MECI analysis is sensitive to color
biases caused by stellar reddening and blending.

We reduce both these biases by incorporating 2MASS colors (see
step 8), which are both less suspectable to reddening than optical
colors, and suffers from significantly less blending than TrES, as
the radius of the 2MASS photometric aperture is $\sim$20 times
smaller that that of TrES. We then attempt to further mitigate
this problem by analyzing each EB twice, using different relative
LC/color information weighting values [see \citet{Devor06b} for
further details]. We first run MECI with the default weighting
value ($w = 10$), and then run MECI again with an increased LC
weighting ($w = 100$) thereby decreasing the relative color
weighting. Finally, we adopt the solution that has a smaller
reduced chi-square. Typically, the results of the two MECI
analyses are very similar, indicating that the observed colors are
consistent with the ones predicted by the theoretical isochrones.
In such cases, the color information provides an important
constraint, which significantly reduces the parameter
uncertainties. However, when there is a significant color bias,
the default model will not fit the observed data as well as the
model that uses a reduced weighting of the color information. In
such a case, the reduced color information model, which has a
smaller chi-squared, is adopted. Following this procedure, we find
that in $\sim$9\% of our EBs, the reduced color information model
provided a better fit, indicating that while significant
color-bias is uncommon, it is a source of error that should not be
ignored.

By default, we had MECI use the Yonsei-Yale \citep{Yi01, Kim02}
isochrone tables of solar metallicity stars. Although they
successfully describe stars in a wide range of masses, these
tables become increasingly inaccurate for low-mass stars, as the
stars become increasingly convective. For this reason we
re-analyze EBs for which both components were found to have masses
below $0.75M_{\sun}$, using instead the \citet{Baraffe98}
isochrone tables, assuming a convective mixing length equal to the
pressure scale height. Our EB models also take into account the
effects of the limb darkening of each of the stellar components.
To this end we employ the ATLAS \citep{Kurucz92} and PHOENIX
\citep{Claret98, Claret00} tables of quadratic limb-darkening
coefficients.

\placefigure{figMECI1}
\placefigure{figMECI2}

As previously mentioned, once we know the absolute properties of
an EB system, we are able to estimate its distance
\citep{Stebbing10, Paczynski97}, and thus such systems can be
considered standard candles. We use \citet{Cox00} for the
extinction coefficients, assuming the standard Galactic ISM
optical parameter, $R_V = 3.1$, to create the following system:

\begin{eqnarray}
mag_J - Mag_J & = & \Delta Mag + 0.282 \cdot A(V)\\
mag_H - Mag_H & = & \Delta Mag + 0.176 \cdot A(V)\\
mag_K - Mag_K & = & \Delta Mag + 0.108 \cdot A(V)
\end{eqnarray}

Where $\Delta Mag$ is the extinction-corrected distance modulus,
and $A(V)$ is the V-mag absorption due to Galactic interstellar
extinction. The estimated distance can then be solved using: $D =
10pc \cdot 10^{\Delta Mag / 5}$. Because we have three equations
for only two unknowns, we adopt the solution that minimizes the
sum of the squares of the residuals. In some cases we remove one
of the bands as being an outlier (i.e. if it would have resulted
in a negative absorption), after which we are still able to solve
the systems. But in cases where we need to remove two bands, we
set $A(V) = 0$ in order to solve for the distance modulus.
Although this method has a typical uncertainty of 10\% to 20\%, it
can be applied to EBs that are far more distant and dim than are
accessible in other methods, such as parallax measurement. It can
be used to map broad features of the Galaxy, and identify binaries
that are in the Galactic halo. This method can also be used on
large group of systems, so that if the EBs are clustered, one can
average their distances, and thus reduce the cluster distance
uncertainty as the inverse square root of the number of systems.

In step (10) we perform a final quality check for the EB model
fits, and classify them into 7 groups:

\renewcommand{\theenumi}{\Roman{enumi}}
\begin{enumerate}
\item Eccentric: EBs with unequally-spaced eclipses
\item Circular: EBs with equally-spaced but distinct eclipses
\item Ambiguous-unequal: EBs with undetected secondary eclipses
\item Ambiguous-equal: EB with equally-spaced and indistinguishable eclipses
\item Inverted: detached EBs that are not successfully modeled by MECI
\item Roche-lobe-filling: non-detached EBs that are filling at least one Roche-lobe
\item Abnormal: EBs with atypical out-of-eclipse distortions
\end{enumerate}

We list the model parameters for the EBs of groups I-IV in the
electronic version of this catalog (see full description in
appendix \ref{appendixCatalogDescription}). The EBs of groups
V-VII could not be well modeled by MECI, therefore we list only
their coordinates and periods, so that they can be followed-up.

\begin{figure}
\plotone{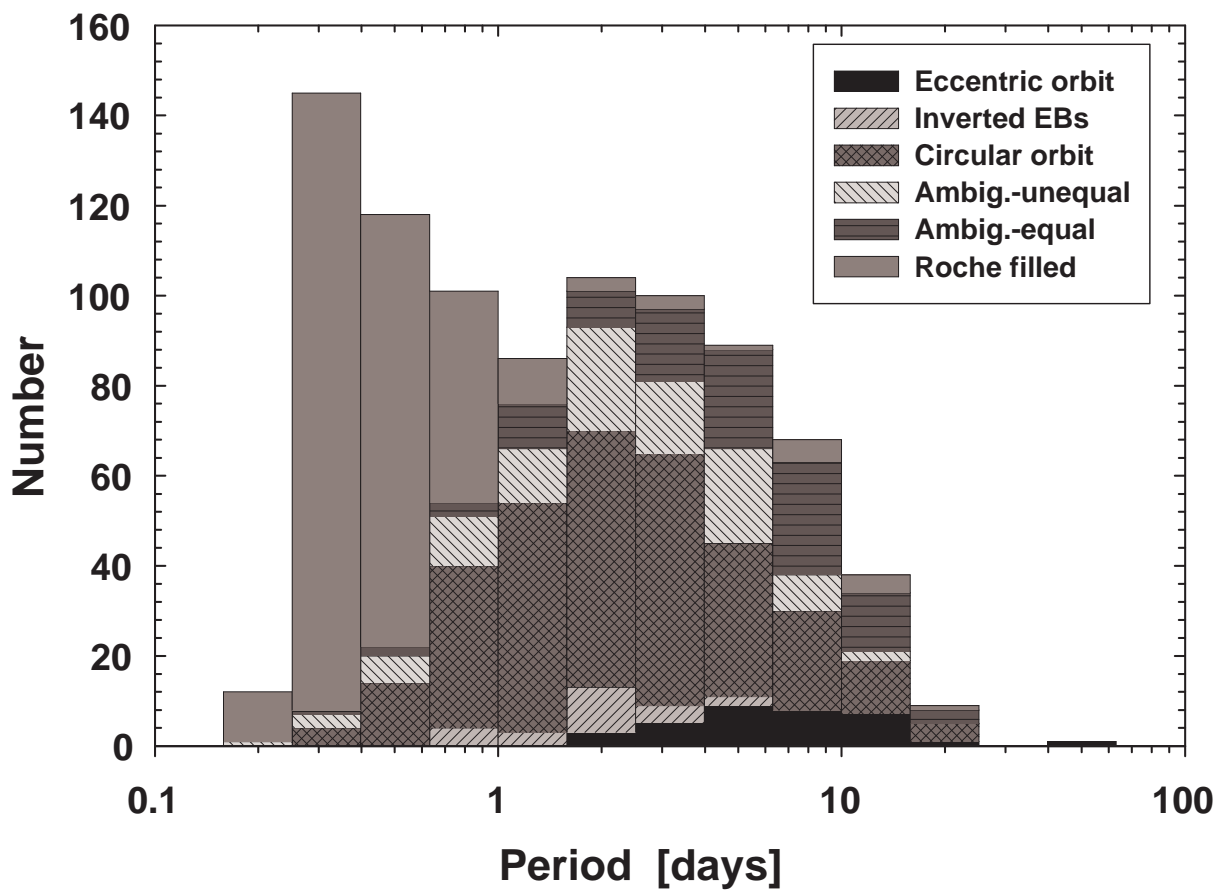}
\caption{The EB orbital period distribution within the catalog.
Each bin is subdivided to show the number of binaries belonging to
each of the classification groups described in \S\ref{secMethod}.
Note that the ambiguous-equal and ambiguous-unequal entries
represent the same stars, with entries in the former group
having double the period of the latter. Notice also how the Roche-lobe-filling
EBs dominate the sub-day bins, and have a long tail stretching well above
10-day periods. Furthermore, the circular orbit EBs have a period
distribution peak of at $\sim$2 days, while the eccentric orbit EBs
peak at $\sim$5 days. This is likely due to the orbital
circularization that occurs preferentially in short-period systems
(see also Figures \ref{figPeriodEcc} and \ref{figPeriodM1}).}
\label{figPeriodDistrib}
\end{figure}

Figure \ref{figPeriodDistrib} illustrates the period distribution
of these seven groups. Note however that both the orbital geometry
of EBs (eclipse probability $\propto P^{-2/3}$), and the limited
duration of the TrES survey data ($\leq 90\:days$ ; varies from
field to field ; see Table \ref{tableFieldsObs}), act to suppress
the detection of binaries with longer periods. An added
complication for single-telescope surveys is that about half of
the EBs with periods close to an integer number of days will not
be detectable, as they eclipse only during the daytime. This EB
distribution is consistent with the far deeper OGLE II field
catalog \citep{Devor05}, where the long tail of Roche-lobe-filling
systems has recently been explained by \citet{Derekas07} as being
the result of a strong selection towards detecting eclipsing giant
stars.

\placefigure{figPeriodDistrib}

\begin{figure}
\plotone{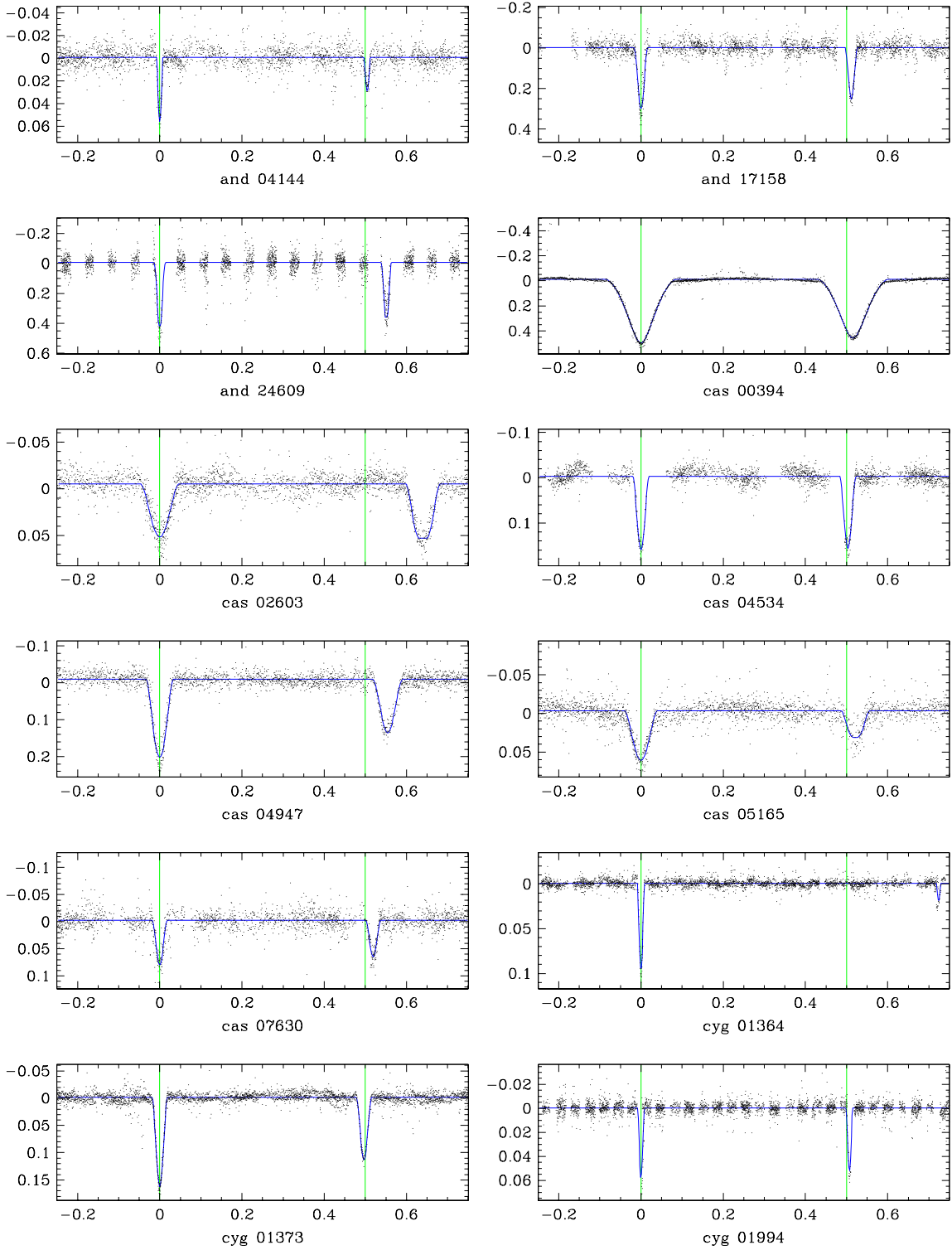}
\caption{Eccentric EBs (panel 1).
Note how the secondary eclipse is not at phase 0.5, as would be in circular orbit EBs.}
\label{figEcc1}
\end{figure}

\begin{figure}
\plotone{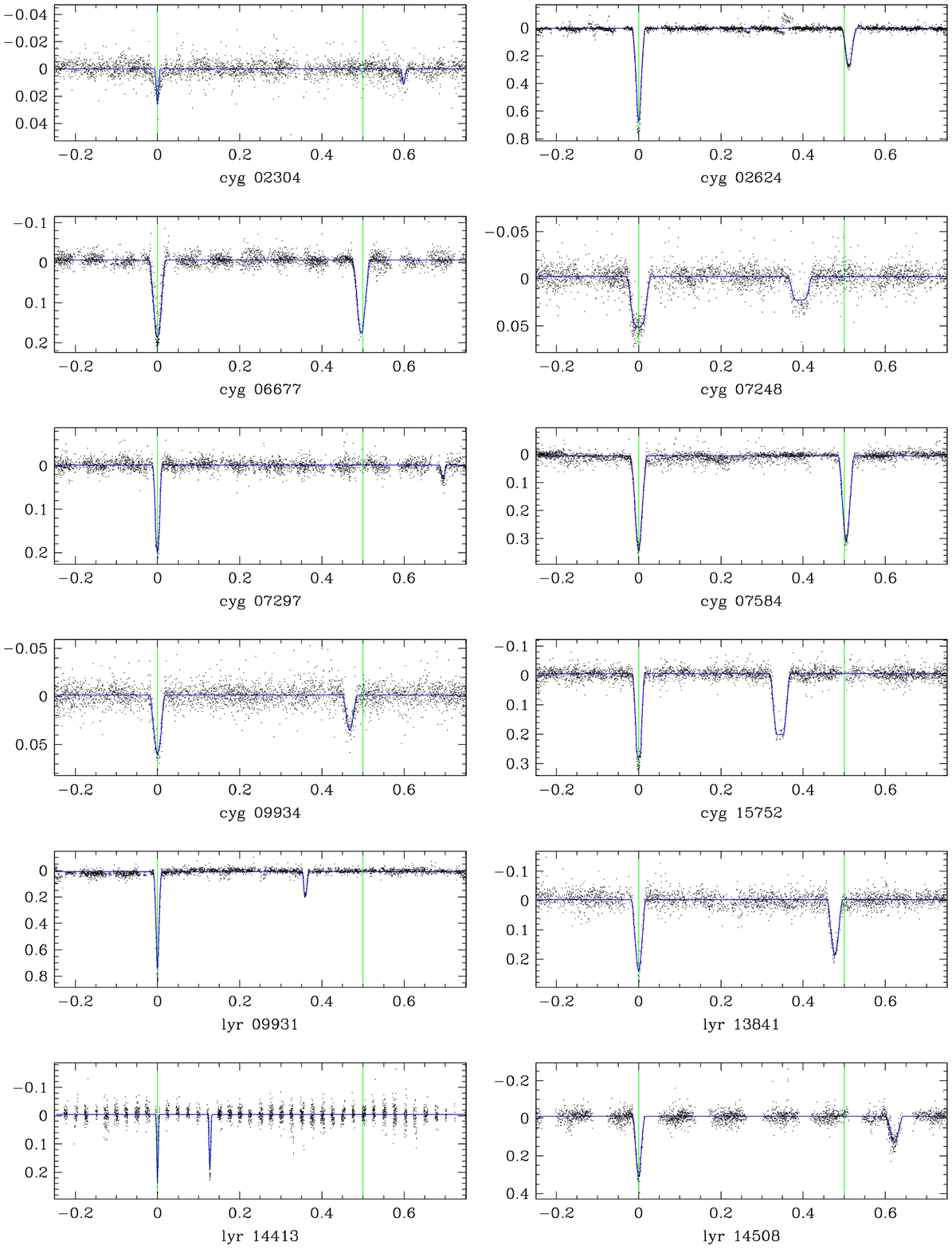}
\caption{Eccentric EBs (panel 2).}
\label{figEcc2}
\end{figure}

\begin{figure}
\plotone{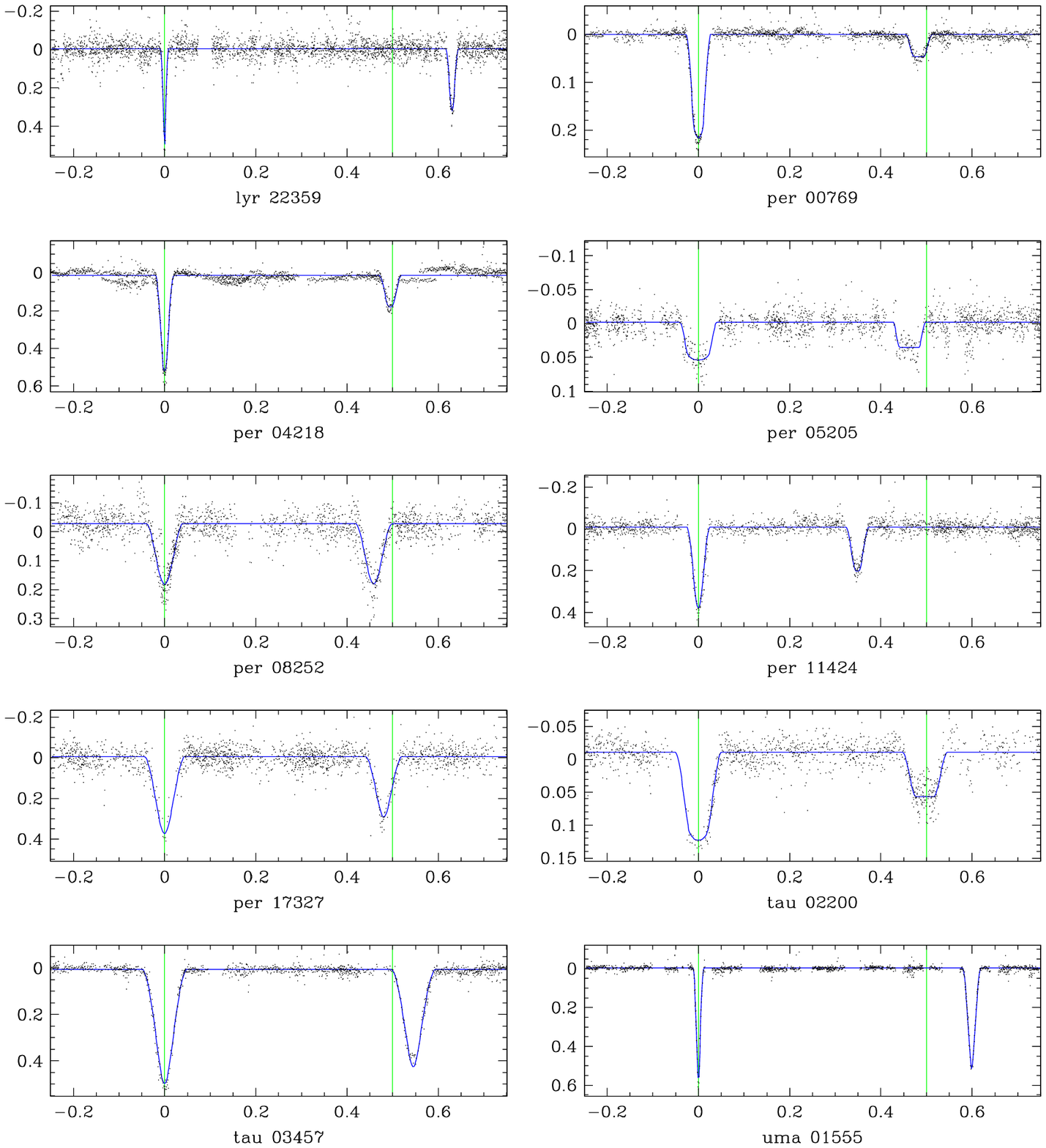}
\caption{Eccentric EBs (panel 3).}
\label{figEcc3}
\end{figure}

\begin{figure}
\plotone{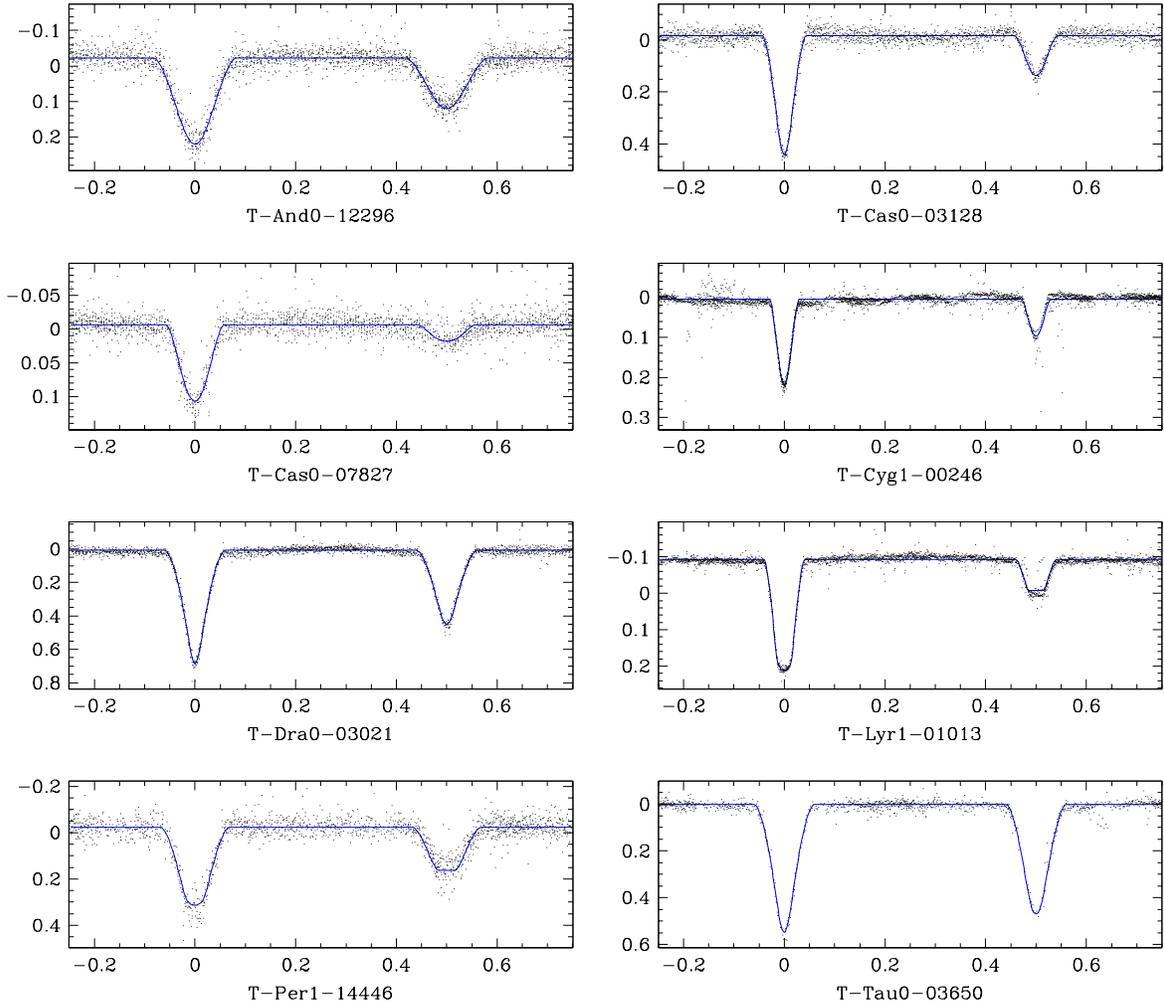}
\caption{Examples of unambiguous
EBs with circular orbits, with their best-fit MECI models (solid line).}
\label{figE0Cat}
\end{figure}

Group [I] contains the eccentric EBs identified in step (4) as
having centers of eclipse that are separated by a duration
significantly different from half an orbital period (see Figures
\ref{figEcc1}, \ref{figEcc2}, and \ref{figEcc3}). This criterion
is sufficient for demonstrating eccentricity, but not necessary,
since we miss systems for which $\cos \omega \simeq 0$ (see
equation \ref{eqOmC}). Fortunately, we are able to detect
eccentricities in well-detached EBs with $|e \cos \omega| \geq
0.005$, using eclipse timing. Therefore, assuming that $\omega$ is
uniformly distributed, we are approximately 67\% complete for $e
= 0.01$, and over 92\% complete for $e = 0.04$. In
principle, it would be possible to be 100\% complete for these
systems by measuring the differences in their eclipse durations,
however this measurement is known to be unreliable \citep{Etzel91}
and so would likely contaminate this group with false positives.
Group [II] consists of all such circular-orbit EBs that were
successfully fit by a single MECI model (see Figure
\ref{figE0Cat}).

\placefigure{figEcc1}
\placefigure{figEcc2}
\placefigure{figEcc3}
\placefigure{figE0Cat}

\begin{figure}
\plotone{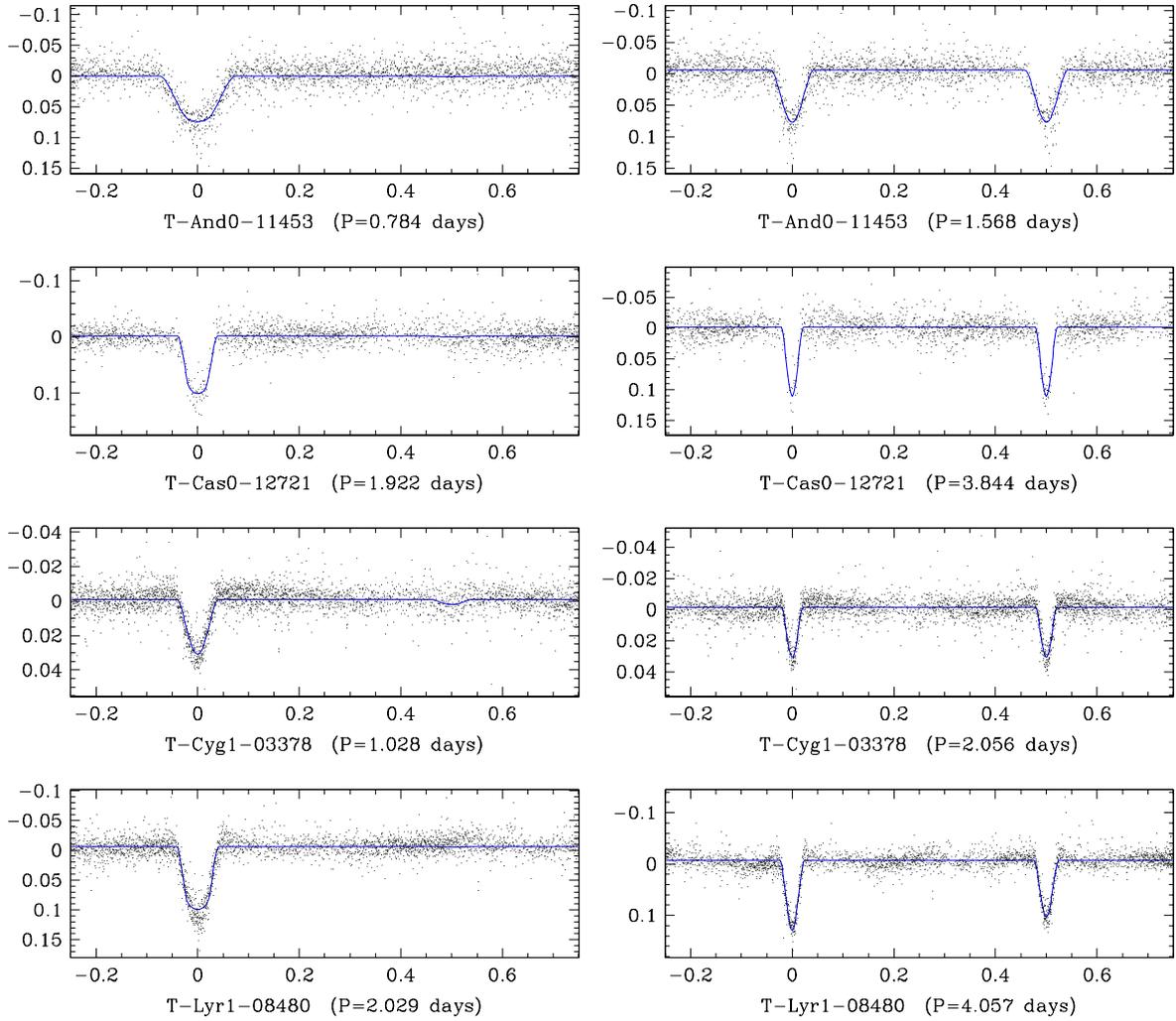}
\caption{Examples of ambiguous EBs. Left
column: assuming very unequal components. Right column: assuming
approximately equal components with double the period.}
 \label{figAmbigCat}
\end{figure}

EBs with only one detectable eclipse can potentially be modeled in
two alternative ways. One way is to assume very unequal stellar
components, which have a very shallow undetected secondary eclipse
(group [III]). Since we cannot estimate the eccentricity of such
systems, we assume that they have circular orbits. The other way
is to assume that the period at hand is twice the correct value,
and that the components are nearly equal (group [IV]). The entries
of such ambiguous LCs were doubled in step (2), so that these two
solutions would be independently processed through the pipeline
(see Figure \ref{figAmbigCat}). Therefore, these two groups have a
one-to-one correspondence between them, although only one entry of
each pair can be correct. Resolving this ambiguity may not always
be possible without spectroscopic data. In some cases we were able
to resolve this ambiguity using either a morphological or a
physical approach. The morphological approach consists of manually
examining the LCs of group [IV] for any asymmetries in the two
eclipses (e.g. width, depth, or shape), or in the two plateaus
between the eclipses (e.g. perturbations due to tidal effects,
refections, or the ``O'Connell effect''). The physical approach
consists of applying our understanding of stellar evolution in
order to exclude entries that cannot be explained through any
coeval star pairing (see appendix \ref{appendixSingleEclipse}).
Either way, once one of the two models has been eliminated, the
other model is moved into group [II] and is adopted as a
non-ambiguous solution. It is interesting to note that when
analyzing the two models with MECI, the equal-component solution
(group [IV]) has masses approximately equal to the primary
component of the unequal-component solution (group [III]). The
mass of the unequal-component solution's secondary component will
typically be the smallest value listed in the isochrone table, as
this configuration will produce the least detectable secondary
eclipse.

\placefigure{figAmbigCat}

\begin{figure}
\plotone{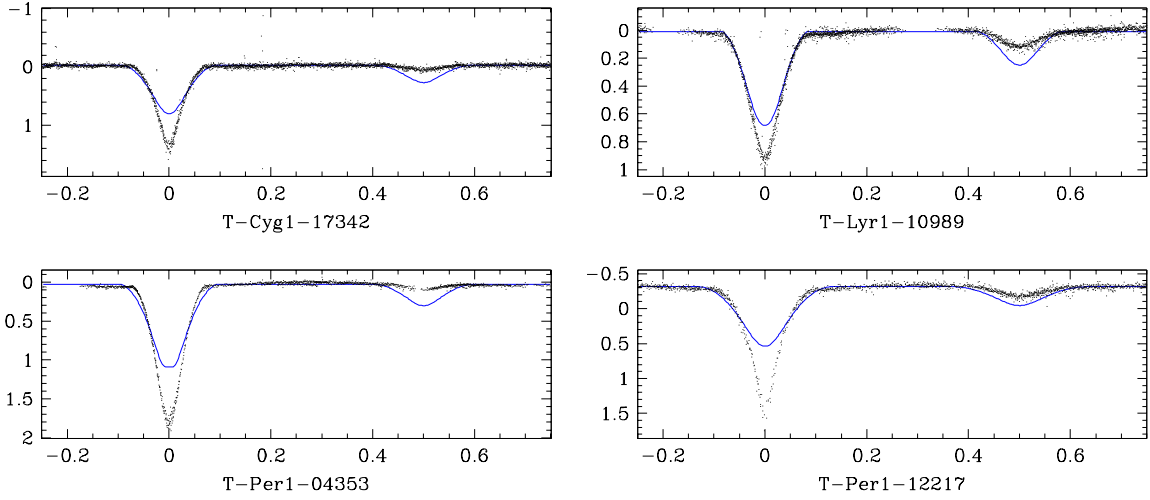}
\caption{Examples of EBs classified as inverted EBs. We included the
unsuccessful best-fit MECI model (solid curve) as an approximate reference to
illustrate the LC of a corresponding binary that has had no mass transfer.
Notice how the model LC is unable to achieve a sufficiently deep primary
eclipse, while producing a secondary eclipse that is too deep.}
\label{figInvertedCat}
\end{figure}

Group [V] consists of detached EBs that cannot be modeled by two
coeval stellar components. As mentioned earlier, we can reject the
single-eclipse solution for EBs with sufficiently deep eclipses
(see appendix \ref{appendixSingleEclipse}). This argument can be
further extended to cases where we can detect both eclipses in the
LC, but where one is far shallower than the other. In some cases,
no two coeval main-sequence components will reproduce such a LC,
but unlike the previous case, since both eclipses are seen, we
cannot conclude that the period needs to be doubled. Such systems
are likely to have had mass transfer from a sub-giant component
onto a main-sequence component through Roche-lobe overflow, to the
point where currently the main-sequence component has become
significantly more massive and brighter than it was originally
\citep{Crawford55}. This process will cause the components to
effectively behave as non-coeval stars, even though they have in
fact the same chronological age. In extreme cases, the originally
lower-mass main-sequence component can become more massive than
the sub-giant, and thus swap their original primary/secondary
designations, so that the main-sequence component is now the
primary component. We call such systems ``inverted'' EBs, and
place them into group [V] (see Figure \ref{figInvertedCat}). This
phenomenon is often referred to in the literature as the ``Algol
paradox'', though we chose not to adopt this term so as to avoid
confusing it with the term ``Algol-type EB'' (EA), which is
defined by the General Catalogue of Variable Stars [GCVS ;
\citep{Kukarkin48, Samus06}] as being the class of all
well-detached EBs.

\placefigure{figInvertedCat}

\begin{figure}
\plotone{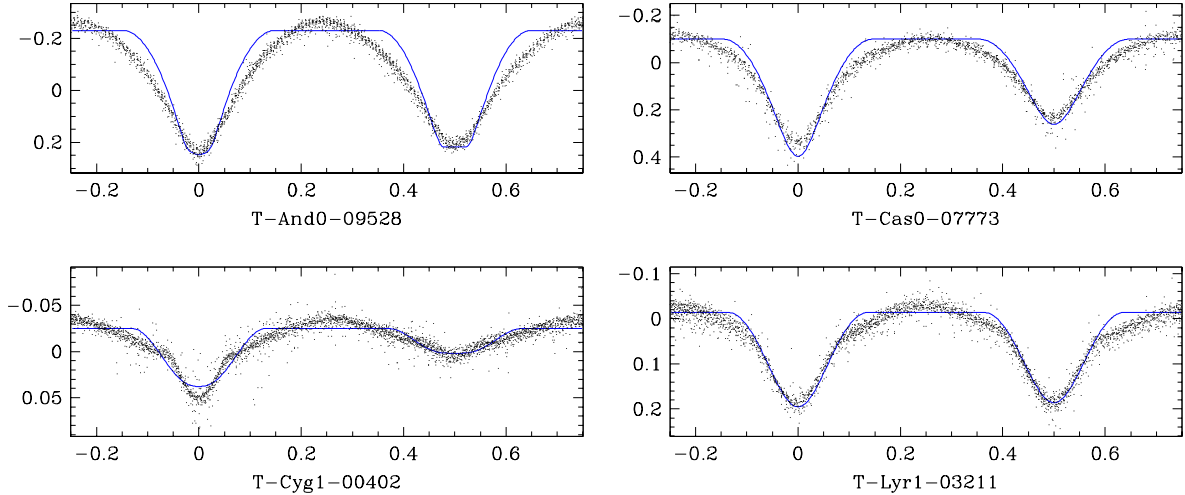}
\caption{Examples of EBs that are assumed to have filled at least one of their Roche lobes.
We included, for illustration purposes only, their best-fit MECI models (solid line).
These models were not adopted since they neglect tidal distortions, reflections, and gravity
darkening effects, and so produce a poor fit to the data.}
\label{figRocheFilledCat}
\end{figure}

\begin{figure}
\plotone{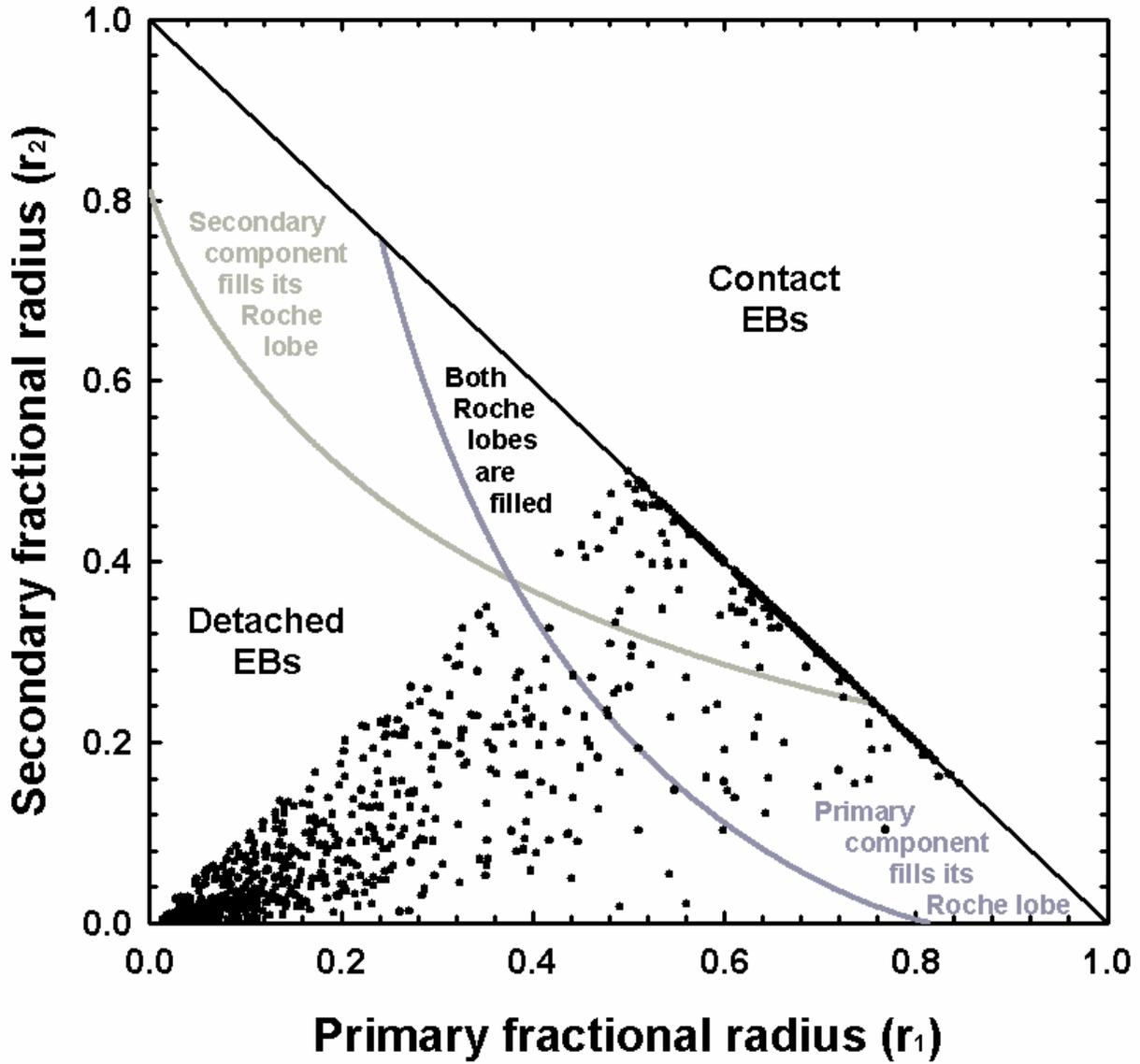}
\caption{The criterion applied in
equations \ref{eqRoche1} and \ref{eqRoche2} to determine whether one or both the EB
components have filled their Roche lobe, and thus need to be
placed into group (6).}
\label{figRoche}
\end{figure}

Group [VI] contains the EBs that have at least one component
filling its Roche-lobe (see Figure \ref{figRocheFilledCat}). Such
system cannot be well-fit by either DEBiL or MECI since they
assume that the binary components are detached, and so neglect
tidal and rotational distortions, gravity darkening, and
reflection effects. These systems must be separated from the rest
of the catalog since their resulting best-fit models will be poor
and therefore their evaluated physical attributes will likely be
erroneous. In a similar fashion to \citet{Tamuz06}, we detect
these systems automatically by applying the \citet{Eggleton83}
approximation for the Roche-lobe radius, and place in group [VI]
all the systems for which at least one of the EB components has
filled its Roche-lobe (see Figure \ref{figRoche}), that is, if
either one of the following two inequalities occurs:

\begin{eqnarray}
\label{eqRoche1}
r_1 &>& \frac{0.49\: q^{-2/3}}{0.6\: q^{-2/3} + \ln \left(1 + q^{-1/3}\right)} {\rm \ \ \ or}\\
\label{eqRoche2}
r_2 &>& \frac{0.49\: q^{2/3}}{0.6\: q^{2/3} + \ln \left(1 + q^{1/3}\right)}\ ,
\end{eqnarray}

where $q = M_2 / M_1$ is the EB components' mass ratio. Since we
expect non-detached EBs to be biased towards evolving, higher mass
stellar components, we estimated $q$ using the early-type
mass-radius power law relation found in binaries \citep{Gorda98}:
$q \simeq (r_2/r_1)^{1.534}$. Although in principle, we could have
estimated $q$ directly from the EB component masses resulting from
the MECI analysis, we chose not to, since as stated above, the
analysis of such systems is inaccurate. The analytic approximation
we used, though crude, proved to be remarkably robust, as we found
only 5 false negatives and no false positives when visually
inspecting the LCs. We found many more false positives/negatives
when using the alarm criteria suggested by \citet{Devor05} or
\citet{Mazeh06}, both of which attempt to identify bad model fits
by evaluating spatial correlations of the model's residuals.

\placefigure{figRocheFilledCat}
\placefigure{figRoche}

\begin{figure}
\plotone{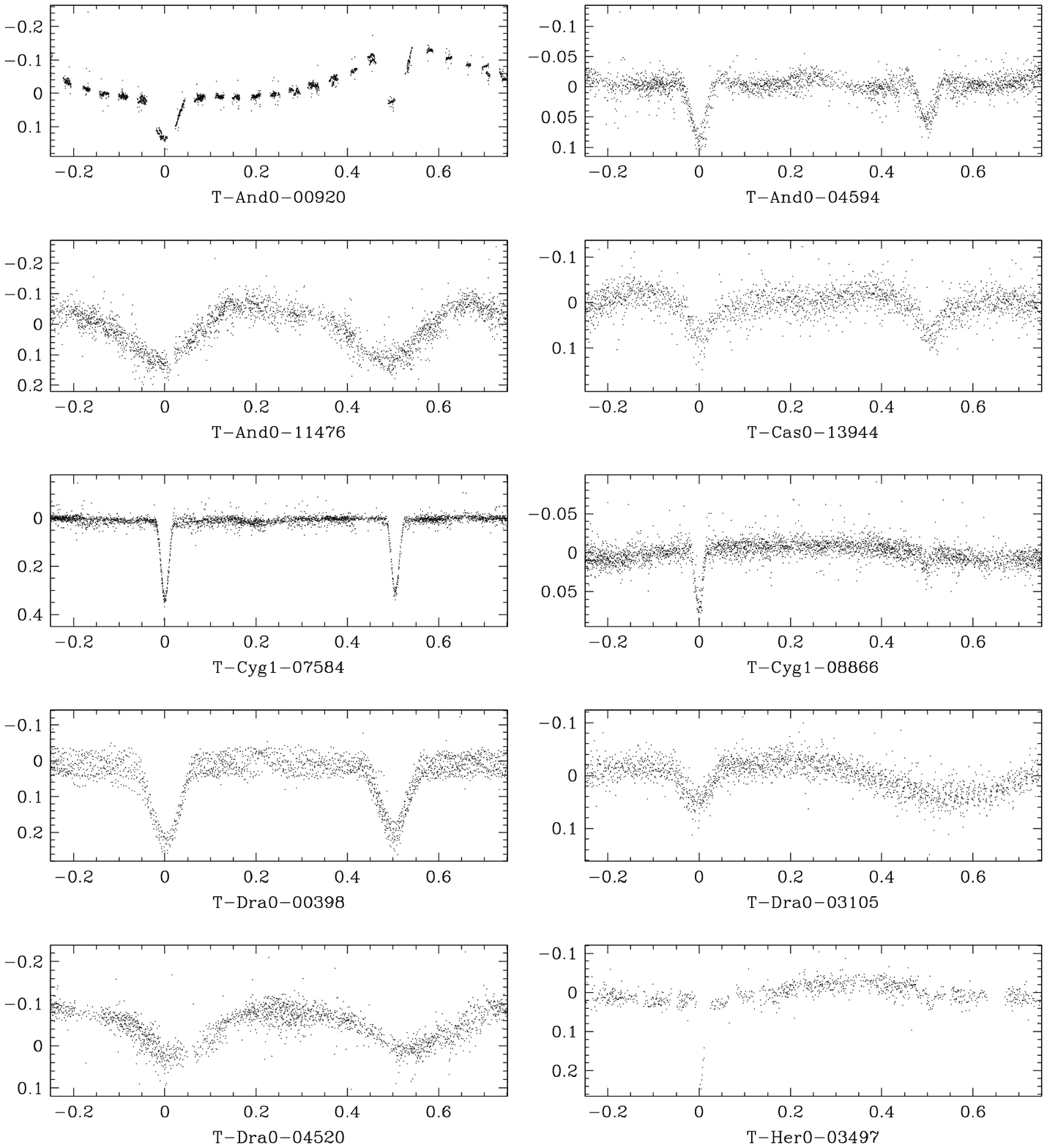}
\caption{LCs of abnormal EBs (panel 1).}
\label{figAbnormalEBs1}
\end{figure}

\begin{figure}
\plotone{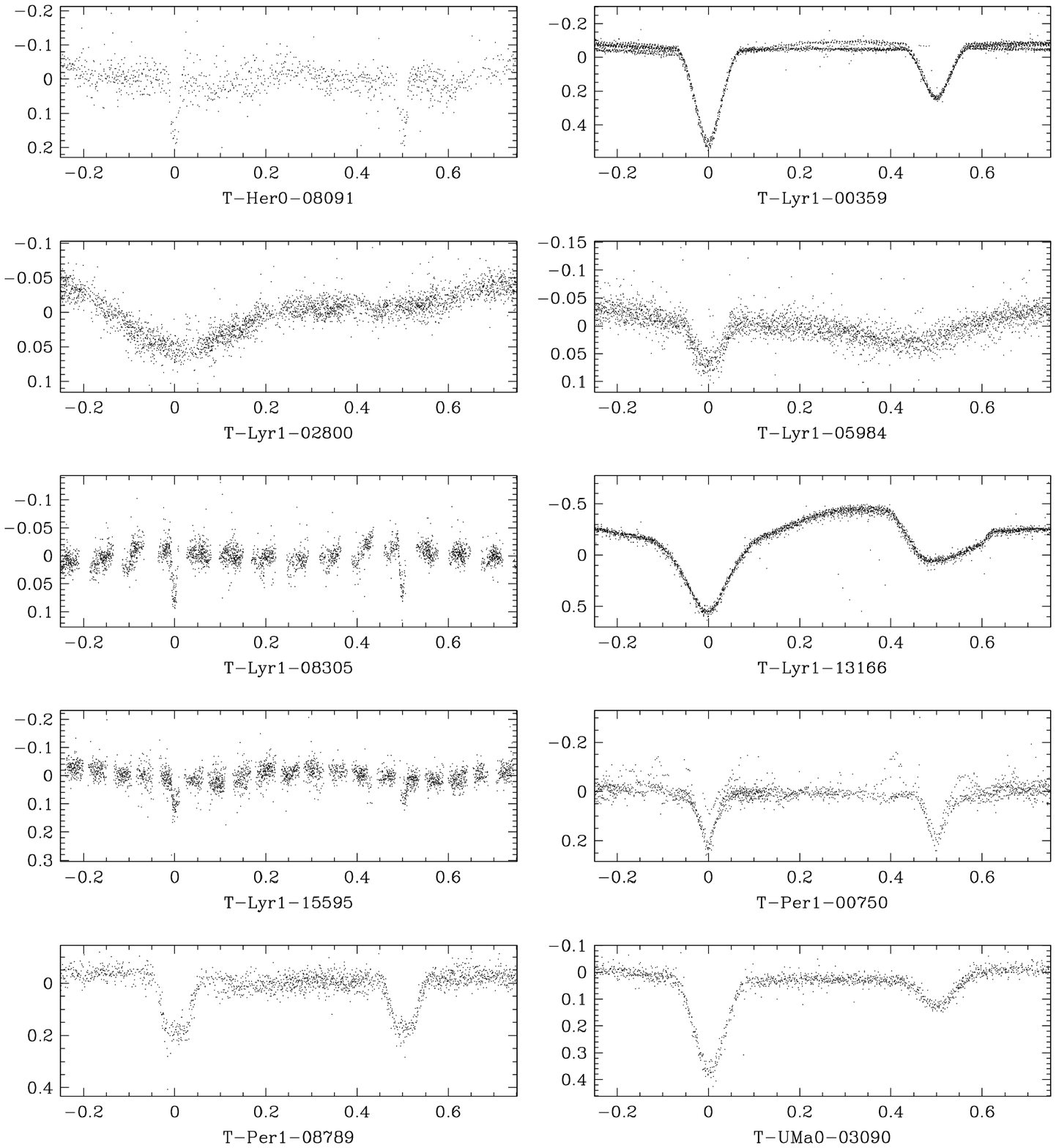}
\caption{LCs of abnormal EBs (panel 2).}
\label{figAbnormalEBs2}
\end{figure}

Finally, group [VII] contains systems visually identified as EBs
(i.e. having LCs with periodic flux dips), yet having atypical LC
perturbations that indicate the existence of additional physical
phenomena (see Figures \ref{figAbnormalEBs1} and
\ref{figAbnormalEBs2}). For lack of a better descriptor, we call
such systems ``abnormal'' (see further information in
\S\ref{subsecAbnomalEBs}). This group is different from the
previous six in that we cannot automate their classification, and
their selection is thus inherently subjective. In 15 of the 20
systems, we were able to approximately model the LCs, and included
them in one of the aforementioned groups. In these cases users
should be aware that these model may be biased by the phenomenon
that brought about their LC distortion.

\placefigure{figAbnormalEBs1}
\placefigure{figAbnormalEBs2}

\section{Results}
\label{secResults}

\begin{figure}
\plotone{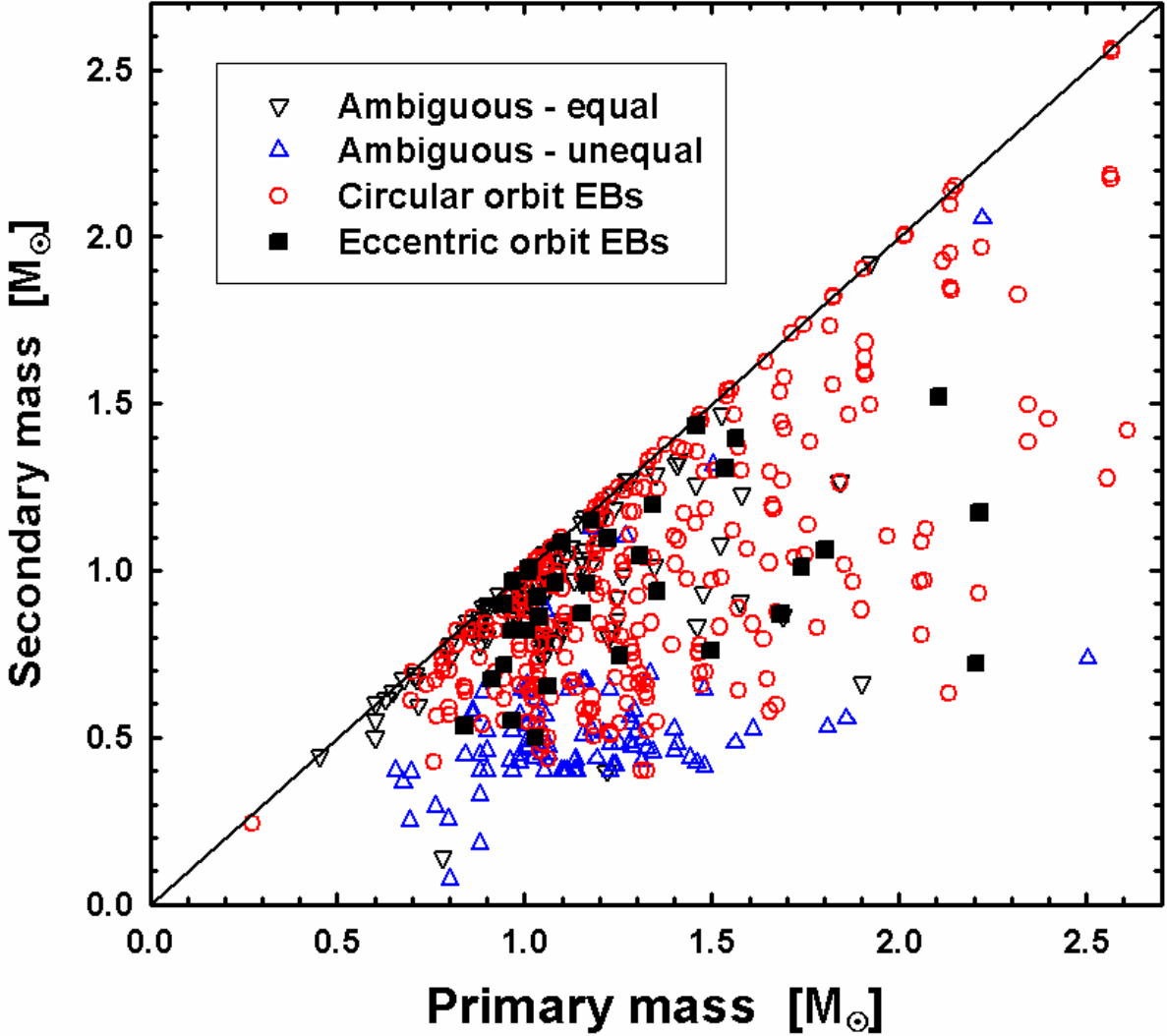}
\caption{The mass-mass relation for the
detached EBs of the TrES dataset. Each category is represented by
a different symbol. Note that the ambiguous EBs are plotted twice,
where only one of the solution can be correct. Notice also that in
the equal-component solutions are clustered along the diagonal,
while the unequal-component solutions are clustered along the
minimum available mass of the Yonsei-Yale isochrones ($0.4
\:M_\sun$). Some of the ambiguous solutions deviate from these
clusters due to poor constraints on the secondary eclipse, which
brings about a large uncertainty. Finally, notice the sparsity of
EBs populating the low-mass corner of this plot ($M_{1,2} <
0.75\:M_\sun$). These systems, whose importance is outlined in
\S\ref{subsecLowMassEBs}, were modeled using the Baraffe
isochrones. CM Draconis (T-Dra0-01363) clearly sets itself apart,
being the lowest-mass binary in the catalog (circle at
bottom-left).}
\label{figMassMass}
\end{figure}

We identified and classified a total of 773 EBs\footnote{The
observed LCs, fitted models, and model residuals of each of these
EBs are shown at:\newline
http://www.cfa.harvard.edu/$\sim$jdevor/Catalog.html}. These
systems consisted of 734 EBs with circular orbits, 34 detached EBs
with eccentric orbits (group [I] ; Table \ref{tableEccentric}),
and 5 unclassified abnormal EBs (group [VII] ; Table
\ref{tableAbnormal}). We marked 15 of the detached EBs with
circular orbits as also being abnormal. Of the 734 EBs with
circular orbits, we classify 290 as unambiguous detached EBs
(group [II] ; Table \ref{tableCircular}), 103 as ambiguous
detached EBs, for which we could not determine photometrically if
they consisted of equal or disparate components (groups [III] and
[IV] ; Table \ref{tableAmbig}), 23 as inverted EBs (group [V] ;
Table \ref{tableInverted}), and 318 as non-detached (group [VI] ;
Table \ref{tableFillRoche}). With the exception of the abnormal
EBs, which were selected by eye, we use an automated method to
classify each of these groups (see \S\ref{secMethod} for details).
Our mass estimates for the primary and secondary components are
plotted in Figure \ref{figMassMass}.

\placefigure{figMassMass}

\begin{figure}
\plotone{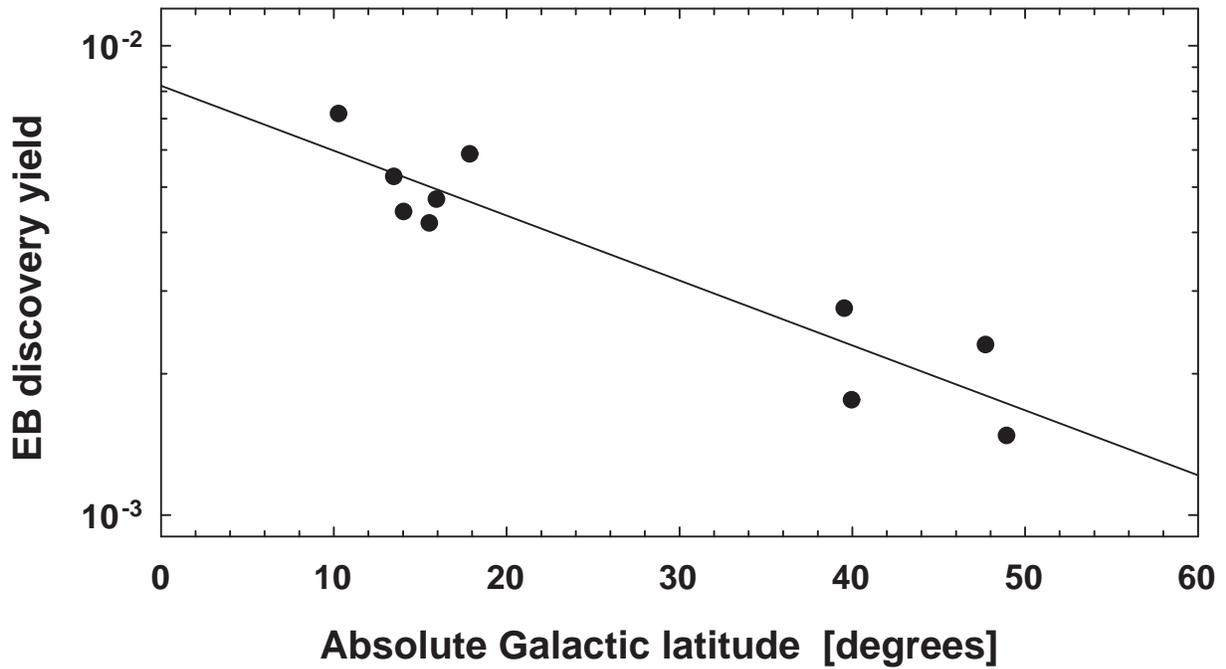}
\caption{The relation between the EB
discovery yield (the fraction of LCs found to be EBs) and the
absolute value of the Galactic latitude, or $|b|$, for the 10 TrES
fields used in this catalog (see Tables \ref{tableFieldsObs} and
\ref{tableFieldsYield}). The solid line is the linear regression
of the log of the EB discovery yield ($r^2 = 0.867$). Some of the
residual scatter can be explained as being due to differences in
the duration of observations in each field. By including the
duration in a bi-linear regression, we get a substantially improved
fit ($r^2 = 0.911$).}
\label{figDiscoveryYield}
\end{figure}

The EB discovery yield (the fraction of LCs found to be EBs),
varies greatly from field to field, ranging from 0.72\% for
Cygnus, to 0.15\% for Corona Borealis (see Table
\ref{tableFieldsYield}). This variation is strongly correlated
with Galactic latitude, where fields near the Galactic plane have
larger discovery yields than those that are farther from it (see
Figure \ref{figDiscoveryYield}). This effect is likely due to the
fact that fields closer to the Galactic plane contain a higher
fraction of early-type stars. These early-type stars are both
physically larger, making them more likely to be eclipsed, and are
more luminous, which has them produce brighter and less noisy LCs,
thereby enabling the detection of EBs with shallower eclipses.
Furthermore, much of the residual scatter can be attributed to the
variation in the observed duration of each field (see Table
\ref{tableFieldsObs}). That is, we find additional EBs, with
longer periods, in fields that were observed for a longer
duration.

\placefigure{figDiscoveryYield}

Currently, 88 of the cataloged EBs (11\%) appear in either the
International Variable Star Index\footnote{Maintained by the
American Association of Variable Star Observers (AAVSO).} (VSX),
or in the SIMBAD\footnote{Maintained by the Centre de Données
astronomiques de Strasbourg (CDS).} astronomical database (Table
\ref{tableSIMBAD}). However, only 49 systems (6\%) have been
identified as being variable. Not surprisingly, with few
exceptions, these targets were among the brightest sources of the
catalog. Using only photometry, it is often notoriously difficult
to distinguish non-detached EBs from pulsating variables that vary
sinusoidally in time, such as type-C RR Lyrae. Furthermore,
unevenly spotted stars may also cause false positive
identifications, especially in surveys with shorter durations.
Ultimately, spectroscopic follow-up will always be necessary to
confirm the identification of such variables.

We highlight three groups of EBs as potentially having special
importance as test beds for current theory. For more accurate
properties, these EBs will likely need to be followed up both
photometrically and spectroscopically. The brightness of these EBs
will considerably facilitate their follow-up.

\subsection{Low-mass EBs}
\label{subsecLowMassEBs}

\begin{figure}
\plotone{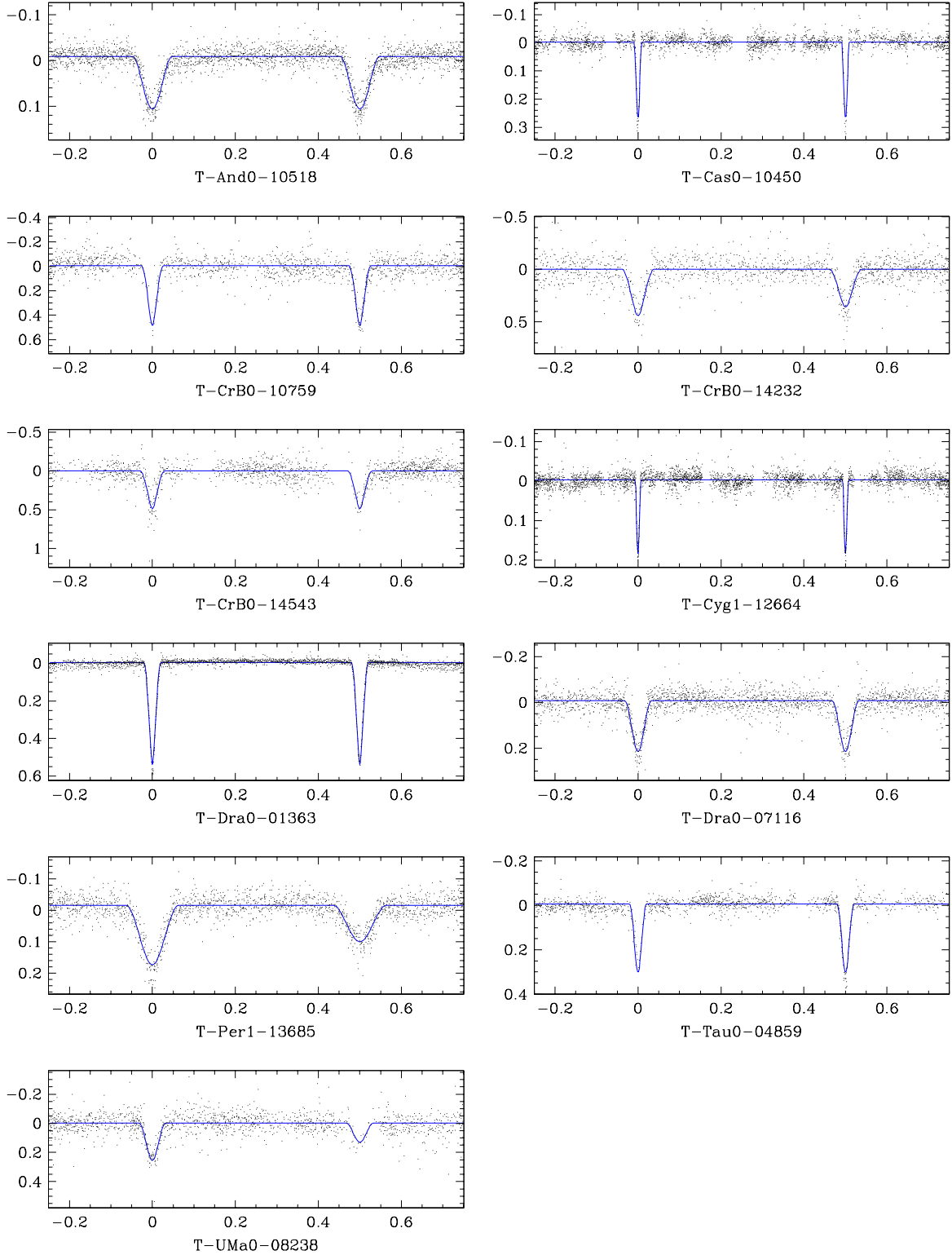}
\caption{Low mass candidates ($M_1 < 0.75 M_\sun$), with their best-fit MECI models (solid line).}
\label{figLowMassCat}
\end{figure}

The first group consists of 11 low-mass EB candidates, including
10 newly discovered EBs with either K or M-dwarf stellar
components. Our criteria for selecting these binaries were that
they be well-detached, and that both components have estimated
masses below $0.75M_\sun$ (see Table \ref{tableLowMass} and Figure
\ref{figLowMassCat}). Currently, only 7 such detached low-mass EBs
have been confirmed [YY~Gem \citep{Kron52, Torres02}, CM~Dra
\citep{Lacy77a, Metcalfe96}, CU~Cnc \citep{Delfosse99, Ribas03},
T-Her0-07621 \citep{Creevey05}, GU~Boo \citep{LopezMorales05},
NSVS01031772 \citep{LopezMorales06}, and UNSW-TR-2
\citep{Young06}].

Despite a great deal of work that has been done to understand the
structure of low-mass stars [e.g. \citet{Chabrier00}], models
continue to underestimate their radii by as much as $15$\%
\citep{Lacy77b, Torres02, Creevey05, Ribas06}, a significant
discrepancy considering that for solar-type stars the agreement
with the observations is typically within $1-2$\%
\citep{Andersen91, Andersen98}. In recent years an intriguing
hypothesis has been put forward that strong magnetic fields may
have bloated these stars through chromospheric activity
\citep{Ribas06, Torres06, LopezMorales07, Chabrier07}.
Furthermore, \citet{Torres06} find that such bloating occurs even
for stars with nearly a solar mass, and suggest that this effect
may also be due to magnetically induced convective disruption. In
either case, these radius discrepancies should diminish for widely
separated binaries with long periods, as they become
non-synchronous and thus rotate slower, which according to dynamo
theory would reduce the strength of their magnetic fields.

Unfortunately, the small number of well characterized low-mass EBs
makes it difficult to provide strong observational constraints to
theory. Despite the fact that such stars make up the majority of
the Galactic stellar population, their intrinsic faintness renders
them extremely rare objects in magnitude-limited surveys. In
addition, once found, their low flux severely limits the ability
of observing their spectra with both sufficiently high resolution
and a high signal-to-noise ratio. To this end, the fact that the
TrES survey was made with small-aperture telescopes is a great
advantage, as any low-mass EB candidate found is guaranteed to be
bright, and thus require only moderate-aperture telescopes for
their follow-up. Thus we propose multi-epoch spectroscopic study
of the systems listed here, in order to confirm their low mass and
to estimate their physical properties with an accuracy sufficient
to test models of stellar structure. Moreover, two of our
candidates (T-Cyg1-12664 and T-Cas0-10450), if they are in fact
ambiguous-equal (group [IV]), have periods greater than 8 days,
making them prime targets for testing the aforementioned
magnetic-bloating hypothesis.

\placefigure{figLowMassCat}

\subsection{Eccentric EBs}
\label{subsecEccentricEBs}

The second group of EBs consists of 34 binaries with eccentric
orbits (see Table \ref{tableEccentric}, and Figures \ref{figEcc1},
\ref{figEcc2}, and \ref{figEcc3}). We were able to reliably
measure values of $|e \cos \omega|$ as low as $\sim0.005$ by using
the eclipse timing technique (see \S\ref{secMethod} and Figure
\ref{figTimingVariationsEcc}). Since this measure provides a lower
limit to the eccentricity, it is well-suited to identify eccentric
EBs, even though the actual value of the eccentricity may be
uncertain. As mentioned earlier, in an effort to avoid
false-positives, we do not include in this group EBs whose eclipse
timing measures $|e \cos \omega| < 0.005$, or EBs with an
eccentricity consistent with zero.

Our interest in these eccentric binaries stems from their
potential to constrain tidal circularization theory
\citep{Darwin1879}. This theory describes how the eccentricity of
a binary orbit decays over time due to tidal dissipation, with a
characteristic timescale ($t_{circ}$) that is a function of the
components' stellar structure and orbital separation. As long as
the components' stellar structure remains unchanged, the orbital
eccentricity is expected to decay approximately exponentially over
time [$e \propto \exp (-t/t_{circ})$]. However, once the
components evolve off the main sequence, this time scale may vary
considerably \citep{Zahn89}. Thus, to understand the
circularization history of binaries with circularization
timescales similar to or larger than their evolutionary
timescales, one must integrate over the evolutionary tracks of
both stellar components.

Three alternate tidal dissipation mechanisms have been proposed:
dynamical tides \citep{Zahn75, Zahn77}, equilibrium tides
\citep{Zahn77, Hut81}, and hydrodynamics \citep{Tassoul88}.
Despite its long period of development, the inherent difficulty of
observing tidal dissipation has prevented definitive conclusions.
\citet{Zahn89} add a further complication by maintaining that most
of the orbital circularization process takes place at the
beginning of the Hayashi phase, and that the eccentricity of a
binary should then remain nearly constant throughout its lifetime
on the main sequence.

Observational tests of these tidal circularization theories,
whereby $t_{circ}$ is measured statistically in coeval stellar
populations, have so far proved inconclusive. \citet{North03}
found that short-period binaries in both the Large and Small
Magellanic Clouds seem to have been circularized in agreement with
the theory of dynamical tides. However, \citet{Meibom05} show that
with the exception of the Hyades, the stars in the clusters that
they observed were considerably more circularized than any of the
known dissipation mechanisms would predict. Furthermore, they find
with a high degree of certainty, that older clusters are more
circularized than younger ones, thereby contradicting the Hayashi
phase circularization model.

\begin{figure}
\plotone{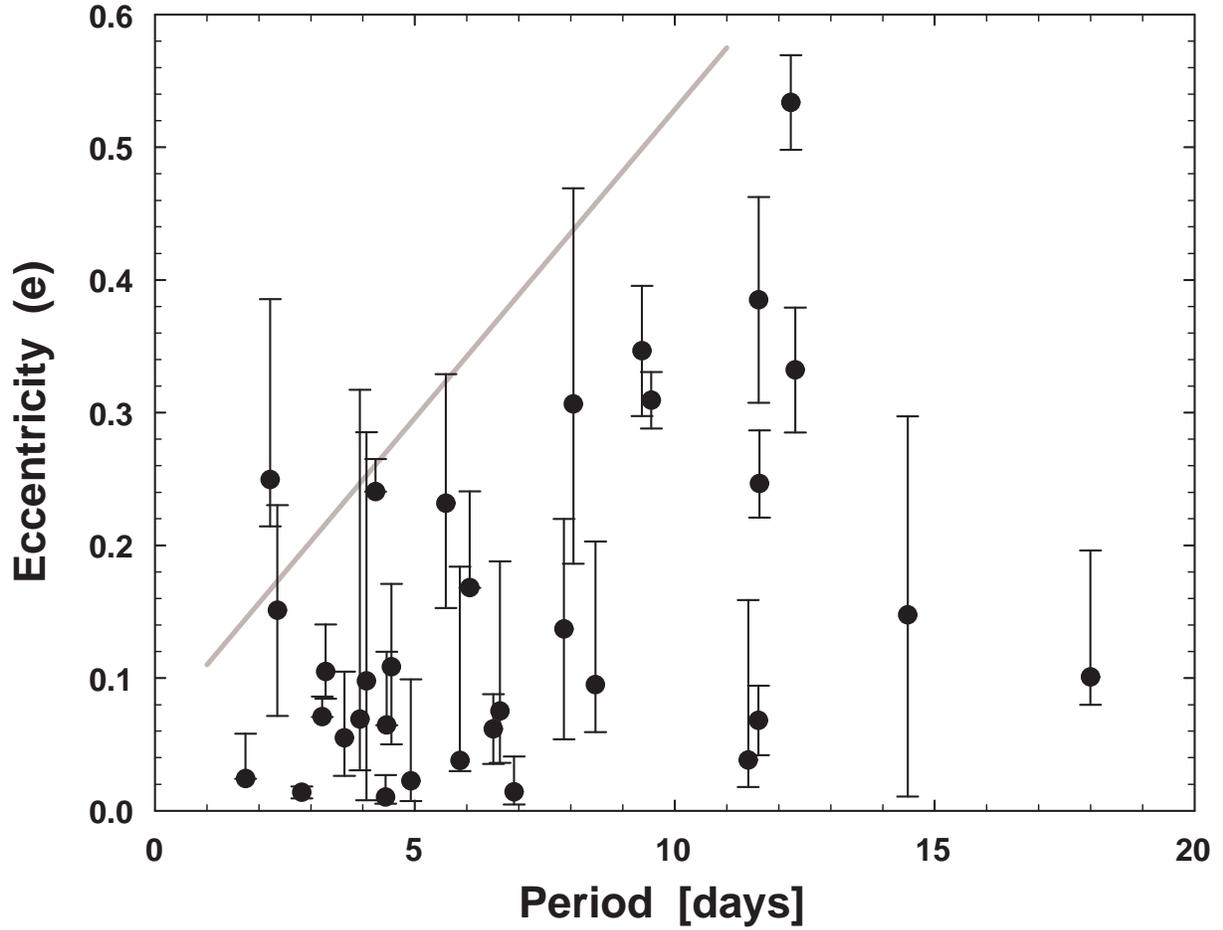}
\caption{The period-eccentricity
relation. The lower end of the error bars were truncated, where
needed, by the measured lower limit, $|e \cos \omega|$. Note the
lack of eccentric short period systems. The diagonal line is
provided to guide the eye.}
\label{figPeriodEcc}
\end{figure}

Encouraged by the statistical effect of circularization that can
be seen in our catalog (Figure \ref{figPeriodEcc}), we further
estimated $t_{circ}$ for each of the eccentric systems as follows.
\citet{Zahn77, Zahn78} provides an estimate for the orbital
circularization timescale due to turbulent dissipation in stars
possessing a convective envelope, assuming that corotation has
been achieved:

\begin{equation}
t_{circ} = \frac{1}{21q(1+q)k_2} \left(\frac{MR^2}{L}\right)^{1/3}\left(\frac{a}{R}\right)^8
\end{equation}

where $M, R, L$ are the star's mass, radius and luminosity, and
$k_2$ is the apsidal motion constant of the star, which is
determined by its internal structure and dynamics.

More massive stars, which do not have a convective envelope but
rather develop a radiative envelope, are thought to circularize
their orbit using radiative damping \citep{Zahn75, Claret97}. This
is a far slower mechanism, whose circularization timescale can be
estimated by:

\begin{equation}
t_{circ} = \frac{2}{21 q (1+q)^{11/6} E_2} \left(\frac{R^3}{GM}\right)^{1/2} \left(\frac{a}{R}\right)^{21/2}
\end{equation}

where $E_2$ is the tidal torque constant of the star, and $G$ is
the universal gravitational constant. We can greatly simplify
these expressions by applying Kepler's law [$a^3 =
GM(1+q)(P/{2\pi})^2$], and adopt the \citet{Cox00} power law
approximations for the main-sequence mass-radius and
mass-luminosity relations. For the convective envelope case, we
adopt the late-type mass-radius relation ($M < 1.3 M_\sun$), and
for the radiative envelope case we adopt the early-type
mass-radius relation ($M \geq 1.3 M_\sun$), thus arriving at:

\begin{equation}
\label{eqCircTime}
t_{circ} \simeq \cases{
0.53 Myr \; (k_2 / 0.005)^{-1}q^{-1}(1+q)^{5/3} \left(P / day\right)^{16/3} \left(M / M_\sun \right)^{-4.99}, \ M < 1.3 M_\sun\cr
1370 Myr \; (E_2 / 10^{-8})^{-1}q^{-1}(1+q)^{5/3} \left(P / day\right)^7 \left(M / M_\sun \right)^{-2.76}, \ M \geq 1.3 M_\sun \cr}
\end{equation}

Determining the value of $k_2$ and $E_2$ is the most difficult
part of this exercise, since their values are a function of the
detailed structure and dynamics of the given star, which in turn
changes significantly as the star evolves \citep{Claret97,
Claret02}. In our calculation, we estimate these values by
interpolating published theoretical tables [$k_2$: \citet{Zahn94},
$E_2$: \citet{Zahn75, Claret97}]. Since both stellar components
contribute to the circularization process, the combined
circularization timescale becomes $t_{circ} = 1/(t_{circ,1}^{-1} +
t_{circ,2}^{-1})$, where the subscripts $1$ and $2$ refer to the
primary and secondary binary components \citep{Claret97}. In Table
\ref{tableEccentric}, we list the combined circularization
timescale for each of the eccentric EBs we identify.

The value of $t_{circ}$ for most of the eccentric systems (21 of
34) is larger than the Hubble time, indicating that no significant
circularization is expected to have taken place since they settled
on the main sequence. About a quarter of the eccentric systems (8
of 34) have a $t_{circ}$ smaller than the Hubble time but larger
than $1\:Gyr$. While circularization is underway, the fact that
they are still eccentric is consistent with theoretical
expectations. The remaining systems (5 of 34) all have $t_{circ} <
1\:Gyr$, have periods less than 3.3 days, and unless they are
extremely young, require an explanation for their eccentric
orbits. Two of these EBs (T-Tau0-02487 and T-Tau0-03916) are
located near the star forming regions of Taurus, supporting the
hypothesis that they are indeed young. However, this hypothesis
does not seem to be adequate for T-Cas0-02603, which has a period
of only 2.2 days and $t_{circ} \simeq 0.26\:Gyr$, while possessing
a large eccentricity of $e \simeq 0.25$. An alternative
explanation is that some of these binaries were once further
apart, having larger orbital periods, and thus larger
circularization timescales. These systems may have been involved
in a comparably recent interaction with a third star (a collision
or near miss), or have been influenced by repeated resonant
perturbations of a tertiary companion.

\begin{figure}
\plotone{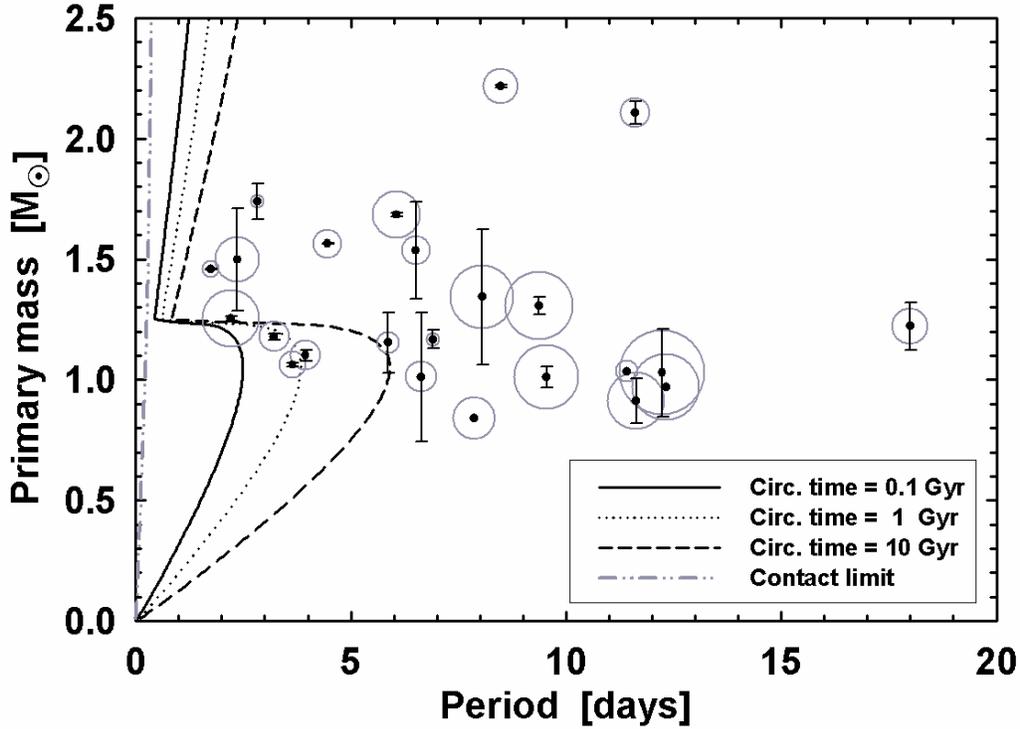}
\caption{The period-primary mass relation
for eccentric EBs. We included all systems with well-determined
masses. The area of the grey circles is proportional to the EB's
eccentricity. All the curves are theoretical boundaries, assuming
that the binary components are both on the main sequence and have
equal masses ($q=1$). The left-most
dot-dash line demarcates the binary contact limit, and the
remaining curves mark systems with increasing circularization time
(see equation \ref{eqCircTime}). Note the abrupt increase in the
circularization time for systems more massive than
$\sim1.25M_\sun$, at which point the stellar convective envelope
becomes radiative, and thus far less efficient at tidal
dissipation.}
\label{figPeriodM1}
\end{figure}

Finally, we would like to draw the reader's attention to our
shortest-period eccentric EB, T-Cas0-00394, whose period is a mere
1.7 days. Notably, this system is entirely consistent with theory,
since its mass falls in a precarious gap, where the stellar
envelopes of its components are no longer convective, yet their
radiative envelopes are not sufficiently extended to produce
significant tidal drag (see Figure \ref{figPeriodM1}).

\placefigure{figPeriodEcc}
\placefigure{figPeriodM1}

\subsection{Abnormal EBs}
\label{subsecAbnomalEBs}

The third group of EBs consists of 20 abnormal systems (see Table
\ref{tableAbnormal}, and Figures \ref{figAbnormalEBs1} and
\ref{figAbnormalEBs2}). While possessing the distinctive
characteristics of EBs, these LCs stood out during manual
inspection for a variety of reasons. These systems underline the
difficulty of fully automating any LC pipeline, as any such system
will inevitably need to recognize atypical EBs that were not
encountered before.

The LCs we listed can be loosely classified into groups according
to the way they deviate from a simple EB model. A few cases
exhibited pulsation-like fluctuations that were not synchronized
with the EB period (shorter-period: T-Dra0-00398, longer-period:
T-Lyr1-00359, T-Per1-00750). These fluctuations may be due either
to the activity of an EB component, or to a third star whose light
is blended with the binary. In principle, one can identify the
active star by examining the amplitude of the fluctuations during
the eclipses. If the fluctuations originate from one of the
components, their observed amplitude will be reduced when the
component is being eclipsed. In such a case, if the fluctuations
are due to pulsations, they can further provide independent
constraints to the stellar properties through astro-seismological
models \citep{Mkrtichian04}. To identify such fluctuating EBs one
must subtract the fitted EB model from the LC, and evaluate the
residuals [e.g. \citet{Pilecki07}]. When the fluctuation period is
fixed, one can simply search the residual LC using a periodogram,
as was done in step (1) of our pipeline (see \S\ref{secMethod}).
However, when the fluctuation period varies (i.e. non-coherent),
as in the aforementioned LCs, one must employ alternate methods,
since simply phasing their LC will not produce any discernable
structure. For LCs with long-period fluctuations one can directly
search the residuals for time dependencies, while for LCs with
short-period fluctuations one can search the residuals for
non-Gaussian distributions. However, in practice these
measurements will likely not be robust, as there are many
instrumental effects that can produce false positives. Thus, we
employ a search for auto-correlations in the residual time series,
which overcomes most instrumental effects, while providing a
reliable indicator for many types of pseudo-periodic fluctuations.

The remaining systems had LC distortions that appear to be
synchronized with the orbital period. The source of these
fluctuations is likely due to long-lasting surface inhomogeneities
on one or both of the rotationally synchronized components. When
the LC has brief periodic episodes of darkening (T-And0-11476,
T-Cas0-13944, T-Cyg1-07584, T-Dra0-04520), they can usually be
explained as stable star spots, but brief periodic episodes of
brightening (T-And0-04594, T-Her0-08091), which may indicate the
presence of stable hot-spots, are more difficult to interpret.
This phenomenon is especially puzzling in the aforementioned two
cases, in which the brightening episodes are briefer than one
would expect from a persistent surface feature and repeat at the
middle of both plateaus.

When the two plateaus of a LC are not flat, they are usually
symmetric about the center of the eclipses. This is due to the
physical mirror symmetry about the line intersecting the binary
components' centers. When the axis of symmetry does not coincide
with the center of eclipse (T-And0-00920, T-Cyg1-08866,
T-Dra0-03105, T-Lyr1-07584, T-Lyr1-15595), a phenomenon we term
``eclipse offset'', we conclude that this symmetry must somehow be
broken. This may occur if the EB components are not rotationally
synchronized, or have a substantial tidal lag. Another form of
this asymmetry can appear as an amplitude difference between the
two LC plateaus (T-Her0-03497, T-Lyr1-13166, T-Per1-08789,
T-UMa0-03090). This phenomenon, which was originally called the
``periastron effect'' and has since been renamed the ``O'Connell
effect'', has been known for over a century, and has been
extensively studied [e.g. \citet{OConnell51, Milone86}]. Classic
hypotheses suggest an uneven distribution of circumstellar
material orbiting with the binary \citep{Struve48} or surrounding
the stars \citep{Mergentaler50}, either of which could induce a
preferential $H^-$ absorption on one side. \citet{Binnendijk60}
was the first of many to suggested that the this asymmetry is due
to subluminous regions of the stellar surface (i.e. star spots).
However, this explanation also requires the stars to be
rotationally synchronized, and for the spots to be stable over the
duration of the observations. Alternate models abound, including a
hot spot on one side of a component brought about through mass
transfer from the other component, persistent star spots created
by an off-axis magnetic field, and circumstellar material being
captured by the components and heating one side of both stars
\citep{Liu03}. As with many phenomena that have multiple possible
models, the true answer may involve a combination of a number of
these mechanisms, and will likely vary from system to system
\citep{Davidge84}.

Finally, a few particularly unusual LCs (T-Dra0-03105,
T-Lyr1-05984) display a very large difference between their
eclipse durations. Although a moderate difference could be
explained by an eccentric orbit, such extreme eccentricities in
systems with such short orbital periods (0.5 and 1.5 days) are
highly unlikely.

\section{Conclusions}

We presented a catalog of 773 eclipsing binaries found in 10
fields of the TrES survey, identified and analyzed using an
automated pipeline. We described the pipeline we used to identify
and model them. The pipeline was designed to be mostly automated,
with manual inspections taking place only once the vast majority
of non-EB LCs had been automatically filtered out. At the final
stage of the pipeline we classified the EBs into 7 groups:
eccentric, circular, ambiguous-equal, ambiguous-unequal, inverted,
Roche-lobe-filling, and abnormal. The former four groups were all
successfully modeled with our model fitting program. However, the
latter three groups possessed significant additional physical
phenomena (tidal distortions, mass-transfer, and surface
activity), which did not conform to the simple detached-EB model
we employed.

We highlighted three groups of binaries, which may be of
particular interest and warrant follow-up observations. These
groups are: low-mass EBs, EBs with eccentric orbits, and abnormal
EBs. The low-mass EBs (both components $< 0.75M_{\sun}$) allow one
to probe the mass-radius relation at the bottom of the
main-sequence. Only 7 such EBs have previously been confirmed, and
the physical properties of many of them are inconsistent with
current theoretical models. Our group of 10 new candidates will
likely provide considerable additional constraints to the models,
and the discovery of 2 long period systems could help confirm a
recent hypothesis that this inconsistency is due to stellar
magnetic activity. The eccentric-orbit EBs may help confirm and
constrain tidal circularization theory, as many of them have
comparably short circularization timescales. We demonstrated that,
as one would predict from the theory, the shortest period systems
fall within a narrow range of masses, in which their stellar
envelopes cease to be convective yet their envelopes are not
extended enough to produce significant tidal drag. The abnormal
EBs seem to show a plethora of effects that are indicative of
asymmetries, stellar activity, persistent hot and cold spots, and
a host of other physical phenomena. Some of these systems may
require dedicated study to be properly understood.

In the future, as LC datasets continue to grow, it will become
increasingly necessary to use such automated pipelines to identify
rare and interesting targets. Such systematic searches promise a
wealth of data that can be used to test and constrain theories in
regions of their parameter-space that were previously
inaccessible. Furthermore, even once the physics of ``vanilla''
systems has been solved, more complex cases will emerge to
challenge us to achieve a better understanding of how stars form,
evolve and interact.

\acknowledgments

We would like to thank Tsevi Mazeh for many useful discussions, as
well as S{\o}ren Meibom for his repeated help. We would also like
to thank Sarah Dykstra for her continuous support throughout the
preparation of this paper. We are thankful to the staff of the
Palomar Observatory for their assistance in operating the Sleuth
instrument, and we acknowledge support from NASA through grant
NNG05GJ29G issued through the Origins of Solar Systems Program.
This research has made use of the NASA's Astrophysics Data System
Bibliographic Services, the SIMBAD database, operated at CDS,
Strasbourg, and the VSX database, which was created by Christopher
Watson for the AAVSO. This publication also utilizes data products
from the Two Micron All Sky Survey, which is a joint project of
the University of Massachusetts and the Infrared Processing and
Analysis Center/California Institute of Technology, funded by NASA
and NSF. Finally, we would like to thanks the anonymous referee
for very insightful comments and suggestions, which significantly
improved this manuscript.

\appendix

\section{Rejecting single-eclipse EB Models}
\label{appendixSingleEclipse}

An EB LC comprised of a deep eclipse and a very shallow eclipse,
can occur in one of two ways. Either the secondary component is
luminous but extremely small (e.g. a white dwarf observed in UV),
thus producing a shallow primary eclipse, or the secondary
component is comparably large but extremely dim, thus producing a
shallow secondary eclipse. The first case, though possible [e.g.
\citet{Maxted04}], is extremely rare, and will have a signature
``flat bottom'' to the eclipse. We have not encountered such a LC
in our dataset. The second case will have a rounded eclipse
bottom, due to the primary component's limb darkening. Assuming
this latter contingency, in which the secondary component is dark
in comparison to the primary component, we can place a lower bound
to its radius ($R_2$):

\begin{equation}
R_2 \geq R_1 \sqrt{1 - 10^{-0.4 \Delta mag_1}}\ ,
\end{equation}

where $R_1$ is the radius of the primary component, and $\Delta
mag_1$ is the magnitude depth of the primary eclipse. Thus, if the
eclipse is very deep, the size of the secondary component must
approach the size of the primary component. However, coeval
short-period detached EBs with components of similar sizes yet
desperate luminosities are expected to be very rare, assuming they
follow normal stelar evolution. Therefore, if only one eclipse is
detected, and it is both rounded and sufficiently deep, we may
conclude that this configuration entry is likely to be incorrect,
and that the correct configuration has double the orbital period
and produces two equal eclipses. Only when we cannot apply such a
period doubling solution (i.e. when the secondary eclipse is
detectable), do we resort to questioning our assumption of normal
stelar evolution (see classification group V, described in
\S\ref{secMethod}).

\section{Description of the Catalog Fields}
\label{appendixCatalogDescription}

Due to the large size of the catalog we were only able to list
small excerpts of it in the body of this paper. Readers interested
in viewing the catalog in its entirety can download it
electronically. Note that although the catalog lists 773 unique
systems, each of the 103 ambiguous EBs appear in both possible
configurations (see \S\ref{secMethod}), raising the total number
of catalog entries to 876. Below, we briefly describe the
catalog's 38 columns. The column units, if any, are listed in
square brackets.

\renewcommand{\theenumi}{\arabic{enumi}}
\begin{enumerate}
\item $Category$ - the EB's classification (see \S\ref{secMethod}).
\item $Binary\ name$ - the EB's designation, which is composed of its TrES field (see Table \ref{tableFieldsObs}) and index.
\item $\alpha$ - the EB's right ascension (J2000).
\item $\delta$ - the EB's declination (J2000).
\item $Period$ [days] - the EB's orbital period.
\item $Period \ uncertainty$ [days] - the uncertainty in the EB's orbital period.
\item $Mass_1$ [$M_\sun$] - the mass of the EB's primary (more massive) component.
\item $Mass_1\ uncertainty$ [$M_\sun$] - the uncertainty in the primary component's mass.
\item $Mass_2$ [$M_\sun$] - the mass of the EB's secondary (less massive) component.
\item $Mass_2\ uncertainty$ [$M_\sun$] - the uncertainty in the secondary component's mass.
\item $Age$ [Gyr] - the age of the EB (assumed to be coeval).
\item $Age\ uncertainty$ [Gyr] - the uncertainty in the EB's age.
\item $Score$ - a weighted reduced $\chi^2$ of the MECI model fit [see \citet{Devor06b} for further details].
\item $Isochrone\ source$ - isochrone tables used [Y2: \citet{Kim02}, or Baraffe: \citet{Baraffe98}].
\item $Color\ weighting$ - the relative weight ($w$) of the LC fit, compared to the color fit [see \citet{Devor06b} for further details].
\item $PM\ source$ - the database that provided the proper motion measurement [UCAC: \citet{Zacharias04}, USNO-B: \citet{Monet03}, or Salim03: \citet{Salim03}].
\item $PM_{\alpha}$ [mas/yr] - the right ascension component of the EB's proper motion.
\item $PM_{\delta}$ [mas/yr] - the declination component of the EB's proper motion.
\item $Location\ error$ [arcsec] - the distance between our listed location (columns 3 and 4) and the location listed by the proper motion database.
\item $mag_B$ - the USNO-B $B$-band observational magnitude of the EB (average of both magnitude measurements, if available).
\item $mag_R$ - the USNO-B $R$-band observational magnitude of the EB (average of both magnitude measurements, if available).
\item $Third-light\ fraction$ - the fraction of third-light flux ($R$-band) blended into the LC (i.e. the flux within 30", excluding the target, divided by the total flux within 30").
\item $mag_J$ - the 2MASS observational $J$-band magnitude of the EB, converted to ESO $J$-band.
\item $mag_H$ - the 2MASS observational $H$-band magnitude of the EB, converted to ESO $H$-band.
\item $mag_K$ - the 2MASS observational $K_s$-band magnitude of the EB, converted to ESO $K$-band.
\item $Mag_J$ - the absolute ESO $J$-band magnitude of the EB listed in the isochrone tables.
\item $Mag_H$ - the absolute ESO $H$-band magnitude of the EB listed in the isochrone tables.
\item $Mag_K$ - the absolute ESO $K$-band magnitude of the EB listed in the isochrone tables.
\item $Distance$ [pc] - the distance to the EB, as calculated from the extinction-corrected distance modulus.
\item $A(V)$ - the EB's V-mag absorption due to Galactic interstellar extinction (assuming $R_V = 3.1$).
\item $\sin(i)$ - the sine of the EB's orbital inclination.
\item $|e \cos(\omega)|$ - a robust lower limit for the EB's eccentricity (see equation \ref{eqOmC}).
\item $Eccentricity$ - the orbital eccentricity of the EB.
\item $Eccentricity\ uncertainty$ - the uncertainty in the orbital eccentricity of the EB.
\item $\Delta mag_1$ - the $r$-band primary (deeper) eclipse depth in magnitudes.
\item $Epoch_1$ - the Heliocentric Julian date (HJD) at the center of a primary eclipse, minus 2400000.
\item $\Delta mag_2$ - the $r$-band secondary (shallower) eclipse depth in magnitudes.
\item $Epoch_2$ - the Heliocentric Julian date (HJD) at the center of a secondary eclipse, minus 2400000.
\end{enumerate}

Note that the value of the uncertainties (columns 6, 8 10, 12, and
34), were calculated by measuring the curvature of the
parameter-space $\chi^2$ contour, near its minimum. This method
implicitly assumes a Gaussian distribution of the parameter
likelihood. If the likelihood distribution not Gaussian, but
rather has a flattened (boxy) distribution, then the computed
uncertainty becomes large. In extreme cases the estimated formal
uncertainty can be larger than the measurement itself.

{}

\clearpage

\begin{deluxetable}{cccccccc}
\tabletypesize{\tiny}
\rotate
\tablecaption{Observational parameters of the TrES fields}
\tablewidth{0pt}
\tablehead{\colhead{Field} &
           \colhead{Constellation} &
           \colhead{\begin{tabular}{c} $\alpha$\\ (J2000)\tablenotemark{a}\end{tabular}} &
           \colhead{\begin{tabular}{c} $\delta$\\ (J2000)\end{tabular}} &
           \colhead{\begin{tabular}{c} Galactic\\ coordinates (l,b)\end{tabular}} &
           \colhead{\begin{tabular}{c} Starting\\ epoch (HJD)\end{tabular}} &
           \colhead{\begin{tabular}{c} Ending\\ epoch (HJD)\end{tabular}} &
           \colhead{\begin{tabular}{c} Duration\\ (days)\end{tabular}}}
\startdata
And0& Andromeda      & 01 09 30.1255& +47 14 30.453& (126.11, -015.52)& 2452878.9& 2452934.9& 56.0\\
Cas0& Cassiopeia     & 00 39 09.8941& +49 21 16.519& (120.88, -013.47)& 2453250.8& 2453304.6& 53.8\\
CrB0& Corona Borealis& 16 01 02.6616& +33 18 12.634& (053.49, +048.92)& 2453493.8& 2453536.8& 43.0\\
Cyg1& Cygnus         & 20 01 21.5633& +50 06 16.902& (084.49, +010.28)& 2453170.7& 2453250.0& 79.3\\
Dra0& Draco          & 16 45 17.8177& +56 46 54.686& (085.68, +039.53)& 2453093.8& 2453163.0& 69.2\\
Her0& Hercules       & 16 49 14.2185& +45 58 59.963& (071.61, +039.96)& 2452769.9& 2452822.0& 52.1\\
Lyr1& Lyra           & 19 01 26.3713& +46 56 05.325& (077.15, +017.86)& 2453541.8& 2453616.7& 74.9\\
Per1& Perseus        & 03 41 07.8581& +37 34 48.712& (156.37, -014.04)& 2453312.8& 2453402.8& 90.0\\
Tau0& Taurus         & 04 20 21.2157& +27 21 02.713& (169.83, -015.94)& 2453702.7& 2453770.9& 68.2\\
UMa0& Ursa Major     & 09 52 06.3560& +54 03 51.596& (160.87, +047.70)& 2453402.9& 2453487.8& 84.9\\
\enddata
\tablenotetext{a}{ICRS 2000.0 coordinates of the guide star, which is located at the center of the field of view.}
\label{tableFieldsObs}
\end{deluxetable}

\begin{deluxetable}{ccccccc}
\tabletypesize{\tiny}
\rotate
\tablecaption{The number of sources and yield of the TrES fields}
\tablewidth{0pt}
\tablehead{\colhead{Field} &
           \colhead{\begin{tabular}{c} Number\\ of LCs\end{tabular}} &
           \colhead{\begin{tabular}{c} Number of observations\\ in each LC\end{tabular}} &
           \colhead{\begin{tabular}{c} Fraction\\ RMS $<$ 1\%\end{tabular}} &
           \colhead{\begin{tabular}{c} Fraction\\ RMS $<$ 2\%\end{tabular}} &
           \colhead{\begin{tabular}{c} Found\\ EBs\end{tabular}} &
           \colhead{\begin{tabular}{c} EB discovery\\ yield\end{tabular}}}
\startdata
And0& 26495& 2357& 16.5\%& 40.4\%& 111& 0.42\%\\
Cas0& 22615& 2069& 11.0\%& 38.2\%& 119& 0.53\%\\
CrB0& 18954& 1287& 11.0\%& 22.4\%&  28& 0.15\%\\
Cyg1& 17439& 3256& 30.3\%& 65.7\%& 125& 0.72\%\\
Dra0& 15227& 2000& 11.8\%& 26.4\%&  42& 0.28\%\\
Her0& 15916&  974& 16.8\%& 35.0\%&  28& 0.18\%\\
Lyr1& 22964& 2815& 19.4\%& 49.0\%& 135& 0.59\%\\
Per1& 20988& 1647& 15.9\%& 38.4\%&  93& 0.44\%\\
Tau0& 14442& 1171& 13.1\%& 32.5\%&  68& 0.47\%\\
UMa0& 10405& 1343& 13.6\%& 29.5\%&  24& 0.23\%\\
\enddata
\label{tableFieldsYield}
\end{deluxetable}

\begin{deluxetable}{ccccccccccc}
\tabletypesize{\tiny}
\rotate
\tablecaption{Eccentric EBs}
\tablewidth{0pt}
\tablehead{\colhead{Object} & \colhead{$\alpha$ (J2000)} & \colhead{$\delta$ (J2000)} & \colhead{Period $[days]$\tablenotemark{a}} & \colhead{$|e \cos\omega|_{timing}$\tablenotemark{b}} & \colhead{$|e \cos\omega|_{adopted}$\tablenotemark{c}} & \colhead{$e$\tablenotemark{d}} & \colhead{$M_1/M_\sun$} & \colhead{$M_2/M_\sun$}& \colhead{age $[Gyr]$} & \colhead{$t_{circ}$ $[Gyr]$}}
\startdata
T-And0-04144& 01 17 35.247& 49 46 16.97&  7.869& 0.0072& 0.0068& $0.14^{+0.08}_{-0.08}$  & 0.84 (-1)\tablenotemark{e}      & 0.54 (-1)      & 10.0 (-3)     & 140\\
T-And0-17158& 01 10 09.143& 48 18 19.68& 11.415& 0.0182& 0.0180& $0.038^{+0.12}_{-0.02}$ & 1.03 (-1)      & 0.92 (-1)      & 10.0 (-3)     & 370\\
T-And0-24609& 00 58 29.826& 49 25 08.88& 17.997& 0.0794& 0.0799& $0.10^{+0.10}_{-0.02}$  & 1.22 $\pm$ 0.10& 1.10 $\pm$ 0.30& 5.4 $\pm$ 12.0& 6400\\
T-Cas0-00394& 00 32 51.608& 49 19 39.36&  1.746& 0.0235& 0.0242& $0.024^{+0.03}_{-0.001}$& 1.46 $\pm$ 0.01& 1.44 $\pm$ 0.01& 3.4 $\pm$ 0.3 & 260\\
T-Cas0-02603& 00 47 08.610& 50 37 19.32&  2.217& 0.2098& 0.2143& $0.25^{+0.14}_{-0.04}$  & 1.25 $\pm$ 0.01& 0.75 $\pm$ 0.04& 5.4 $\pm$ 5.0 & 0.26\\
T-Cas0-04534& 00 31 04.585& 51 52 10.88&  6.909& 0.0057& 0.0048& $0.014^{+0.03}_{-0.01}$ & 1.17 $\pm$ 0.04& 0.96 $\pm$ 0.15& 6.4 $\pm$ 6.9 & 29\\
T-Cas0-04947& 00 47 10.336& 50 45 12.36&  3.285& 0.0845& 0.0845& $0.10^{+0.04}_{-0.02}$  & 1.04 (-1)      & 0.86 (-1)      & 10.0 (-3)     & 0.53\\
T-Cas0-05165& 00 43 59.256& 51 14 00.07&  2.359& 0.0311& 0.0327& $0.15^{+0.08}_{-0.08}$  & 1.50 $\pm$ 0.21& 0.76 $\pm$ 0.17& 2.7 $\pm$ 2.9 & 0.34\\
T-Cas0-07630& 00 37 23.347& 47 19 20.68&  5.869& 0.0200& 0.0298& $0.038^{+0.15}_{-0.008}$& 1.15 $\pm$ 0.12& 0.87 $\pm$ 0.34& 5.9 $\pm$ 9.7 & 13\\
T-Cyg1-01364& 20 09 38.211& 49 05 08.02& 12.233&   N/A & 0.3254& $0.53^{+0.04}_{-0.04}$  & 1.03 $\pm$ 0.18& 0.50 $\pm$ 0.09& 0.4 $\pm$ 1.2 & 1100\\
T-Cyg1-01373& 19 55 44.105& 52 13 34.61&  4.436& 0.0059& 0.0054& $0.010^{+0.02}_{-0.005}$& 0.97 (-1)      & 0.82 (-1)      & 10.0 (-3)     & 3.0\\
T-Cyg1-01994& 20 03 03.111& 52 42 04.17& 14.482&   N/A & 0.0107& $0.15^{+0.15}_{-0.14}$  & 1.80 (-1)      & 1.06 (-1)      & 0.20 (-2)     & 2300\\
T-Cyg1-02304& 20 02 04.388& 47 34 14.75&  5.596& 0.1549& 0.1529& $0.23^{+0.10}_{-0.08}$  & 2.20 $\pm$ 1.28& 0.72 $\pm$ 0.41& 0.7 $\pm$ 4.8 & 46\\
T-Cyg1-02624& 19 59 25.926& 52 23 59.91& 11.608& 0.0172& 0.0172& $0.068^{+0.03}_{-0.03}$ & 2.11 $\pm$ 0.05& 1.52 $\pm$ 0.03& 0.3 $\pm$ 0.1 & $10^7$\\
T-Cyg1-06677& 20 07 25.526& 52 22 00.54&  6.512& 0.0077& 0.0069& $0.062^{+0.03}_{-0.03}$ & 1.54 $\pm$ 0.20& 1.31 $\pm$ 0.22& 1.6 $\pm$ 1.9 & $10^6$\\
T-Cyg1-07248& 19 54 45.937& 50 24 05.32&  6.058& 0.1674& 0.1681& $0.17^{+0.07}_{-0.001}$ & 1.68 $\pm$ 0.01& 0.87 $\pm$ 0.20& 2.0 $\pm$ 2.1 & 33\\
T-Cyg1-07297& 20 10 46.910& 49 09 29.42& 11.613& 0.3019& 0.3010& $0.38^{+0.08}_{-0.08}$  & 0.97 (-1)      & 0.55 (-1)      & 10.0 (-3)     & 830\\
T-Cyg1-07584& 19 58 58.012& 47 38 19.26&  4.925& 0.0074& 0.0074& $0.022^{+0.08}_{-0.01}$ & 0.94 (-1)      & 0.90 (-1)      & 10.0 (-3)     & 4.7\\
T-Cyg1-09934& 20 10 44.209& 51 07 51.77&  4.549& 0.0505& 0.0501& $0.11^{+0.06}_{-0.06}$  & 1.35 $\pm$ 0.64& 0.94 $\pm$ 0.41& 3.5 $\pm$ 5.6 & 5.6\\
T-Cyg1-15752& 20 13 52.454& 50 52 23.12&  9.372& 0.2402& 0.2402& $0.35^{+0.05}_{-0.05}$  & 1.31 $\pm$ 0.04& 1.05 $\pm$ 0.11& 3.6 $\pm$ 4.9 & 230\\
T-Lyr1-09931& 18 59 08.441& 48 36 00.04& 11.632& 0.2207& 0.2209& $0.25^{+0.04}_{-0.03}$  & 0.91 $\pm$ 0.09& 0.67 $\pm$ 0.08& 2.7 $\pm$ 3.3 & 730\\
T-Lyr1-13841& 19 06 26.558& 48 28 47.04&  6.640& 0.0362& 0.0362& $0.075^{+0.11}_{-0.04}$ & 1.01 $\pm$ 0.27& 1.01 $\pm$ 0.24& 8.7 $\pm$ 13.1 & 19\\
T-Lyr1-14413& 19 03 41.143& 47 36 55.78& 39.861& 0.5922& 0.6240& $0.64^{+0.006}_{-0.006}$& 1.08 $\pm$ 0.34& 0.96 $\pm$ 0.26& 6.4 $\pm$ 18.9 & $10^5$\\
T-Lyr1-14508& 18 57 40.271& 48 40 51.28&  8.050& 0.1861& 0.1862& $0.31^{+0.16}_{-0.12}$  & 1.34 $\pm$ 0.28& 1.20 $\pm$ 0.78& 2.9 $\pm$ 8.4 & 220\\
T-Lyr1-22359& 19 10 54.290& 49 26 06.95& 12.319& 0.1990& 0.1984& $0.33^{+0.05}_{-0.05}$  & 0.97 $\pm$ 0.48& 0.97 $\pm$ 0.46& 6.9 $\pm$ 29.3 & 550\\
T-Per1-00769& 03 31 43.915& 36 31 52.36&  3.648& 0.0248& 0.0263& $0.055^{+0.05}_{-0.03}$ & 1.06 $\pm$ 0.01& 0.65 $\pm$ 0.03& 7.6 $\pm$ 2.1 & 1.4\\
T-Per1-04218& 03 35 33.667& 40 00 49.18&  4.070& 0.0072& 0.0079& $0.10^{+0.19}_{-0.09}$  & 0.94 (-1)      & 0.72 (-1)      & 10.0 (-3)     & 2.4\\
T-Per1-05205& 03 34 19.432& 39 32 44.41&  8.472& 0.0558& 0.0592& $0.095^{+0.11}_{-0.04}$ & 2.22 $\pm$ 0.01& 1.17 $\pm$ 0.28& 0.9 $\pm$ 1.7 & 210\\
T-Per1-08252& 03 52 00.670& 40 03 47.73&  4.457& 0.0656& 0.0645& $0.065^{+0.06}_{-0.001}$& 1.56 $\pm$ 0.01& 1.40 $\pm$ 0.34& 2.4 $\pm$ 2.5 & $10^5$\\
T-Per1-11424& 03 47 56.473& 37 31 31.83&  4.247& 0.2403& 0.2404& $0.24^{+0.02}_{-0.006}$ & 1.01 (-1)      & 0.82 (-1)      & 10.0 (-3)     & 2.3\\
T-Per1-17327& 03 40 45.644& 34 47 57.26&  3.946& 0.0332& 0.0305& $0.069^{+0.25}_{-0.04}$ & 1.10 $\pm$ 0.02& 1.09 $\pm$ 0.09& 8.4 $\pm$ 16.4& 1.2\\
T-Tau0-02487& 04 21 55.933& 25 35 49.28&  2.826& 0.0125& 0.0054& $0.014^{+0.005}_{-0.005}$&1.74 $\pm$ 0.07& 1.01 $\pm$ 0.08& 1.6 $\pm$ 0.7 & 0.39\\
T-Tau0-03916& 04 23 37.351& 25 46 36.00&  3.217& 0.0713& 0.0706& $0.071^{+0.02}_{-0.004}$& 1.18 $\pm$ 0.01& 1.15 $\pm$ 0.03& 6.0 $\pm$ 4.4 & 0.56\\
T-UMa0-01822& 09 53 37.710& 52 45 44.72&  9.551& 0.1502& 0.1503& $0.31^{+0.02}_{-0.02}$  & 1.01 $\pm$ 0.04& 1.00 $\pm$ 0.05& 8.3 $\pm$ 4.8 & 130\\
\enddata
\tablenotetext{a}{The full precision of the measured period is
listed in the electronic version of the catalog, together with its
uncertainty and the epoch of the center of eclipse (see appendix
\ref{appendixCatalogDescription}).}
\tablenotetext{b}{Measurements
made using the eclipse timing of step (4). Although these values
are approximations, they do not suffer from nearly as much
numerical error as the DEBiL measurement, and are therefore
usually accurate. ``N/A'' marks LCs for which there were too few
eclipses to be able to apply the timing method.}
\tablenotetext{c}{The adopted value is a combination of the values
measured with the timing method and with DEBiL.}
\tablenotetext{d}{The uncertainties of the eccentricities are
non-Gaussian, since they have a strict lower bound ($e \geq |e
\cos\omega|$). We truncated the quoted lower uncertainties at this
value, though even at this truncated value the real uncertainty is
beyond $1\sigma$.}
\tablenotetext{e}{When the most likely model is
at the edge of the parameter space, MECI is not able to bound the
solution, and therefore cannot estimate the uncertainties. We mark
(-3) when the upper limit was reached, (-2) when the lower limit
was reached, and (-1) if one of the other parameters is at its
limit.} \label{tableEccentric}
\end{deluxetable}

\begin{deluxetable}{cccc}
\tabletypesize{\tiny}
\rotate
\tablecaption{EBs that fill at least one of their Roche-lobes (first 20)}
\tablewidth{0pt}
\tablehead{\colhead{Object} & \colhead{$\alpha$ (J2000)} & \colhead{$\delta$ (J2000)} & \colhead{Period $[days]$}}
\startdata
T-And0-03774& 00 59 01.029& 46 47 17.08&  1.362\\
T-And0-04813& 01 16 37.880& 47 33 23.43&  0.552\\
T-And0-05140& 01 03 22.258& 44 56 24.31&  0.981\\
T-And0-05153& 01 18 48.278& 49 39 36.86&  0.492\\
T-And0-05343& 00 52 55.122& 48 01 37.68&  0.824\\
T-And0-07638& 01 09 27.871& 49 20 33.81&  0.403\\
T-And0-07892& 00 56 15.567& 48 39 10.73&  0.380\\
T-And0-08330& 01 19 15.949& 48 00 17.45&  0.630\\
T-And0-08652& 00 56 58.855& 49 05 05.00&  0.335\\
T-And0-09528& 01 22 09.328& 47 14 29.86&  0.918\\
T-And0-10071& 01 14 50.412& 49 17 46.28&  0.387\\
T-And0-10206& 00 55 55.724& 49 49 46.56&  0.859\\
T-And0-10511& 01 19 16.430& 47 07 46.27&  0.563\\
T-And0-10722& 01 04 03.859& 48 37 13.04&  1.062\\
T-And0-11354& 01 18 05.168& 46 10 14.66&  0.331\\
T-And0-11476& 01 07 32.106& 45 55 44.93&  6.380\\
T-And0-11599& 01 09 28.113& 46 18 24.85&  0.280\\
T-And0-11617& 01 07 28.020& 45 22 40.35&  0.503\\
T-And0-12453& 01 17 12.316& 46 42 35.43&  0.448\\
T-And0-12769& 00 52 58.164& 44 44 11.26&  0.325\\
\enddata
\label{tableFillRoche}
\end{deluxetable}

\begin{deluxetable}{ccccccl}
\tabletypesize{\tiny}
\rotate
\tablecaption{Abnormal EBs}
\tablewidth{0pt}
\tablehead{
\colhead{Object} &
\colhead{$\alpha$ (J2000)} &
\colhead{$\delta$ (J2000)} &
\colhead{\begin{tabular}{c} Period\\ (days) \end{tabular}} &
\colhead{\begin{tabular}{c} Classified\\ in catalog?\end{tabular}} &
\colhead{\begin{tabular}{c} In SIMBAD/VSX?\\ (see table \ref{tableSIMBAD}) \end{tabular}} &
\colhead{Notes}}
\startdata
T-And0-00920& 01 17 30.677& 47 03 31.61&24.073& no & no & Large asymmetric reflection ($0.1$ mag) \ offset eclipse\\
T-And0-04594& 01 16 10.713& 48 52 18.97& 3.910& yes& no & Spots / active\\
T-And0-11476& 01 07 32.106& 45 55 44.93& 6.380& yes& no & Tilted plateaus (spots?)\\
T-Cas0-13944& 00 29 48.990& 50 49 54.06& 1.739& yes& no & Irregular eclipse depths\\
T-Cyg1-07584& 19 58 58.012& 47 38 19.26& 4.925& yes& no & Large persistent spot\\
T-Cyg1-08866& 20 08 36.448& 49 29 35.79& 2.876& yes& no & Offset eclipse\tablenotemark{a}\\
T-Dra0-00398& 16 57 33.875& 59 31 51.98& 1.046& yes& yes& Active (has $0.2$ mag fluctuations with periods of a few hours)\\
T-Dra0-03105& 16 23 02.558& 59 27 23.44& 0.485& no & yes& Unequal eclipses\tablenotemark{b} / semi-detached\\
T-Dra0-04520& 16 49 57.960& 56 26 45.56& 3.113& yes& no & Tilted plateaus (spots?)\\
T-Her0-03497& 16 52 28.391& 44 51 29.63& 7.853& yes& no & Unequal plateaus\tablenotemark{c}\\
T-Her0-08091& 16 51 52.608& 47 01 47.98& 2.694& yes& no & Offset eclipse\\
T-Lyr1-00359& 19 15 33.695& 44 37 01.30& 1.062& yes& yes& Large recurring spots ($\sim0.05$ mag)\\
T-Lyr1-02800& 19 08 18.809& 47 12 48.16& 4.876& no & no & Semi-detached / unequal plateaus (spots?)\\
T-Lyr1-05984& 18 53 50.481& 45 33 20.90& 1.470& no & no & Unequal eclipses\tablenotemark{b} / semi-detached\\
T-Lyr1-08305& 18 56 43.798& 48 07 02.86&14.081& yes& no & Large asymmetric reflection ($0.05$ mag) ; offset eclipse\\
T-Lyr1-13166& 19 02 28.120& 46 58 57.75& 0.310& no & no & Unequal plateaus ; misshapen eclipse (persistent spot?)\\
T-Lyr1-15595& 19 06 05.267& 49 04 08.95& 9.477& yes& no & Offset eclipse\\
T-Per1-00750& 03 47 45.543& 35 00 37.08& 1.929& yes& yes& Spots / active\\
T-Per1-08789& 03 54 33.282& 39 07 41.53& 2.645& yes& no & Tilted plateaus\\
T-UMa0-03090& 10 08 52.180& 52 45 52.49& 0.538& yes& yes& Unequal plateaus\\
\enddata
\tablenotetext{a}{Even when the LC plateaus are
not flat, due to tidal distortion or reflections, the system's
mirror symmetry normally guarantees that the eclipses will occur
during a plateau minimum or maximum. When, as in these cases, the
eclipses are significantly offset from the plateau minima/maxima
we can conclude that some mechanism, perhaps severe tidal lag, is
breaking the system's symmetry.}
\tablenotetext{b}{Might not be an EB. This LC could be due to
non-sinusoidal pulsations.}
\tablenotetext{c}{The two LC plateaus
between the eclipses, have a significantly different mean
magnitude. This may be due to one or both components being tidally
locked, and having a persistent spot or surface temperature
variation at specific longitudes.}
\label{tableAbnormal}
\end{deluxetable}

\begin{deluxetable}{cccccccc}
\tabletypesize{\tiny}
\rotate
\tablecaption{Ambiguous EBs (first 10)}
\tablewidth{0pt}
\tablehead{\colhead{Ver.} & \colhead{Object} & \colhead{$\alpha$ (J2000)} & \colhead{$\delta$ (J2000)} & \colhead{Period $[days]$} & \colhead{$M_1/M_\sun$} & \colhead{$M_2/M_\sun$}& \colhead{age $[Gyr]$}}
\startdata
A& T-And0-00657& 01 06 06.159& 47 31 59.37&  6.725&  2.50       (-1)& 0.74       (-1)& 0.20       (-2)\\
B& T-And0-00657& 01 06 06.159& 47 31 59.37& 13.456&  1.92       (-1)\tablenotemark{c}& 1.92       (-1)& 0.20       (-2)\\
A& T-And0-01203& 01 03 34.745& 48 32 39.27&  3.505&  1.86 $\pm$ 0.09& 0.56 $\pm$ 0.10& 0.89 $\pm$ 0.83\\
B& T-And0-01203& 01 03 34.745& 48 32 39.27&  7.011&  1.90 $\pm$ 0.12& 0.66 $\pm$ 0.19& 0.80 $\pm$ 1.13\\
A& T-And0-06017& 01 12 48.217& 49 58 07.16&  2.543&  1.40 $\pm$ 0.35& 0.52 $\pm$ 0.77& 3.49 $\pm$ 4.28\\
B& T-And0-06017& 01 12 48.217& 49 58 07.16&  5.085&  1.18 $\pm$ 0.71& 1.12 $\pm$ 0.85& 3.12 $\pm$ 11.15\\
A& T-And0-06500& 01 25 56.083& 49 23 31.74&  5.337&  0.97 $\pm$ 0.20& 0.49 $\pm$ 0.53& 7.71 $\pm$ 16.33\\
B& T-And0-06500& 01 25 56.083& 49 23 31.74& 10.674&  1.01 $\pm$ 0.30& 0.93 $\pm$ 0.45& 0.74 $\pm$ 1.60\\
A& T-And0-06680& 00 55 48.153& 45 02 48.57&  4.551&  1.16 $\pm$ 0.04& 0.51 $\pm$ 0.20& 6.09 $\pm$ 8.91\\
B& T-And0-06680& 00 55 48.153& 45 02 48.57&  9.104&  1.16 $\pm$ 0.09& 0.96 $\pm$ 0.29& 6.24 $\pm$ 10.78\\
A& T-And0-08053& 01 13 59.402& 45 51 43.43&  4.116&  1.14       (-1)& 0.40       (-2)& 6.00       (-1)\\
B& T-And0-08053& 01 13 59.402& 45 51 43.43&  8.231&  1.09 $\pm$ 0.55& 1.05 $\pm$ 0.64& 3.22 $\pm$ 16.37\\
A& T-And0-08417& 01 01 39.041& 45 03 32.98&  2.053&  1.01       (-1)& 0.47       (-1)& 10.00       (-3)\\
B& T-And0-08417& 01 01 39.041& 45 03 32.98&  4.106&  1.01       (-1)& 0.90       (-1)& 10.00       (-3)\\
A& T-And0-09365& 01 01 00.459& 45 14 24.77&  1.887&  1.05 $\pm$ 0.03& 0.43 $\pm$ 0.39& 8.74 $\pm$ 16.06\\
B& T-And0-09365& 01 01 00.459& 45 14 24.77&  3.774&  1.05 $\pm$ 0.05& 0.93 $\pm$ 0.52& 9.47 $\pm$ 23.55\\
A& T-And0-10518& 01 07 44.417& 48 44 58.11&  0.194&  0.90       (-1)& 0.40       (-2)& 0.40       (-1)\\
B& T-And0-10518& 01 07 44.417& 48 44 58.11&  0.387&  0.45 $\pm$ 0.27& 0.45 $\pm$ 0.28& 0.27 $\pm$ 0.54\\
A& T-And0-11453& 01 05 42.744& 44 54 02.26&  0.784&  1.12       (-1)& 0.40       (-2)& 7.00       (-1)\\
B& T-And0-11453& 01 05 42.744& 44 54 02.26&  1.568&  1.02 $\pm$ 0.43& 1.01 $\pm$ 0.32& 8.81 $\pm$ 14.54\\
\enddata
\tablenotetext{A}{Unequal eclipse model, assuming an unseen secondary eclipse.}
\tablenotetext{B}{Equal eclipse model, with double the period of the unequal model.}
\tablenotetext{c}{When the most likely model is at the edge of the parameter space, MECI is not
able to bound the solution, and therefore cannot estimate the uncertainties. We mark (-3) when
the upper limit was reached, (-2) when the lower limit was reached, and (-1) if one of the other
parameter is at its limit.}
\label{tableAmbig}
\end{deluxetable}

\begin{deluxetable}{cccccccccc}
\tabletypesize{\tiny}
\rotate
\tablecaption{Circular EBs (first 20)}
\tablewidth{0pt}
\tablehead{\colhead{Object} & \colhead{$\alpha$ (J2000)} & \colhead{$\delta$ (J2000)} & \colhead{Period $[days]$} & \colhead{$M_1/M_\sun$} & \colhead{$M_2/M_\sun$}& \colhead{age $[Gyr]$} &
\colhead{\begin{tabular}{c} Proper motion \\ source catalog \end{tabular}} & \colhead{\begin{tabular}{c} $PM_{\alpha}$ \\ $[MAS/year]$ \end{tabular}} & \colhead{\begin{tabular}{c} $PM_{\delta}$\\ $[MAS/year]$\end{tabular}}}
\startdata
T-And0-00194& 01 20 12.816& 48 36 41.36&  2.145&  2.07 $\pm$ 0.02& 0.97 $\pm$ 0.02& 0.59 $\pm$ 0.12& UCAC& 28.4& -12.2\\
T-And0-00459& 01 11 24.845& 46 57 49.44&  3.655&  1.20 $\pm$ 0.01& 1.19 $\pm$ 0.01& 5.35 $\pm$ 1.13& UCAC& -1.6& -20.6\\
T-And0-00745& 01 03 45.076& 44 50 41.14&  2.851&  1.86 $\pm$ 0.23& 1.02 $\pm$ 0.22& 1.04 $\pm$ 0.72& UCAC& -6.6& -4.8\\
T-And0-01461& 01 06 15.353& 45 08 25.66&  5.613&  1.47 $\pm$ 0.01& 1.45 $\pm$ 0.08& 2.76 $\pm$ 2.72& UCAC& -11.4& 2.8\\
T-And0-01554& 01 17 04.999& 45 54 06.20&  1.316&  0.90    (-1)\tablenotemark{a}   & 0.84     (-1)  & 10.00    (-3)  & UCAC& -44.6& -40.8\\
T-And0-01597& 01 10 32.071& 46 49 53.18&  3.503&  1.55 $\pm$ 0.03& 1.54 $\pm$ 0.01& 2.37 $\pm$ 0.76& UCAC& 2.9& -5.5\\
T-And0-02462& 01 18 00.594& 49 27 12.47&  3.069&  1.97 $\pm$ 0.69& 1.10 $\pm$ 1.31& 1.02 $\pm$ 1.58& UCAC& 5.8& -1.1\\
T-And0-02699& 01 06 44.813& 47 31 08.61&  1.759&  1.18 $\pm$ 0.02& 0.53 $\pm$ 0.07& 5.21 $\pm$ 3.37& UCAC& 0.2& -6.8\\
T-And0-02798& 01 21 18.345& 48 48 05.63&  2.860&  1.04 $\pm$ 0.10& 0.65 $\pm$ 0.13& 6.14 $\pm$ 9.51& UCAC& 6.3& -8.1\\
T-And0-03526& 01 20 17.451& 47 39 23.32&  1.536&  1.04 $\pm$ 0.02& 0.84 $\pm$ 0.02& 6.29 $\pm$ 2.37& UCAC& 17.9& -11.1\\
T-And0-04046& 00 55 20.157& 47 44 53.20&  3.916&  1.30 $\pm$ 0.09& 1.25 $\pm$ 0.12& 3.10 $\pm$ 4.31& UCAC& -3.8& -7.3\\
T-And0-04594& 01 16 10.713& 48 52 18.97&  3.910&  1.05    (-1)   & 0.82    (-1)   & 10.00    (-3)  & UCAC& 1.5& -1.9\\
T-And0-04829& 01 15 15.228& 47 45 58.97&  0.678&  0.99    (-1)   & 0.92    (-1)   & 10.00    (-3)  & UCAC& -23.8& 44.4\\
T-And0-05241& 00 56 34.679& 46 37 02.91&  1.454&  1.56 $\pm$ 0.01& 1.47 $\pm$ 0.31& 2.69 $\pm$ 7.01& UCAC& -4.5& -0.5\\
T-And0-05375& 01 10 58.225& 49 52 48.69&  1.640&  2.13    (-1)   & 1.85    (-2)   & 1.00    (-1)   & UCAC& -6.3& 0.1\\
T-And0-05794& 01 12 11.763& 47 32 30.94&  1.053&  2.06 $\pm$ 0.19& 1.08 $\pm$ 0.54& 1.08 $\pm$ 1.92& UCAC& -0.4& -1.4\\
T-And0-06039& 01 23 37.548& 48 25 37.73&  4.923&  1.22 $\pm$ 0.05& 1.08 $\pm$ 0.31& 5.33 $\pm$ 7.17& UCAC& -2.5& -5.0\\
T-And0-06340& 01 01 55.269& 49 18 38.23&  5.437&  1.33    (-1)   & 0.40    (-2)   & 4.00    (-1)   & UCAC& 0.3& -2.9\\
T-And0-06538& 01 20 58.907& 49 29 08.89& 18.669&  1.33 $\pm$ 0.15& 0.97 $\pm$ 0.17& 3.38 $\pm$ 3.45& UCAC& 1.1& -6.8\\
T-And0-06632& 01 22 36.840& 47 52 53.29&  1.669&  1.69 $\pm$ 0.01& 1.45 $\pm$ 0.24& 2.21 $\pm$ 1.02& UCAC& -7.2& -7.6\\
\enddata
\label{tableCircular}
\tablenotetext{a}{When the most likely model is at the edge of the parameter space, MECI is not
able to bound the solution, and therefore cannot estimate the uncertainties. We mark (-3) when
the upper limit was reached, (-2) when the lower limit was reached, and (-1) if one of the other
parameter is at its limit.}
\end{deluxetable}

\begin{deluxetable}{cccc}
\tabletypesize{\tiny}
\rotate
\tablecaption{Inverted EBs}
\tablewidth{0pt}
\tablehead{\colhead{Object} & \colhead{$\alpha$ (J2000)} & \colhead{$\delta$ (J2000)} & \colhead{Period $[days]$}}
\startdata
T-And0-13653& 00 59 57.881& 45 03 41.53&  3.342\\
T-Cas0-02069& 00 49 17.959& 50 39 02.92&  2.830\\
T-Cas0-03012& 00 45 41.832& 51 01 35.40&  1.108\\
T-Cas0-04618& 00 46 22.661& 50 39 17.57&  2.798\\
T-Cas0-07780& 00 34 18.779& 52 00 35.72&  1.852\\
T-Cas0-19045& 00 21 44.707& 50 32 29.55&  0.785\\
T-Cas0-19668& 00 48 01.342& 47 06 11.58&  1.848\\
T-Cas0-21651& 00 26 34.895& 46 38 42.69&  1.155\\
T-Cyg1-01956& 19 53 29.106& 47 48 49.86&  2.045\\
T-Cyg1-02929& 20 11 57.009& 48 07 03.59&  4.263\\
T-Cyg1-17342& 19 49 54.197& 50 53 28.08&  2.220\\
T-Her0-05469& 16 54 51.245& 43 20 35.89&  0.899\\
T-Lyr1-04431& 19 12 16.047& 49 42 23.58&  0.903\\
T-Lyr1-05887& 18 52 10.489& 47 48 16.67&  1.802\\
T-Lyr1-07179& 18 49 14.039& 45 24 38.61&  1.323\\
T-Lyr1-10989& 19 06 22.791& 45 41 53.82&  2.015\\
T-Lyr1-11067& 18 52 53.489& 47 51 26.58&  2.241\\
T-Per1-04353& 03 45 04.887& 37 47 15.91&  2.953\\
T-Per1-06993& 03 40 59.668& 39 12 35.90&  2.125\\
T-Per1-09366& 03 49 20.305& 39 55 41.97&  2.374\\
T-Per1-12217& 03 28 59.454& 37 37 42.14&  1.690\\
T-Tau0-00686& 04 07 13.870& 29 18 32.44&  5.361\\
T-UMa0-00127& 09 38 06.716& 56 01 07.32&  0.687\\
\enddata
\label{tableInverted}
\end{deluxetable}

\begin{deluxetable}{cccccll}
\tabletypesize{\tiny}
\rotate
\tablecaption{EBs that appear in either the VSX or the SIMBAD astronomical databases}
\tablewidth{0pt}
\tablehead{\colhead{Category} & \colhead{Object} & \colhead{$\alpha$ (J2000)} & \colhead{$\delta$ (J2000)} & \colhead{Spectral type} & \colhead{Classification} & \colhead{Identifiers}}
\startdata
Circular  & T-And0-00194& 01 20 12.816& 48 36 41.36& A5  & Star                  & BD+47 378 ; GSC 03269-00662 ; SAO 37126 ; AG+48 143\\
          &             &             &            &     &                       & PPM 43886 ; TYC 3269-662-1\\
Circular  & T-And0-00459& 01 11 24.845& 46 57 49.44& F8  & EB of Algol type      & CO And ; GSC 03268-00398 ; TYC 3268-398-1 ; BD+46 281 ; BV 74\\
Ambiguous & T-And0-00657& 01 06 06.159& 47 31 59.37& K0  & Star                  & BD+46 254 ; GSC 03267-01349 ; TYC 3267-1349-1\\
          &             &             &            &     &                       & AG+47 120 ; PPM 43637\\
Circular  & T-And0-00745& 01 03 45.076& 44 50 41.14&     & Star                  & TYC 2811-470-1 ; GSC 02811-00470\\
Ambiguous & T-And0-01203& 01 03 34.745& 48 32 39.27&     & Star                  & TYC 3267-1176-1 ; GSC 03267-01176\\
Circular  & T-And0-04046& 00 55 20.157& 47 44 53.20&     & Star                  & GPM 13.833991+47.748193\\
Roche-fill& T-And0-05153& 01 18 48.278& 49 39 36.86&     & EB of W UMa type      & QW And\\
Roche-fill& T-And0-05343& 00 52 55.122& 48 01 37.68&     & Star                  & GPM 13.232700+48.019757\\
Roche-fill& T-And0-07892& 00 56 15.567& 48 39 10.73&     & EB                    & NSVS 3757820\\
Circular  & T-And0-23792& 00 54 09.254& 47 45 19.91&     & Star                  & GPM 13.538629+47.755510\\
Roche-fill& T-Cas0-00170& 00 53 37.847& 48 43 33.83&     & Star                  & TYC 3266-195-1 ; GSC 03266-00195\\
Eccentric & T-Cas0-00394& 00 32 51.608& 49 19 39.36& B3  & EB of $\beta$ Lyr type& V381 Cas ; BD+48 162 ; BV 179\\
Roche-fill& T-Cas0-00430& 00 40 06.247& 50 14 15.64& K4  & EB of W UMa type      & V523 Cas ; GSC 03257-00167 ; WR 16 ; CSV 5867\\
          &             &             &            &     &                       & 1RXS J004005.0+501414 ; TYC 3257-167-1\\
Circular  & T-Cas0-00640& 00 47 06.277& 48 31 13.14&     & Star                  & TYC 3266-765-1 ; GSC 03266-00765\\
Circular  & T-Cas0-00792& 00 48 26.554& 51 35 02.52&     & Star                  & TYC 3274-664-1 ; GSC 03274-00664\\
Roche-fill& T-Cas0-02013& 00 40 46.427& 46 56 57.41&     & Star                  & TYC 3253-1767-1 ; GSC 03253-01767\\
Inverted  & T-Cas0-02069& 00 49 17.959& 50 39 02.92&     & EB                    & V385 Cas\\
Roche-fill& T-Cas0-08802& 00 51 32.351& 47 16 42.57&     & Star                  & GPM 12.884787+47.278540\\
Roche-fill& T-CrB0-00654& 16 00 14.507& 35 12 31.56&     & EB of W UMa type      & AS CrB ; GSC 02579-01125 ; NSVS 7847829\\
          &             &             &            &     &                       & ROTSE1 J160014.54+351228.4\\
Roche-fill& T-CrB0-00705& 15 55 51.838& 33 11 00.39&     & EB of W UMa type      & ROTSE1 J155551.87+331100.5\\
Roche-fill& T-CrB0-01589& 16 10 09.313& 35 57 30.57&     & Variable of $\delta$ Sct type& ROTSE1 J161009.33+355730.8\\
Roche-fill& T-CrB0-01605& 16 00 58.472& 34 18 54.34&     &EB of W UMa or RR Lyr-C& NSVS 7848126 ; ROTSE1 J160058.45+341854.5\\
Roche-fill& T-CrB0-04254& 16 09 19.589& 35 32 11.48&     & EB of W UMa type      & ROTSE1 J160919.62+353210.8\\
Circular  & T-Cyg1-00246& 19 44 01.777& 50 13 57.42&     & Star                  & TYC 3565-643-1 ; GSC 03565-00643\\
Roche-fill& T-Cyg1-00402& 19 54 39.939& 50 36 41.91&     & Star                  & TYC 3566-606-1 ; GSC 03566-00606\\
Ambiguous & T-Cyg1-01385& 20 15 21.936& 48 17 14.14&     & Star                  & TYC 3576-2035-1 ; GSC 03576-02035\\
Circular  & T-Cyg1-01627& 19 45 20.426& 51 35 07.22&     & Star                  & TYC 3569-1752-1 ; GSC 03569-01752\\
Roche-fill& T-Cyg1-04652& 20 07 07.305& 50 34 01.34&     &EB of W UMa type       & GSC 03567-01035\\
Roche-fill& T-Cyg1-04852& 19 51 59.208& 50 05 29.61&     &EB of W UMa type       & NSVS 5645908\\
Circular  & T-Cyg1-09274& 20 16 06.814& 51 56 26.07&     & EB of W UMa type      & V1189 Cyg ; CSV 8488 ; GSC 03584-01600 ; SON 7885\\
Roche-fill& T-Cyg1-11279& 19 59 53.377& 49 23 27.86&     & X-ray source          & 1RXS J195954.0+492318\\
Roche-fill& T-Cyg1-12518& 19 58 15.339& 48 32 15.79&     & Variable star         & Mis V1132\\
Roche-fill& T-Cyg1-14514& 19 48 05.077& 52 51 16.25&     &EB of W UMa or RR Lyr-C& V997 Cyg ; GSC 03935-02233 ; ROTSE1 J194804.79+525117.6 ; SON 7839\\
Ambiguous & T-Dra0-00240& 17 03 52.919& 57 21 55.54&     & Star                  & TYC 3894-898-1  ; GSC 03894-00898\\
Ambiguous & T-Dra0-00358& 16 45 38.339& 54 31 32.02&     & Star                  & TYC 3879-2689-1 ; GSC 03879-02689\\
Circular  & T-Dra0-00398& 16 57 33.875& 59 31 51.98&     & EB of Algol type/X-ray source& RX J1657.5+5931 ; 1RXS J165733.5+593156\\
          &             &             &            &     &                       & VSX J165733.8+593151 ; GSC 03898-00272\\
Roche-fill& T-Dra0-00405& 16 27 49.103& 58 50 23.30&     & Star                  & TYC 3884-1488-1 ; GSC 03884-01488\\
Roche-fill& T-Dra0-00959& 16 27 44.159& 56 45 59.30&     & EB of W UMa type/X-ray source& NSVS 2827877 ; 1RXS J162743.9+564557\\
Circular  & T-Dra0-01363& 16 34 20.417& 57 09 48.95&M4.5V& EB of BY Dra type     & CM Dra ; CSI+57-16335 1 ; LSPM J1634+5709 ; G 225-67 ; G 226-16\\
          &             &             &            &     &High proper-motion Star& IDS 16326+5721 A ; [RHG95] 2616 ; SBC7 580 ; CCDM J16343+5710A\\
          &             &             &            &     &                       & GJ 630.1 A ; LP 101-15 ; IDS 16325+5721 A ; [GKL99] 324 ; LHS 421\\
          &             &             &            &     &                       & 2MASS J16342040+5709439 ; CCABS 108 ; CABS 134 ; GEN\# +9.80225067\\
          &             &             &            &     &                       & RX J1634.3+5709 ; 1RXH J163421.2+570941 ; 1RXS J163421.2+570933\\
          &             &             &            &     &                       & PM 16335+5715 ; USNO 168 ; USNO-B1.0 1471-00307615 ; NLTT 43148\\
Roche-fill& T-Dra0-01346& 16 52 12.345& 57 43 31.70&     & EB of Algol type      & BPS BS 16080-0095 ; VSX J165212.3+574331 ; GSC 03885-00583\\
Roche-fill& T-Dra0-02224& 16 30 01.408& 54 45 55.80&     & Star                  & BPS BS 16084-0159\\
Circular  & T-Dra0-03021& 17 01 03.618& 55 14 54.70&     & EB of Algol type      & VSX J170103.5+551455 ; GSC 03890-01216 \\
Abnormal  & T-Dra0-03105& 16 23 02.558& 59 27 23.44&     & X-ray source          & 1RXS J162303.6+592717\\
Roche-fill& T-Dra0-05259& 16 41 48.751& 56 22 34.40&     & EB of W UMa type      & VSX J164148.7+562234 ; GSC 03882-02264 ; USNO-B1.0 1463-0278621\\
Ambiguous & T-Her0-00274& 17 00 51.150& 45 25 35.94&     & Star                  & TYC 3501-2245-1 ; GSC 03501-02245\\
Roche-fill& T-Her0-01086& 16 48 15.539& 44 44 28.73&     & EB of W UMa type      & GSC 03082-00896 ; NSVS 5252572 ; 1RXS J164817.3+444430\\
Roche-fill& T-Her0-03579& 16 35 47.390& 45 24 58.19&     & EB of W UMa type      & GSC 03499-01631\\
Inverted  & T-Her0-05469& 16 54 51.245& 43 20 35.89&     & EB                    & V747 Her ; SVS 2066\\
Circular  & T-Lyr1-00359& 19 15 33.695& 44 37 01.30& G0V & EB                    & V2277 Cyg ; GSC 03133-01149 ; ROTSE1 J191533.92+443704.9\\
          &             &             &            &     & X-ray source          & BD+44 3087 ; ILF1+44 155 ; 1RXS J191533.7+443704\\
Circular  & T-Lyr1-00687& 18 55 27.911& 47 13 41.76&     & Star                  & TYC 3544-1392-1 ; GSC 03544-01392\\
Circular  & T-Lyr1-01013& 18 55 03.963& 47 49 08.39&     & Star                  & TYC 3544-2565-1 ; GSC 03544-02565\\
Circular  & T-Lyr1-01439& 19 06 13.439& 46 57 26.42&     & Star                  & TYC 3545-2716-1 ; GSC 03545-02716\\
Circular  & T-Lyr1-02109& 18 57 35.415& 45 07 44.10&     & Cepheid variable star & ROTSE1 J185735.99+450752.5\\
Roche-fill& T-Lyr1-02166& 19 05 07.448& 46 15 07.51&     & X-ray source          & 1RXS J190504.8+461512\\
Roche-fill& T-Lyr1-03173& 18 59 45.531& 47 20 07.34&     & EB of W UMa type      & ROTSE1 J185945.43+472007.0\\
Roche-fill& T-Lyr1-03211& 18 45 56.939& 47 19 09.54&     & EB of W UMa type/X-ray source& ROTSE1 J184556.86+471914.4 ; 1RXS J184557.9+471906\\
Roche-fill& T-Lyr1-03270& 18 57 33.098& 48 05 22.49&     & EB of W UMa type      & ROTSE1 J185733.12+480522.5\\
Roche-fill& T-Lyr1-03783& 18 50 12.684& 45 35 44.05&     & Star                  & GPM 282.552858+45.595521\\
Inverted  & T-Lyr1-04431& 19 12 16.047& 49 42 23.58&     & EB of Algol type      & NSV 11822 ; GSC 03550-01770 ; NSVS 5578839 ; SON 9371\\
Roche-fill& T-Lyr1-05706& 18 47 57.211& 44 38 11.30&     & EB of W UMa type      & ROTSE1 J184757.18+443810.8\\
Inverted  & T-Lyr1-05887& 18 52 10.489& 47 48 16.67&     & EB of Algol type      & WX Dra ; AN 24.1925\\
Roche-fill& T-Lyr1-06583& 18 52 26.837& 44 55 20.86&     & EB                    & ROTSE1 J185226.53+445527.8\\
Inverted  & T-Lyr1-07179& 18 49 14.039& 45 24 38.61&     & Star                  & GPM 282.308454+45.410868\\
Roche-fill& T-Lyr1-08406& 18 50 06.942& 45 41 05.95&     & Star                  & GPM 282.528833+45.685035\\
Roche-fill& T-Lyr1-10276& 18 46 55.088& 45 00 52.27&     & EB of W UMa type      & V596 Lyr ; GPM 281.729421+45.014635 ; GSC 03540-00085\\
          &             &             &            &     &                       & ROTSE1 J184654.98+450054.7\\
Inverted  & T-Lyr1-10989& 19 06 22.791& 45 41 53.82&     & EB of Algol type      & V512 Lyr ; SON 10931\\
Roche-fill& T-Lyr1-11226& 18 45 21.748& 45 53 28.79&     & EB of W UMa type or $\delta$ Sct & V594 Lyr ; GPM 281.340617+45.891326 ; GSC 03540-01842\\
          &             &             &            &     &                       & ROTSE1 J184522.47+455321.0\\
Roche-fill& T-Lyr1-12772& 18 52 25.096& 44 55 40.23&     & EB of W UMa type      & ROTSE1 J185226.53+445527.8\\
Abnormal  & T-Lyr1-13166& 19 02 28.120& 46 58 57.75& F9V & EB                    & V361 Lyr ; SON 9349\\
Roche-fill& T-Per1-00328& 03 41 57.108& 39 07 29.60& G5  & EB of Algol type      & HD 275743 ; BD+38 787 ; GSC 02863-00755 ; TYC 2863-755-1\\
Circular  & T-Per1-00459& 03 34 57.745& 39 33 18.70& G5  & Star                  & HD 275547 ; GSC 02866-01995 ; TYC 2866-1995-1\\
Circular  & T-Per1-00750& 03 47 45.543& 35 00 37.08&     &Double or multiple star& TYC 2364-2327-1 ; GSC 02364-02327 ; CCDM J03478+3501BC\\
          &             &             &            &     &                       & ADS 2771 BC ; BD+34 732B ; CSI+34 732 2 ; NSV 1302\\
Roche-fill& T-Per1-00974& 03 34 43.738& 38 40 22.22&  A  & Star                  & HD 275481\\
Circular  & T-Per1-01218& 03 42 33.165& 39 06 03.63&  A  & EB                    & HU Per ; HD 275742 ; SVS 922\\
Roche-fill& T-Per1-01482& 03 48 45.999& 35 14 10.05& F0  & Star                  & HD 279025\\
Circular  & T-Per1-02597& 03 44 32.202& 39 59 34.94& K4V & T Tau type Star       & [LH98] 94 ; 1RXS J034432.1+395937 ; 1SWASP J034433.95+395948.0\\
Inverted  & T-Per1-04353& 03 45 04.887& 37 47 15.91&     & EB of Algol type      & HV Per ; SVS 368 ; P 107\\
Roche-fill& T-Tau0-00397& 04 30 09.466& 25 32 27.05& A3  & EB of $\beta$ Lyr type& GW Tau ; SVS 1421 ; HD 283709 ; ASAS 043009+2532.4\\
Inverted  & T-Tau0-00686& 04 07 13.870& 29 18 32.44&     & EB of Algol type      & IL Tau ; SON 9543\\
Roche-fill& T-Tau0-00781& 04 12 51.218& 24 41 44.26& G9  &Eruptive/T Tau-type Star& V1198 Tau ; NPM2+24.0013 ; 1RXS J041250.9+244201\\
          &             &             &            &     &                       & GSC 01819-00498 ; RX J0412.8+2442 ; [WKS96] 14\\
Roche-fill& T-Tau0-01262& 04 16 28.109& 28 07 35.81& K7V &Variable Star of Orion Type& V1068 Tau ; EM StHA 25 ; JH 165 ; EM LkCa 4\\
          &             &             &            &     &                       & HBC 370 ; ASAS 041628+2807.6\\
Roche-fill& T-Tau0-01715& 04 19 26.260& 28 26 14.30& K7V &T Tau-type Star/X-ray source& V819 Tau ; HBC 378 ; NAME WK X-Ray 1 ; 1E 0416.3+2830\\
          &             &             &            &     &                       & IRAS C04162+2819 ; TAP 27 ; [MWF83] P1 ; WK81 1\\
          &             &             &            &     &                       & 1RXS J041926.1+282612 ; X 04163+283\\
Roche-fill& T-Tau0-06463& 04 07 27.415& 27 51 06.36&     & EB of W UMa type      & V1022 Tau ; HV 6199 ; NSV 1464\\
Inverted  & T-UMa0-00127& 09 38 06.716& 56 01 07.32& A2V & EB of Algol type      & VV UMa ; GEN\# +0.05601395 ; HIP 47279 ; TYC 3810-1290-1\\
          &             &             &            &     &                       & GSC 03810-01290 ; SBC7 384 ; GCRV 6211 ; BD+56 1395\\
          &             &             &            &     &                       & HIC 47279 ; SVS 770 ; AAVSO 0931+56\\
Circular  & T-UMa0-00222& 10 07 18.023& 56 12 37.12& A0  & Star                  & HD 237866 ; GSC 03818-00504 ; SAO 27524 ; AG+56 778 ; HIC 49581\\
          &             &             &            &     &                       & BD+56 1432 ; HIP 49581 ; YZ 56 6209 ; TYC 3818-504-1\\
Roche-fill& T-UMa0-01701& 10 03 02.856& 55 47 53.34&     & X-ray source          & RX J100303.4+554752 ; [PTV98] H22 ; [PTV98] P29\\
Circular  & T-UMa0-03090& 10 08 52.180& 52 45 52.49& K2e & Star                  & GSC 03815-01151 ; RIXOS 229-302 ; RX J100851.6+524553\\
Roche-fill& T-UMa0-03108& 10 04 16.780& 54 12 02.83&     & EB of W UMa type      & NSVS 2532137\\
\enddata
\label{tableSIMBAD}
\end{deluxetable}

\begin{deluxetable}{lcccccccccc}
\tabletypesize{\tiny}
\rotate
\tablecaption{Low-mass EB candidates  ($M_{1,2} < 0.75M_\sun$ ; sorted by mass)}
\tablewidth{0pt}
\tablehead{\colhead{Category} & \colhead{Object} & \colhead{$\alpha$ (J2000)} & \colhead{$\delta$ (J2000)} & \colhead{Period $[days]$} & \colhead{$M_1/M_\sun$} & \colhead{$M_2/M_\sun$} & \colhead{age $[Gyr]$} &
 \colhead{\begin{tabular}{c} Proper motion \\ source catalog\tablenotemark{a} \end{tabular}} & \colhead{\begin{tabular}{c} $PM_{\alpha}$ \\ mas/year \end{tabular}} & \colhead{\begin{tabular}{c} $PM_{\delta}$ \\ mas/year \end{tabular}}}
\startdata
Circular& T-Dra0-01363\tablenotemark{b}& 16 34 20.417& 57 09 48.95&  1.268&   0.27 $\pm$ 0.02&  0.24 $\pm$ 0.03&  1.6 $\pm$ 1.6  & \citet{Salim03} & -1121& 1186\\
AmbigEq\tablenotemark{c}&  T-And0-10518& 01 07 44.417& 48 44 58.11&  0.387&   0.45 $\pm$ 0.27&  0.45 $\pm$ 0.28&  0.3 $\pm$ 0.5  & UCAC  &  2.7& -2.0\\
AmbigEq&  T-Cyg1-12664& 19 51 39.824& 48 19 55.38&  8.257&   0.50 $\pm$ 0.20&  0.48 $\pm$ 0.19&  0.3 $\pm$ 0.4  & USNO-B&  -18&  -6\\
AmbigEq&  T-CrB0-14232& 16 10 22.495& 33 57 52.33&  0.971&   0.60 $\pm$ 0.24&  0.55 $\pm$ 0.29&  4.4 $\pm$ 8.8  & UCAC  &-15.2& -24.2\\
AmbigEq&  T-CrB0-14543& 15 57 45.926& 33 56 07.28&  1.506&   0.60 (-1)\tablenotemark{d}      &  0.60 (-1)      &  0.2 (-2)       & UCAC&  -13.9&  13.3\\
Circular& T-Per1-13685& 03 53 51.217& 37 03 16.73&  0.384&   0.60 (-1)      &  0.50 (-1)      &  10.0 (-3)      & UCAC  &-24.1&-15.9\\
AmbigEq&  T-CrB0-10759& 15 52 18.455& 30 35 32.13&  1.901&   0.63 $\pm$ 0.24&  0.62 $\pm$ 0.21&  7.3 $\pm$ 49.6 & UCAC  &  3.6&-19.4\\
AmbigEq&  T-UMa0-08238& 10 09 25.384& 53 57 01.31&  1.250&   0.69 $\pm$ 0.54&  0.61 $\pm$ 0.51&  4.1 $\pm$ 15.0 & USNO-B&    6&   -4\\
AmbigEq&  T-Cas0-10450& 00 29 16.288& 50 27 38.58&  8.656&   0.71 $\pm$ 0.21&  0.67 $\pm$ 0.20&  0.3 $\pm$ 0.4  & UCAC  & -3.1& -4.2\\
AmbigEq&  T-Dra0-07116& 17 02 53.025& 55 07 47.44&  1.369&   0.71 $\pm$ 0.22&  0.69 $\pm$ 0.22&  2.1 $\pm$ 3.6  & USNO-B&   -2&  -16\\
Circular& T-Tau0-04859& 04 08 11.608& 24 51 10.18&  3.068&   0.74 $\pm$ 0.10&  0.66 $\pm$ 0.10&  8.8 $\pm$ 14.8 & UCAC  &  3.4& -8.0\\
\enddata
\tablenotetext{a}{Where possible, we used the more accurate UCAC
catalog, otherwise we reverted to the USNO-B catalog. Since they
are dim and nearby, we expect most of the low-mass binaries to
have comparably large proper motions.}
\tablenotetext{b}{This binary is CM Draconis, which has been extensively
studied and found to have a masses of $M_1 = 0.2307 \pm 0.0010 M_\sun$ and
$M_2 = 0.2136 \pm 0.0010 M_\sun$ \citep{Lacy77a, Metcalfe96}. For consistency,
we listed the MECI results, which are off by less than $0.04 M_\sun$
($\sim1.5\sigma$). We also adopted an alternate proper motion
estimate, as its USNO-B values seems to be erroneous, probably due
to its very high angular velocity.}
\tablenotetext{c}{For clarity
we list for the ambiguous systems, only the solution with
approximately equal components. But it is likely that at least a
few of the ambiguous systems may be unequal, with half the period.
Such cases can be identified as single-line spectroscopic
binaries, with the secondary component being no larger than a few
$0.1M_\sun$.}
\tablenotetext{d}{When the most likely model is at
the edge of the parameter space, MECI is not able to bound the
solution, and therefore cannot estimate the uncertainties. We mark
(-3) when the upper limit was reached, (-2) when the lower limit
was reached, and (-1) if one of the other parameter is at its
limit.}
\label{tableLowMass}
\end{deluxetable}

\clearpage

\clearpage

\clearpage

\clearpage

\end{document}